\documentstyle[12pt,epsf]{article}
\def\half{\frac{1}{2}}

\def\eq{\begin{equation}}
\def\eqe{\end{equation}}
\def\eqa{\begin{eqnarray}}
\def\eqae{\end{eqnarray}}

\def\be{\begin{equation}}
\def\ee{\end{equation}}
\def\bea{\begin{eqnarray}}
\def\ena{\end{eqnarray}}

\def\to{\rightarrow}

\def\be{\begin{equation}}
\def\ee{\end{equation}}
\def\bea{\begin{eqnarray}}
\def\ena{\end{eqnarray}}

\def\a{\alpha}
\def\b{\beta}

\def\d{\delta}
\def\e{\epsilon}           
\def\f{\phi}               
\def\g{\gamma}
\def\h{\eta}

\def\j{\psi}
\def\k{\kappa}                    
\def\l{\lambda}
\def\m{\mu}
\def\n{\nu}
\def\o{\omega}
\def\p{\pi}                
\def\q{\theta}                    
\def\r{\rho}                      
\def\s{\sigma}                    
\def\t{\tau}

\def\x{\xi}
\def\z{\zeta}
\def\D{\Delta}

\def\G{\Gamma}

\def\L{\Lambda}

\def\P{\Pi}

\def\del{\partial}

\def\cf{{\cal F}}

\def\cl{{\cal L}}

\def\co{{\cal O}}

\def\cs{{\cal S}}

\catcode`@=11

\def\un#1{\relax\ifmmode\@@underline#1\else
$\@@underline{\hbox{#1}}$\relax\fi}

{\catcode`\'=\active \gdef'{{}^\bgroup\prim@s}}

\def\magstep#1{\ifcase#1 \@m\or 1200\or 1440\or 1728\or 2074\or 2488\or
       2986\fi\relax}   

\catcode`@=12

\def\bop#1{\setbox0=\hbox{$#1M$}\mkern1.5mu
	\vbox{\hrule height0pt depth.04\ht0
	\hbox{\vrule width.04\ht0 height.9\ht0 \kern.9\ht0
	\vrule width.04\ht0}\hrule height.04\ht0}\mkern1.5mu}

\def\sl#1{\rlap{\hbox{$\mskip 1 mu /$}}#1}
\def\leftrightarrowfill{$\mathsurround=0pt \mathord\leftarrow \mkern-6mu
       \cleaders\hbox{$\mkern-2mu \mathord- \mkern-2mu$}\hfill
       \mkern-6mu \mathord\rightarrow$}
\def\dvec#1{\vbox{\ialign{##\crcr
       \leftrightarrowfill\crcr\noalign{\kern-1pt\nointerlineskip}
       $\hfil\displaystyle{#1}\hfil$\crcr}}}          
\def\hook#1{{\vrule height#1pt width0.4pt depth0pt}}
\def\leftrighthookfill#1{$\mathsurround=0pt \mathord\hook#1
       \hrulefill\mathord\hook#1$}
\def\underhook#1{\vtop{\ialign{##\crcr                 
       $\hfil\displaystyle{#1}\hfil$\crcr
       \noalign{\kern-1pt\nointerlineskip\vskip2pt}
       \leftrighthookfill5\crcr}}}
\def\smallunderhook#1{\vtop{\ialign{##\crcr      
       $\hfil\scriptstyle{#1}\hfil$\crcr
       \noalign{\kern-1pt\nointerlineskip\vskip2pt}
       \leftrighthookfill3\crcr}}}

\def\sfrac#1#2{{\vphantom1\smash{\lower.5ex\hbox{\small$#1$}}\over
       \vphantom1\smash{\raise.4ex\hbox{\small$#2$}}}} 
\def\bfrac#1#2{{\vphantom1\smash{\lower.5ex\hbox{$#1$}}\over
       \vphantom1\smash{\raise.3ex\hbox{$#2$}}}}      
\def\afrac#1#2{{\vphantom1\smash{\lower.5ex\hbox{$#1$}}\over#2}}  
\def\on#1#2{{\buildrel{\mkern2.5mu#1\mkern-2.5mu}\over{#2}}}
\def\ddt#1{\on{\hbox{\LARGE .\kern-2pt.}}#1}             
\def\tdt#1{\on{\hbox{\LARGE .\kern-2pt.\kern-2pt.}}#1}   

\def\boxes#1{
       \newcount\num
       \num=1
       \newdimen\downsy
       \downsy=-1.5ex
       \mskip-2.8mu
       \bo
       \loop
       \ifnum\num<#1
       \llap{\raise\num\downsy\hbox{$\bo$}}
       \advance\num by1
       \repeat}
\def\boxup#1#2{\newcount\numup
       \numup=#1
       \advance\numup by-1
       \newdimen\upsy
       \upsy=.75ex
       \mskip2.8mu
       \raise\numup\upsy\hbox{$#2$}}

\newskip\humongous \humongous=0pt plus 1000pt minus 1000pt

\newif\ifdtup

\parskip=\medskipamount  
\thicklines              
\begin{document}

\begin{flushright}
ITP-SB-96-4\\
January, 1996\\
hep-ph/9606312
\end{flushright}
\vbox{\vskip 0.75 true in}

\centerline{\large \bf PARTONS, FACTORIZATION AND RESUMMATION}
\centerline{\large \bf TASI 95\ \footnote{Based on 
seven lectures at the Theoretical Advanced Study Institute,
{\it QCD and Beyond},
Boulder, Colorado, June 1995.}}
\vbox{\vskip 0.25 true in}

\centerline{\large  GEORGE STERMAN}
\vbox{\vskip 0.25 true in}

\centerline{Institute for Theoretical Physics} 
\centerline{State University of New York at Stony Brook}
\centerline{Stony Brook, NY 11794-3840, USA}
\vbox{\vskip 2.0 true in}

\begin{abstract}
I review the treatment of high-energy QCD in Minkowski
space, with an emphasis on factorization theorems
as extensions of the operator product expansion.
I discuss how the factorization properties
of high-energy cross sections and amplitudes 
lead to evolution equations that 
resum large logarithms for two-scale problems.
\end{abstract}

\newpage

\tableofcontents

\section{Introduction:  Fundamental Issues and Problems}

Many of the fundamental properties of field theory,
such as the operator product expansion, are
best developed in Euclidean space.  The very
nature of space-time, however, dictates 
that an understanding of field theory directly in
Minkowski space is also indispensable.  I hope these
lectures will arm the reader with general
insights into how  perturbative quantum chromodynamics (QCD)
manifests itself
at high energy
in Minkowski space.
I will discuss the calculational challenges that 
result from
singularities on the light cone, and develop
some of the methods that we currently possess to deal with them.

The unifying thread that runs through these lectures is 
 factorization \cite{jccrv}, the systematic separation of dynamics 
associated with short and long distance scales \cite{Wilsonope,KogutWilson}. 
 This is a recurring theme
in modern theoretical physics, which can also  be found 
as a central idea, sometimes under
the names effective actions or field theories,
 in other lecture series in this school. 
One aim of these lectures is to identify
quantities in QCD that are genuinely short-distance
dominated, which will lead us to the concept of
infrared safety, and the ubiquity of jet cross sections \cite{StermanWeinberg}.
Another is to show
 how one of the great phenomenological successes of high energy
physics, the parton model \cite{partonmodel1,partonmodel2}, emerges as a consequence of
the factorization properties of QCD in Minkowski space,
which lead as well to its systematic improvement,
including evolution \cite{dglap}.  

Resummation refers below to the summation of
enhancements (usually logarithmic) 
in ratios of kinematic variables, such as energy
and momentum transfer,
to all orders in field-theoretic
perturbation theory.  Rather than try to review all
recent progress in this large and growing field, I
will emphasize two representative and 
classic examples,
Sudakov \cite{collinsrv} resummation in ${\rm e}^+{\rm e}^-$ annihilation 
and BFKL \cite{bfkl} resummation in deeply inelastic
scattering, stressing how they 
may be regarded as consequences of the underlying factorization 
properties of field theory.

These lectures are concerned primarily with the theoretical
foundation of perturbative QCD.
Of necessity, much introductory material involves quite
general properties of quantum field theory.
Much of this material may be found in field theory textbooks \cite{qfttexts,Stermanbook},
as well as in the books by Yndur\'ain and Muta on QCD \cite{MutaYndurain}.
although some developments are given below which
may be less widely familiar \cite{Stermanbook,Sterman78}.
In particular, I have tried to outline
a general analysis of long-distance behavior in
perturbation theory, based upon the analytic structure
of Feynman diagrams \cite{Edenetal}
and on an infrared power-counting procedure. 
The reader will find many points of contact, but many differences
in emphasis, with my TASI 91 lectures \cite{StermanTASI91}.

Many of the important and innovative 
calculational techniques and the phenomenological
analyses that realize the program described here, as
well as other equally important aspects of the theory, are treated in
other lectures at this school.  Very useful recent reviews that  
treat these topics in a more directly phenomenological
manner include the TASI 94 lectures of Ellis \cite{EllisTASI}, and
a ``Handbook" by the CTEQ Collaboration \cite{cteqhb}.   
As a review
of perturbative QCD, the collection of monographs
edited by Mueller \cite{edMueller} is extermely helpful, and
introduces and
reviews subjects, particularly elastic scattering \cite{elastic}
and QCD coherence \cite{cohere,Ciaf}, that are closely connected to
the discussion in these lectures.
Finally, for a theoretical introduction from a complementary point of view,
see the book of 
Dokshitser, Khoze, Mueller and Troian \cite{basicspQCD}.
 
So much of the terminology of QCD is intertwined with
its sources in experiment and theory, that 
it seems appropriate to begin with a very brief review of
the strong interaction physics
that led to QCD.  This will be followed by an introduction
to deeply inelastic scattering, the pivotal 
experiments whose outcome made it possible to identify
quantum chromodynamics as a promising theory of the strong interactions
over twenty years ago.   The balance of these lectures will
discuss how this promise has been realized.

\subsection{A Prehistory of QCD}

It is plausible to identify the birth of strong interaction
physics with the discovery of the neutron, which along with
its beta decay, signaled the existence 
of two new interactions,
the strong and weak.  From this slender thread, physicists
suspended the four-fermion theory of weak 
interactions, and the concepts of
(strong) isospin, the S-matrix, and the Yukawa theory,
in which the strong interaction is mediated by the 
exchange of scalar
particles, now identified as pions, between nucleons.  Beyond this, 
through the thirties and forties into
the fifties, strong interaction
physics remained for the most part nuclear physics, 
with many well-known successes and consequences \cite{Pais}.

With the discovery of the pion and then other mesons, around 1950, and
the development of high energy accelerators, the
substructure of nucleons became for the first time
an object of study.  The earliest information
was on the excited states of mesons and baryons, and
the discovery of the $\D$ in pion-nucleon scattering ushered in a 
bewildering array of resonances.  Order was imposed on this
chaos through symmetry principles, leading eventually to
the concept of quarks as the building blocks of hadrons;
three quarks for baryons and a quark-antiquark pair for mesons.
This ``constituent" quark model still successfully 
describes most of the
qualitative features of baryon spectroscopy \cite{ChengLi,HalzenMartin}.  It was thus
natural to try and ``see"
the quarks in experiments with momentum transfers large
enough to resolve the internal structure of the nucleon,
and to explore the possibility that forces
between quarks are mediated by a field, dubbed, perhaps for want of a better name,
the gluon.

At this time it was by no means obvious, or universally
recognized, that such a picture of strong interactions 
would or could succeed.  Indeed, it was a widely held
view that the idea of elementary particles was 
inappropriate for the strong interactions altogether,
and that theories, particularly those involving perturbative
methods, could serve only as guides to suggest the
mathematical properties of the S-matrix \cite{Edenetal}.

The deeply inelastic scattering experiments, which we shall
discuss shortly, changed all that.  They showed unequivocally
that the proton possessed charged substructure of a spatial size
much smaller than the proton itself.  Indeed, the experiments
also suggested spin one-half for these particles.  At the
same time, it was recognized that the constituent-quark 
model seemed to require a new quantum number for the 
quarks, ``color".  Color, although originally introduced to
solve the problem of Fermi statistics for the spin-1/2 quarks,
provided a natural set of currents to which the gluons
might couple.  A three-color model of quarks has a global
SU(3) symmetry, with currents reflecting its
group structure. 
This suggested a local nonabelian gauge theory of the type
originally introduced by Yang and Mills many years before \cite{YM}.

In a relatively few months \cite{QCD} this theory was recognized
to possess a number of important properties.  At lowest
order, it produces attractive forces in the three-quark 
and quark-antiquark systems, and repulsive forces for a
quark-quark system.  It automatically incorporated the 
well-established successes of current 
algebra\ \footnote{In essence, current algebra follows
from the assumption that the electromagnetic 
and weak interactions couple to hadrons via
point-like operators (the currents) which 
are then assumed to obey commutation relations
consistent with the symmetries of these
interactions (the algebra).  In QCD,
this algebra follows automatically by
identifying the currents with operators
${\bar q}_i\gamma_\mu q_j$, with $q_i$ 
quarks of differing flavors.}.  Perhaps its characteristic
and most crucial 
property, however, is {\it asymptotic freedom},
according to which its coupling decreases with decreases in the
distance scale over which it is measured \cite{AF}.  The
need for asymptotic freedom was signalled by
the very experiments designed
to detect point-like structure in the nucleon.  These are the 
deeply inelastic scattering (DIS) experiments, initiated
at SLAC in the late sixties, to which we now turn. 

\subsection{Deeply inelastic scattering and the parton model}

In deeply inelastic (deep-inelastic) scattering, 
a massive hadronic state $X$ of invariant mass
$M_X^2\gg m_N^2$ is produced by the scattering
of a lepton (for instance, an electron) on a nucleon (or other hadron), 
\eq
e(k) + N (p) \to e (k^\prime) + X_{\rm hadronic}\, .
\eqe
This process in illustrated in fig.\ 1.
\begin{figure}[ht]
\centerline{\epsffile{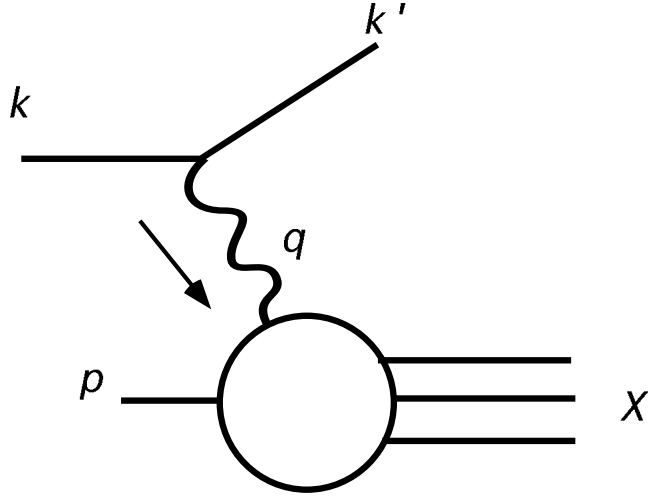}}
\caption{Deeply inelastic scattering.}
\label{fig1}
\end{figure}

{\it Kinematics.}
Because the lepton interacts with the nucleon only through
the exchange of a photon, W or Z, with relatively small
electroweak corrections, the cross section for this process
factors into leptonic and hadronic tensors,
\eq
d \s = {d^3 k^\prime \over 2s | \vec{k}^\prime |} \; {1\over (q^2)^2} \;
L^{\m\n} (k, q) W_{\m\n} (p, q)\, ,
\label{sigLW}
\eqe
where we have taken the example of photon exchange.
The leptonic tensor $L^{\m\n}$ is known from the electroweak Lagrangian,
while the hadronic tensor may be expressed in terms of matrix elements
of the electroweak currents to which the vector bosons couple,
\eqa
L^{\m\n} &\equiv& {e^2 \over 8 \p^2} tr \left[ \sl{k} \g^\m \sl{k}^{\; \prime}
\g^\n \right] \nonumber\\
W_{\m\n} &\equiv& {1\over 8 \p} \sum_{{\rm spins}\ \s} \;
\sum_X  < N (p, \s) \mid J_\m (0) \mid X >\; < X \mid J_\n (0) \mid N (p, \s ) > 
\nonumber \\
&\ & \hbox{\hskip 2.0 true in} \times (2 \p )^4 \d^4 (p_X - q - p )\, .
\eqae
The expression for $L^{\mu\nu}$ is elementary, and the 
expression for $W_{\m\n}$ is quite general, depending only
on the assumption that the electroweak interactions couple to
the hadron via local currents\, \footnote{This assumption, we have observed above,
had considerable experimental and theoretical support by the late 
sixties.  Nevertheless, it is itself a 
nontrivial assertion.}.  

The matrix elements in $W_{\m\n}$ 
include the strong interactions only, and they hence satisfy
symmetry properties of the strong interactions.  For instance,
both electromagnetic and strong interactions enjoy
invariance under parity and time reversal.
This leads to a symmetric $W_{\m\n}$ when the 
hadron is unpolarized, while time reversal invariance of
the strong interactions leads to a real hadronic tensor,
\eq
 W^{\rm (em)}_{\m\n} = W^{\rm (em)}_{\n\m}\quad\
 {(\rm spin-averaged),}\quad\quad\quad W_{\m\n} = W_{\m\n}^*\, .
\eqe
Along with Lorentz invariance, and electromagnetic 
current conservation,
\eq
q^\mu W_{\m\n}=0\, ,
\label{ccons}
\eqe
these constraints may easily be used
to show that the sixteen components of the
hadronic tensor are determined by two independent {\em structure
functions}, $W_1$ and $W_2$,  
\eqa
W_{\m\n}{}^{(em)} &=& - \left ( g_{\m\n} - {q_\m q_\n \over q^2} \right ) W_1
(x, q^2) 
\nonumber\\
&& + \left (p_\m + q_\m \left ( {1\over 2x} \right ) \right ) 
\left (p_\n + q_\n\left  ({1\over 2x} \right ) \right ) W_2 (x, q^2 )\, .
\label{Wdef}
\eqae
The $W$'s are functions of $Q^2$ and the dimensionless ratio,
\eq
x = - {q^2 \over 2 p \cdot q} \equiv {Q^2 \over 2 p \cdot q}\, .
\label{xdef}
\eqe
For reasons which will become clear in a moment, it is also
convenient to introduce dimensionless structure functions,
\eq
F_1\equiv W_1, \quad\quad F_2=p\cdot q\; W_2\, .
\label{Fdef}
\eqe
It is worth noting that in the literature many
 definitions
are given for the $W$'s (usually differing by factors of target
mass), while definitions of the $F$'s are much more standardized.

{\em Scaling}. The striking result of the early deeply inelastic
scattering experiments was that,
for $Q^2\ge$ 1 GeV$^2$, the structure functions $F(x,Q^2)$
become functions of $x$ only, nearly independent of $Q^2$.
This property is called {\em scaling} \cite{scaling1,scaling2}.  In its importance,
it is (and should be) compared to the experiments of
Rutherford and collaborators, who discovered the atomic nucleus
in the wide-angle scattering of alpha particles.  In deeply
inelastic scattering, the wide-angle scattering of the
electron serves to detect point-like structure in the 
nucleon though scaling.  If electic charge were uniformly
distributed within the nucleon, we would expect 
wide-angle scattering to be very rare, giving structure functions that decrease
rapidly with $Q^2$.
  To see how point charges produce scaling, we shall review the 
parton model.

{\em The parton model}.  In the parton model  \cite{partonmodel1,partonmodel2}, we
imagine the proton, or any other hadron, to be made of 
point-like constituents, the partons, through which it
couples to the electroweak interactions, and eventually
to the strong interactions as well.   Deeply inelastic scattering will give information on
the spin of these partons, which, in anticipation,
we take to be one-half.

The fundamental relation of the parton model 
for deeply inelastic scattering may be written 
\eq
 d \s^{(\ell N )} (p, q) = \sum_f \int^1_0 d \x \; d \s_{\rm
Born}{}^{(\ell f)} (\x p,
q) \f_{f/N } (\x)\, ,
\label{pmsigma}
\eqe
where  $d \s^{(\ell N )} (p, q)$ is the {\em inclusive} cross
section for nucleon-electron scattering, while
$d \s_{\rm Born}(\xi p,$ $q)$ is the {\em lowest-order, elastic}
parton-electron cross section, with the parton's
momentum given by $\xi p$, $\xi$ between zero and one.
The functions $\f_{f/N}(\xi)$ are {\em parton distributions}, which
describe the probability of finding a parton (of ``flavor" $f$)
in the ``target" hadron $N$.  There is assumed to be no
interference, either between different flavors or between
different fractions $\xi$.  This ``incoherence" is a hallmark
of the parton model and its extension to QCD.  An important consequence
is that the parton distributions are universal, in the sense
that, since they describe processes that do not interfere with
the hard scattering, they are the same for all inclusive hard
scattering processes, not only for electromagnetic DIS.
The most obvious extension is to neutrino DIS, but there
are many others.
Eq.\ (\ref{pmsigma}) is illustrated by fig.\ 2, in which the
hadronic interaction of fig.\ 1 is broken up into a
parton distribution and parton-electron scattering.
\begin{figure}[ht]
\centerline{\epsffile{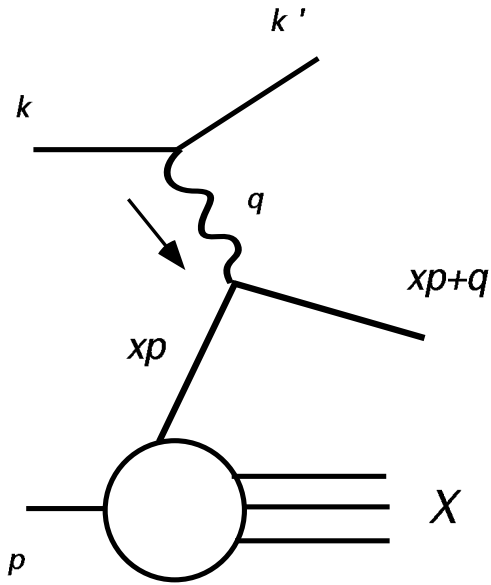}}
\caption{Deeply inelastic scattering in the parton model.}
\label{fig2}
\end{figure}

{\it Scaling and spin in the parton model}.
Both the inclusive hadronic cross section and the elastic (exclusive)
partonic cross section may be factored into leptonic and hadronic
tensors as in eq.\ (\ref{sigLW}). 
The hadronic tensor at Born level is given explicitly by
\eq
W_{\m\n}{}^{(f)} = {1\over 8 \p}
 \int {d^3 p' \over (2\p)^3 2 \o_{p^\prime}}
Q_f{}^2 tr [\g_\m {\rlap{\,/}p}^{\; \prime} \g_\n \rlap{\,/}p ]\, 
(2\p)^4 \d^4  (p^\prime - \x p - q )\, ,
\label{WBorn}
\eqe
with $Q_f$ the electromagnetic charge of quark flavor $f$.
Substituting (\ref{WBorn}) into (\ref{pmsigma}), 
and using the definitions (\ref{Wdef}) and (\ref{Fdef}), 
we find the
following simple
relations between hadronic and partonic structure functions,
\eqa
F_2{}^{(N)} (x) &=& \sum_f \int^1_0 d \x\, F_2{}^{(f)} (x/\x) \f_{f/N} (\x )
\nonumber\\ 
F_1{}^{(N)} (x) &=& \sum_f \int^1_0 \; {d \x \over \x}\,
F_1{}^{(f)} (x / \x)
\f_{f /N} (\x )\, ,
\label{pmF}
\eqae
from which we can read off the partonic structure functions,
\eq
2 F_1{}^{(f)} (z) = F_2{}^{(f)} (z) = Q_f{}^2 \d (1-z)\, .
\eqe
Note the minor difference in the powers of $\x$ in the convolutions
(\ref{pmF}), which
is a reflection of the tensor structure of $W^{\m\n}$.

Turning again to the general relation eq.\ (\ref{pmsigma}),
we derive nucleon structure functions in the parton 
model
\eq
F_2{}^{(N)}(x) = \sum_f Q_f{}^2 x \f_{f/N} (x) = 2x
F_1{}^{(N)} (x)\, .
\label{pmFs}
\eqe
The relation $F_2 = 2 x F_1$ is a direct  consequence of 
the spin of the partons.  This result \cite{CallanGross}, known as the 
``Callan-Gross relation" is important evidence that the 
partons detected in deeply inelastic scattering are indeed
the quarks of hadron spectroscopy.

A crucial feature of eq.\ (\ref{pmFs}) is that the
structure functions $F_i$ are independent of $Q^2$.  This is
the {\em scaling} of experiment, and is a direct result of
the assumed point-like nature of the partons.  In summary,
the parton model serves both to explain scaling and to 
identify partons as quarks.

\begin{figure}[ht]
\centerline{\epsffile{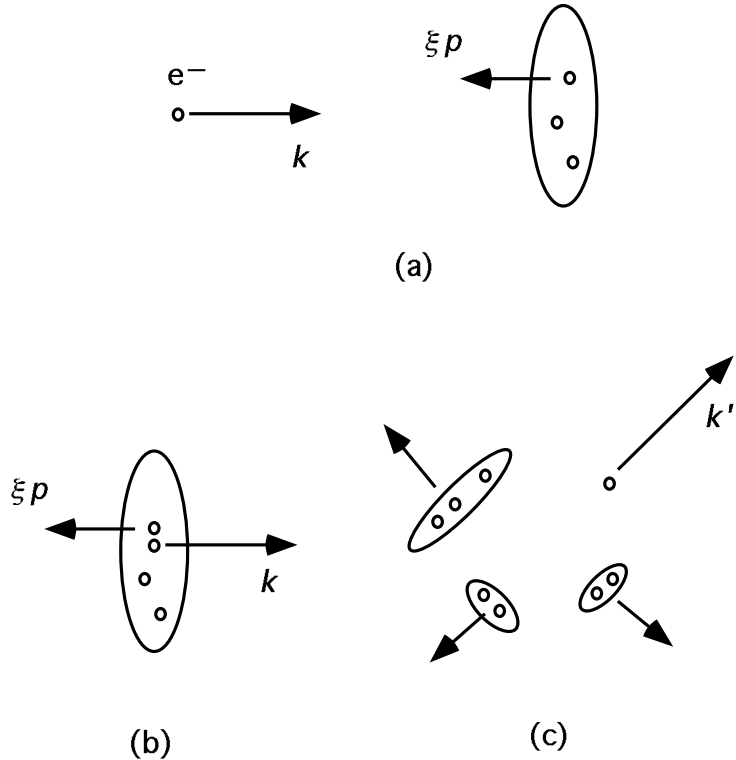}}
\caption{Heuristic picture of parton model DIS.}
\label{dispict}
\end{figure}

{\it Heuristic justification and requirements on a field theory}.
The ultimate justification of the parton model must be in
quantum field theory, but many of its features can be
understood on the basis of compelling heuristic argurments.
Fig.\ \ref{dispict} illustrates the time development of a deeply
inelastic scattering event. 
Fig.\ \ref{dispict}a shows the
system before the collision, as seen in the center of
mass.  The isolated electron approaches from the left,
and from the right the nucleon, of momentum $p$.  
The nucleon is pictured as a set of 
partons, spread out more-or-less evenly in the transverse
direction.  The nucleon is
highly Lorentz contracted in the longitudinal direction.
We must first justify why the nucleon may be treated
as a ``pure" state with a definite number of partons.
For this, we imagine that the wave function of these partons is
formed by interactions which occur on time scales
of the order of 1 GeV$^{-1}$.  
Such scales are long compared with the time
it takes the electron to traverse the nucleon, and hence
remain uncorrelated with hard processes that the electron
may initiate in that time.  During that short time, each of
the partons may be thought of as possessing a definite fraction
of the nucleon's momentum, denoted $\xi p$ in the figure.

To initiate a hard scattering, the electron should pass
very close to one of the partons, 
{\em i.e.} at $O(Q^{-1})$, close enough so that 
they may exchange a photon with an invariant mass 
$Q^2$.  Such a 
collision is shown in fig.\ \ref{dispict}b.  Actually,
in most cases the electron will miss the highly localized
partons altogether, and if it does come close to one, it
is highly unlikely to hit another.  Indeed, the 
probability for encountering $n$ partons at a distance
of $Q^{-1}$ behaves as
\eq
P_{\rm n\ partons} \sim \left ( {1\over R_0^2Q^2} \right )^n\, ,
\label{npartons}
\eqe
with $R_0$ the nucleon radius.  Multi-parton events
are thus power-suppressed in the parton model, a feature
that will turn out to have a natural field-theoretic 
analog.  Fig.\ \ref{dispict}c shows the results of the collision.  
Long-distance forces produce the observed hadrons from the
fragments of the struck nucleon.  These forces, however,
are again on the scale of hadronic internal interactions,
and hence cannot interfere with the short-distance
scattering of fig.\ \ref{dispict}b.  In a rough summary, deeply
inelastic scattering
shows incoherence because initial-state
interactions (which bind the proton) are too early,
and final-state interactions 
(which produce final-state hadrons) are too late relative to
the short time scale of the hard scattering. 

The above arguments rely on very general features
of relativity and quantum mechanics.  
To justify their application within a specific quantum field theory,\
however, is highly nontrivial.  The particular challenge concerns
the separation of short and long time scales.  We have
assumed that at short scales the ``strong" interactions, 
which evidently produce confinement, are weak enough
to be neglected compared to the single, electroweak
interaction.  As we have seen in Collins' lectures at
this school, interaction strengths that change with
distance scales are the rule in quantum field theory.  It is
clear, however, that the successes of the parton model,
and their consistency with long-distance strong interactions,
requires an asymptotically free theory.  Thus, we
are ready to review elements of the asymptotically
free theory of the strong interactions, quantum chromodynamics.

\subsection{QCD, asymptotic freedom and the running mass}

Quantum chromodynamics is specified by its Lagrange density,
as a nonabelian SU(3) gauge theory for Dirac spinors.
Including gauge fixing and ghost terms 
the density for an SU(N) theory may be written as
\eqa
\cl &=& \sum_f \bar{q}_{f,i} (i \rlap{\,/}\del\d_{ij} + 
i g \rlap{\,/}A_a (T_a^{(F)} )_{ij} - M_f\d_{ij} ) 
q_{f,j}
\nonumber\\
&-& {1\over 2} tr (F_{\m\n} F^{\m\n} ) - {\l\over 2} (\h \cdot A_a )^2
\nonumber\\
&+& \h_{\m} \bar{c}_a (\del^\m \d_{a d} - g C_{abd} A_b{}^\m ) c_d\, .
\label{LQCD}
\eqae
Here the (Dirac spinor) quark fields are $q_f$ 
of mass $M_f$ ($f=1,\dots n_f$ labels flavor)
with color index $i$, where $i=1\dots N$
for an SU(N) theory.  The gluon fields are $A^\mu_a$, $a=1\dots N^2-1$, and
ghost
fields $c_a$ (and antighost ${\bar c}_a$), $a=1\dots N^2-1$.
The $T_a^{(F)}$ are $SU(N)$ generators in the $N$-dimensional
defining representation.  In matrix notation,
the nonabelaian ``field strengths" $F_{\m\n}\equiv F_{\m\n,a}T_a^{(F)}$ are given by
\eq
F^{\m\n} = \del^\m A^\n - \del^\n A^\m + ig [A^\m, A^\n ]\, ,
\eqe
where the commutator is in terms of the color matrices.
A number of useful identities and other properties of the
color generators are listed in Appendix A.

In the gauge-fixing term $(\lambda/2)(\eta\cdot A)^2$, $\eta$
is typically chosen as the gradient $\partial$  (covariant gauge)
or as a fixed vector $n$.  For the latter choice, it is sometimes
convenient to distinguish the possibilities $n^2>0$ (temporal gauge),
$n^2<0$ (axial) and $n^2=0$ (light-cone). 
In the limit 
$\lambda\rightarrow \infty$, the condition $\eta\cdot A=0$ is
enforced, and the gluon decouples from the ghost.

The diagrammatic rules for perturbation theory that follow from
Eq.\ (\ref{LQCD}) are derived by standard
techniques \cite{qfttexts,Stermanbook,MutaYndurain}.  Their most characteristic
feature is the self-interaction of the massless gluons and, in covariant
gauges, the necessity of ghost fields.

The theory, of course, must be renormalized, and 
the strength of its coupling
therefore depends upon (``runs with") the momentum scale at
which it is defined,
according to 
\eq
{\partial \over \partial \ln \m}g(\mu) \mid_{g_0}
=\b\left( g(\m)\right )\, ,
\label{effcoup}
\eqe
where the bare coupling $g_0$ held fixed.
The beta function $\b(g)$ is determined to one loop from the
coupling renormalization constant $Z_g$ by
\eqa
\b (g) &=& - g \left (b_2 {\a \over 4\p} +
b_3 ( {\a \over
4 \p} )^2  + \ldots \right )
\nonumber \\
&=& {\del \over \del \ln \m} (Z_g^{-1}g_0) 
\nonumber \\
&=& {\del \over \del \ln \m}\left ( g_0- \ln (\m^2/M^2)\,
 \left \{ {g^3_0 \over 8 \p^2}\left [ {11 N\over
3}  -{2n_f\over 3} \right ]  \right \} 
+ \cdots \right ) \nonumber\\
&=& - {g^3 \over 16 \p^2} \left ( {11\over 3} \; N -{2\over 3}\; n_f\; \right )\, ,
\eqae
where $M^2$ is an ultraviolet cutoff, and where in the final
line we have used $g=g_0$ to lowest order.  
The solution to (\ref{effcoup})
is then the familiar 
\eqa
{g}^2 (\m ) &=& {{g}^2 (\m_0) 
\over 1 + {{g}^2(\m_0) \over 16 \p^2} b_2 \ln {\m^2 \over
\m_0^2} }
\nonumber \\
&\equiv& {16\pi^2 \over b_2 \ln (\m^2/\Lambda^2)}\, ,
\label{runningalpha}
\eqae
with $b_2=11-2n_f/3$ for QCD and 
where in the second form we define
\eq
\Lambda=\mu_0 {\rm e}^{-8\p^2/2b_2g^2(\mu_0)}\, ,
\eqe
independent of $\mu_0$.  Evidently, the coupling $g(\m)$, or
equivalently ``alpha-strong", 
\eq
\alpha_s(\mu^2)\equiv {g^2(\mu)\over 4\pi} = {4\p\over b_2 \ln(\m^2/\L^2)}\, ,
\label{alphadef}
\eqe
decreases for large momenta, or short distances (with corrections
due to higher terms in $\b(g)$).  
At the same time, as the momentum scale decreases, and
distances correspondingly increase, the perturbative
coupling grows.  This is just what we needed for the parton model.
It is, however, still a long way from this observation to
a phenomenology of deeply inelastic and other hard-scattering
 processes.  Nevertheless, it is with asymptotic freedom
that everything begins.  Without it, there is no natural
explanation in field theory of the successes of the parton model.

The last item in this synopsis of QCD and asymptotic freedom is
a brief discussion of quark masses.  Unlike quantum electrodynamics,
where the physical mass of
the electron is directly observable, the
masses of the quarks must remain to a large extent theoretical constructs,
which may be determined from experiment once a scheme for
doing so is defined, but for which there is no unique scheme.  
In this sense, quark masses are much like the gauge coupling itself,
and for many purposes it is useful to define a running quark mass,
\eqa
{ m} (\m^2) &=& m (\m_0{}^2 ) \exp \left\{ - \int^\m_{\m_0} {d \l \over
\l}  \left [ 1+ \g_m ({g} (\l )) \right ] \right\}\, , 
\nonumber\\
&& {{m} (\m^2) \over \m^2} \; \stackrel{\longrightarrow}{\m \to \infty} \; 0\, .
\eqae
Here $\gamma_m(g)$ is a perturbative quantity, much like $\b(g)$, which
therefore vanishes as $\m$ increases and $g(\mu)$ decreases.
Thus, as indicated, the effective mass vanishes compared to the 
renormalization scale when that scale diverges, and the perturbative
theory becomes, effectively, a massless theory.  

An important point for the phenomenology of QCD \cite{LeutGass} is that the 
``light" u and d quarks have masses  of just a few MeV at
scales of $\m\sim$ 1 GeV, so that
\eq 
m_{u,d}(\mu) \ll \Lambda
\eqe
for any scale $\mu$, right down to $\m$'s characteristic of
strong coupling.  For any but the longest scales, the same
holds for the s quark.  Thus, for these quarks, QCD is effectively
a theory of massless particles at any scale where we can hope
to do perturbation theory.  The other three, ``heavy" quarks,
c, b and t have masses whose running must be taken into account.
Depending on the quantity being considered, each requires 
special treatment, and correspondingly offers a rich variety of
theoretical and experimental challenges and opportunities.
For the most part here, however, we shall concentrate on QCD as a
theory of massless quarks, and try
to address the question of how to use its asymptotic
freedom in the ``real world" of Minkowski space.
 
\subsection{Wick rotation}

By itself, asymptotic freedom is a striking result,
and very supportive of the partonic picture of 
scaling in deeply inelastic scattering.  
But the
problem arises of how to separate short distances,
represented by the Born cross section in the parton model,
from long, represented by the parton distributions.
How can we increase our confidence in the heuristic arguments
given above?  The operator product expansion
is designed to organize the relationship between short
distances and long, and indeed, it
will play an important role in our analysis.  This
analysis, however, now will be somewhat more complicated,
just because we want to discuss cross sections in
Minkowski space, rather than Green functions in 
Euclidean space.  
 
Since our aim is to use asymptotic freedom
in Minkowski space, it is worthwhile to recall the
method of Wick rotation, which is used to
define Green functions as they approach physical regions
in external momenta.  This technique is a very
general one, which applies to any relativistic field
theory.  For simplicity, consider a self-interacting scalar field.
The generating functional for its Green functions
may be defined in Euclidean space as the parth integral
\eq
Z_E[J] = \int [ d \f ] \; \exp \left[ - \int d^4 x 
\left(  {1\over 2} \left [ {\sum^4_{i=1}}
(\del_i \f)^2 + m^2\f^2 \right ]
 + V (\f ) + J\f \right) \right ]\, ,
\label{ZE}
\eqe
where $V(\phi)$ is a potential.

Formulating the field theory in Euclidean
space has a number of advantages.  The path integral 
 rapidly damps contributions from large values of
the field.  Correspondingly, the perturbative Green functions
found from $Z_E[J]$ are real functions, free of singularities 
for real external momenta.  In perturbation theory,
their internal propagators are of
the negative definite form
\eq
\Delta(k)={1\over k_E^2-m^2}\ ,\quad k_E^2=-\sum_{i=1}^4 k_i^2\, .
\eqe
Such Green functions, however, cannot describe physical processes.
Indeed, the S-matrix elements necessary to define cross sections
are themselves found from singularities in Green functions when
external momenta approach the mass shell, $p_i^2=m^2$.
Wick rotation makes the connection between these two regions.
We analytically continue $Z_E$ by replacing the real ``time" component
$x_4$ by a complex number $\tau$
in terms of two real parameters $x_0$ and $\q$ as follows: 
\eq
-ix_4\rightarrow \t\equiv 
 x_\q e^{- i \q}\, .
\label{xsubthe}
\eqe
For arbitrary values of $\q$ we
{\it define} a new generating functional $Z_\q$  by
replacing $dx_4$ by $idx_0e^{-i\q}$,
\eq
Z_\q[J]  \equiv  \int \; [ d \f ] \; 
\exp \left [ -i e^{-i\q} \int d x_\q \; d^3 x 
\left ( {1\over 2} \left [ -e^{2 i\q}
\left ( { \del \f \over \del x_\q} \right )^2
 + (\nabla \f)^2  + m^2 \f^2 \right ]  +  V(\f) + J\f \right ) \right ].
\nonumber\\
\eqe
Note that (\ref{xsubthe}) is not a change of
variables, and that $Z_\q$ is not equal to $Z_E$
except for $\q=\p/2$; rather, we
regard it as an analytic continuation 
of $Z_E$ in $\t$.  We must assume that such a continuation 
is possible for the full generating functional, as turns out
to be the case for its perturbative expansion, for which the 
propagator is given at arbitrary $\theta$ by
\eq
\Delta(k,\q)={e^{-i\q}\over -k_\q^2e^{-2i\q}-\sum_{i=1}^3 k_i^2-m^2}\, .
\label{Deltheta}
\eqe
This expression has no singularities for real $k_\theta,\ k_i$,
so long as $\q$ remains finite.

The physical field theory is generated by the limit $\q\rightarrow 0^+$,
in which $\q$ approaches zero from above,
\eq
Z_0[J] = \int [d \f ] \exp \left [
i \int dx_0 d^3 x\, \left ( 
{1\over 2} \left ( ( \del_0 \f)^2 - ({\vec \nabla}
\f)^2 - m^2
\f^2 + i \e \f^2 \right )
 - V (\f ) -J\f \right ) \right ]\, .
\eqe
In this limit $\Delta(k,0^+)$ becomes precisely
the normal Feynman propagator,
\eq
\Delta(k,0^+)={1\over k_0^2 -\sum_{i=1}^3 k_i^2 -m^2 + i\e}\, ,
\eqe
whose singularities are regulated by an ``$i\e$ prescription".

\begin{figure}[ht]
\centerline{\epsffile{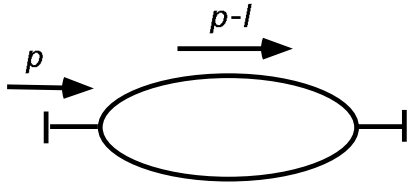}}
\caption{Scalar two-point function.}
\label{s2pf}
\end{figure}

{\em Wick rotation for a two-point function}.  As a practical
example in perturbation theory, let's consider the scalar equal-mass
two point function at one loop, fig.\ \ref{s2pf},
directly in Minkowski space.
Working in $n$ dimensions,
combining denominators by Feynman parameterization and 
completing the square in the loop momentum, we find
\eqa
I (p^2) &=& \int {d^n \ell \over (2 \p)^n} \; {g^2
\m^{2\e}
\over (\ell^2 - m^2 + i \e ) ((p - \ell)^2 - m^2 + i \e )} \nonumber\\
&=& g^2 \m^{2\e} \int {d^n \ell \over (2 \p)^n} \int^1_0 \; {d x \over
(\ell^2 - 2 x
p \cdot \ell + x p^2 - m^2 + i \e)^2} \nonumber\\
&=& g^2 \m^{2\e} \int^1_0  dx \int {d^n \ell^\prime \over (2 \p)^n} \; {1\over
(\ell'{^2} + x (1-x) p^2 - m^2 + i \e)^2}\, .
\label{12int}
\eqae
The factor $\m^{2\e}$ keeps the coupling dimensionless in $n$ dimensions.

\begin{figure}[ht]
\centerline{\epsffile{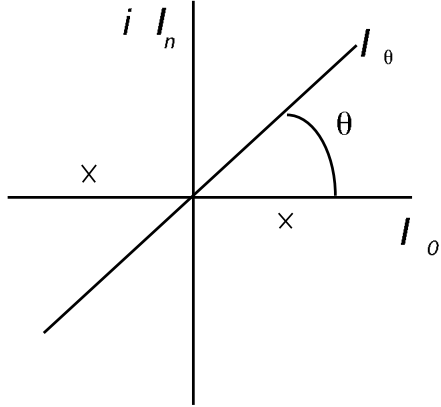}}
\caption{Wick rotation in the energy plane. }
\label{l0plane}
\end{figure}

In Minkowski space, with ${\ell'}^2={\ell'}^2_0-{{\vec \ell}'}^2$, the
integrand in (\ref{12int}) has poles at
\eq
\ell'_0=\pm\sqrt{{{\vec \ell}'}^2+m^2-x(1-x)p^2}\ \mp i\e\, .
\label{ellppoles}
\eqe
The integral, however, is defined by the 
$i\e$ prescription, which keeps the poles off
the real axis, so long as $p^2<4m^2$.  For these values of 
$p^2$, the poles are in the standard arrangement shown in 
fig.\ \ref{l0plane}, and the $\ell'_0$ integration contour may be 
rotated to the vertical (imaginary) axis, according to
\eq
\ell'_\q = \ell'_0e^{i\q}\, \quad 0\le \theta \le \p/2\, ,
\label{ellWR}
\eqe
where $i{\ell'}_n\equiv \ell'_{\p/2}$ is the imaginary energy along the vertical
axis in the $\ell'_0$ plane.  
This rotation is in just the opposite sense from the rotation of time
in eq.\ (\ref{xsubthe}).  For $p^2=-Q^2<0$
(in fact, for $p^2< 4m^2$), the integral
may be rotated as an analytic continuation into a purely 
Euclidean integral, 
\eqa
I (-Q^2 ) &=& g^2 \m^{2 \e} \int^\prime_0 dx \;   \; \int \; { id^n
\ell^\prime \over (2
\p)^n} \; \; {1 \over ({\ell^\prime}_E^2 - x (1-x) Q^2 - m^2 )} \nonumber\\
&=& g^2 \m^{2\e} i \G (2- {n\over 2} ) {\p^{n/2}\over (2 \p)^n} \;
\int^1_0 \;
{dx \over (x (1-x) Q^2 + m^2 )^{2-n/2}}\, ,
\eqae
where $d^n\ell'=d\ell'_nd^3\ell'$.
  Exanding around
$n=4$, and defining \footnote{This $\e$ is one-half of the
convention used by Collins in his lectures.},
\eq
\e=2-n/2\, ,
\label{epdef}
\eqe
we find
\eq
 I (p^2) = {i\over \e} \; {g^2 \over (4 \p)^2} 
\left [1-\e(\g_E - \ln 4 \p) \right]
   - {i g^2 \over (4 \p)^2} \int^1_0
\; dx  \; \ln
\left [(m^2 - x (1-x) p^2)/\m^2 \right ] + \co (\e)\, .
\label{stpintresult}
\eqe
In the full theory,
the pole term is removed by ultraviolet renormalization.
We are interested, rather, in the finite, momentum-dependent remainder.
It is real for $p^2<4m^2$ and is defined 
by analytic continuation for $p^2\ge4m^2$, where, on the real 
axis, it is complex but finite, as long as
$m^2>0$.  Real, negative $p^2$ is equivalent to Euclidean space in which,
as observed above, all Green functions are (relatively) real,
analytic functions of their external momenta.  Our example
shows how branch points develop in their analytic continuation.
Note in particular that the branch point occurs at 
``threshold" for the production of two on-shell particles.
This correspondence of singularities to physical processes is
quite general, as we shall see presently.

Recalling that in QCD light quark masses are small compared to
the natural scale of the coupling, we may take a special interest
in the $m^2\rightarrow 0$ limit of (\ref{stpintresult}).  
The branch point moves to the origin in the $p^2$ plane, where
the function $I(p^2)$ diverges logarithmically.  
This divergence has quite different interpretations in 
Euclidean space, where it occurs at a single point, $p^\mu=0$, and
in Minkowski space, where it is on an entire two-dimensional
subspace, the light-cone $p_0^2={\vec p}{^{\; 2}}$.  Such divergences
in Green functions will recur in QCD, and get worse at higher 
order.  This is a clear difficulty for any program to exploit 
asymptotic freedom - since it indicates that Green functions have,
in general, divergent sensitivity to low-momentum, long-distance
physics.  In fact, since in Minkowski space, small 
invariant mass does not necessarily mean small energy, the 
situation is much more complicated than in Euclidean space.
It is this long-distance behavior whose analysis,
which is still an ongoing topic of research, we shall describe 
in the following.  Before we do so, however, it will be useful
to intoduce a simple rule for identifying quantities that
are {\it not} sensitive to long-distance, nonperturbative physics.

\subsection{Infrared safety}

Asymptotic freedom is useful for quantities that are 
dominated by the short-distance behavior of the theory \cite{StermanWeinberg}.
Such quantities, which are
termed {\em infrared safe}, cannot depend sensitively on the masses
of quarks, nor can they suffer from infrared divergences
of the sort identified in the example above.  
Infrared safety is one of the fundamental concepts of 
perturbative QCD, and it makes essential use of the renormalization
group.  To see how, and to see further why an analysis of 
infrared divergences is necessary, we consider a generic
physical quantity (say, a cross section) $\tau(Q^2/\m^2,\a_s(\mu^2),m^2/\m^2)$,
where $Q$ represents ``large" invariants, much greater than $\Lambda$,
$m$ represents light quark masses (and the vanishing gluon mass),
and $\m$ the renormalization scale.  We assume that $\t$ has been
scaled by an overall factor of $Q$,
to make it dimensionless.  Now because $\t$ is physical, it
cannot depend on $\m$,
\eq
\t \left ({Q^2 \over \m^2} , \a_s(\m^2), {m^2 (\m^2)\over \m^2} \right )
 = \t  \left (1, \a_s(Q^2), {m^2
(Q^2) \over Q^2} \right )\, .
\eqe
This just says that we may, if we like, expand in the coupling at the
scale of the large momenta of the problem, and  use
asymptotic freedom.   In general, however, we pay the price of
introducing dependence on the large ratio $Q/m$, which typically
occurs in powers of logarithms.
The presence of such logarithms (an example of which was illustrated
above) can make the perturbative expansion unusable.  
On the other hand, pertubation theory {\em can} be used if
$\t$ happens to be infrared safe.  To be specific, we shall
demand that $\t$ behave in the large $\m$ limit as
\eq
\t \left ( {Q^2 \over \m^2} , \a_s (\m^2), {m^2 (\m^2) \over \m^2} \right )
\stackrel{\longrightarrow}{\m \to \infty}\quad  
\hat{\t} \left ( {Q^2 \over \m^2} , \a_s (\m^2)
\right ) + \co \left ( ({m^2 \over \m^2})^a \right),\ a > 0\, .
\eqe
That is, $\t$ should approach a limit as $m/\m\rightarrow 0$,
with $Q/\m$ held fixed,
with corrections that vanish as a power of $m$.  Again,
$I(p^2)$ discussed above, is an example of an infrared
safe quantity so long as $p^2\sim Q^2\ne 0$.

Much of the remaining discussion will center on 
identifying infrared safe quantities, and separating
them from long-distance dependence.  Indeed, this is the
essence of the QCD justification of the parton model.  In this
connection we note that infrared safety can apply not only
to cross sections and other direct physical observables, but
also to renormalization-group variant quantities that
obey equations of the form
\eq
\left (\m {d \over d \m} - \g_{\; \G} \right ) \G = 0\, ,
\eqe
with 
$\g_{\; \G}$ an anomalous dimension.

\subsection{QCD in the $n$ plane}

We'll end this introductory section with a few comments on
the double role of dimensional regularization in perturbative
QCD.  Dimensional continuation serves to regulate both 
ultraviolet ($n<4,\e>0$) and infrared ($n>4,\e<0$) divergences.
When the number of dimensions is less than four, the volume of phase
space is decreased.  Because ultraviolet divergences result from
the large number of states at high energy, they are softened as
the number of dimensions is decreased.  In contrast, infrared
divergences result from singularities in the integrands 
of momentum space integrals, and when these singularities are 
spread out over a larger phase space at higher dimension, they
are softened.  The question arises, however, how dimensional
continuation can handle both problems at the same time.

To see how it works, we recall the basic path from a
Lagrangian to cross sections and other physical quantities.
This is illustrated by the sequence below:
\eqa
\cl_{QCD} &\rightarrow& G^{(reg)}(p_1,\dots p_n)\, , \quad n<4\nonumber \\    
     &\rightarrow& G^{(ren)}(p_1,\dots p_n)\, , \quad n<4+\Delta\nonumber \\
     &\rightarrow& S^{(unphys)}(p_1,\dots p_n)\, , \quad 4<n<4+\Delta\nonumber \\
     &\rightarrow& \t^{(unphys)}(p_1,\dots p_n)\, , \quad 4<n<4+\Delta\nonumber \\
     &\rightarrow& \t^{(phys)}(p_1,\dots p_n)\, , \quad n=4\, .
\label{Ltotau}
\eqae
The first step is to generate  regularized off-shell Green functions,
denoted $G^{(reg)}$ from the Feynman rules of the theory.  As we shall see,
these Green functions are free of infrared divergences, and so are
rendered completely finite for $n<4$.  Here is where the theory is
renormalized, producing a set of renormalized Green functions,
$G^{(ren)}$ which are now 
analytic functions of $n$, treated as
a complex number, in some strip in the $n$-plane,
$4-\Delta'\le n\le 4+\Delta$, with $\Delta,\Delta'>0$.  
It is thus possible to analytically continue
the renormalized Green functions to $n$ slightly greater than four.
Because the {\em infrared} divergences of the theory are regulated in
this region of $n$, we can now define S-matrix elements, $S^{(unphys)}$ for the
renormalized theory.  These S-matrix elements, although fully
renormalized are nevertheless unphysical, because
they are only infrared finite for $n>4$.  That is, they are
not infrared safe.  They can, however, be used to compute infrared
safe quantities in $n>4$, $\t^{(unphys)}$, which, finally, may be
continued to $n=4$ to derive finite, physical predictions,
$\t^{(phys)}$  from the
theory.  It is important to stress that QCD for $n>4$ in not
physical QCD, and cannot be used to compute physical quantities 
unless they are infrared safe, and thus have finite limits for
$n\rightarrow 4$.  Having made these observations, we are ready
to begin our analysis of infrared divergences in field theory.

\section{Long and Short Distances in Minkowski Space}

In this section we shall discuss how to analyze and classify
sources of long-distance behavior in perturbation theory.
We do so not because we expect perturbation theory to be
correct at long distances, but rather to identify and eventually
calculate short-distance (infrared safe) quantities 
for which perturbation theory may reasonably be trusted.

\subsection{Example:  IR and CO divergences in the massless vertex
function}
\label{vertexsubsec}

\begin{figure}[ht]
\centerline{\epsffile{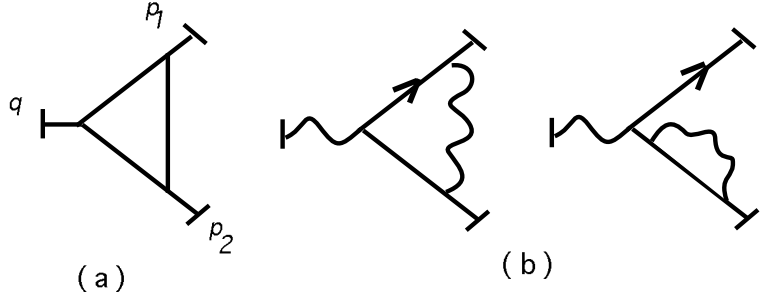}}
\caption{(a) scalar vertex and (b) gauge theory electromagnetic vertex.}
\label{tri}
\end{figure}

A very informative example, which already illustrates many of
the general properties we will identify below, is the fully
massless three-point function at one loop,
with two on-shell external lines.  We shall start, as above, with this
vertex in a scalar theory, shown in fig.\ \ref{tri}a,
\eq
I_\D = g^3 \m^{3\e}   \int {d^n k
\over (2\p)^n} \; {1\over (k^2 + i \e)
 ((p_1 - k)^2 + i \e) \; ((p_2 + k)^2
+ i \e)}\, .
\label{Itri}
\eqe  
This integral may be evaluated just as the two-point integral
above, by Feynman parameterization and $n$-dimensional integration.  The
resulting (two) parametric integrals can then be done in terms of Euler
beta functions, with the result,
\eq
I_\D = (-ig\m^\e)\, {1\over q^2 } \; {g^2 \over (4 \p)^2} \; 
\left ( {4 \p \m^2 \over -q^2 - i \e}
\right )^\e \G (1+ \e) {B (-\e, 1-\e) \over -\e}\, ,
\label{IDel}
\eqe
with $q^2=2p_1\cdot p_2$, and
where again, $\e=2-n/2$.
This scalar integral is ultraviolet-finite, so the double
pole in $\e$ is entirely infrared in origin.  Before discussing
it further, we may exhibit the corresponding result for the
electromagnetic vertex function in
a massless gauge theory, found from the diagrams of fig.\ \ref{tri}b.

For zero-mass fermions (quarks) the electromagnetic vertex 
is given in terms of a single form factor.  For example, in a
quark-antiquark production process we have
\eq
\G_\m (q^2, \e) = - i e \m^\e \; \bar{u} \; 
(p_1) \g_\m v (p_2)\, \r (q^2, \e)\, .
\eqe
At one loop the form factor is given by
\eq
\r (q^2,\e) = - {\a_s \over 2\p} C_F \left ( {4 \p \m^2 \over - q^2 - i \e} \right )^\e \;
 {\G^2 (1-\e) \G (1+\e) \over  \G (1-2\e)}\,
\left\{ {1\over (-\e)^2} - {3\over 2(-\e)} + 4 \right\}\, ,
\label{gammaoneloop}
\eqe
where $C_F=(N^2-1)/2N$ (see Appendix A).
  Up to this group factor, the leading, double
pole term is essentially the same as in the scalar case,
eq.\ (\ref{IDel}).   These double poles are common in 
dimensionally regulated massless integrals.
To understand their origin, we may return to the scalar case.   

Consider the integral $I_\D$ over a region of momentum space
where the loop momentum $k$ is small enough that $k^2\ll p_1\cdot k,\
p_2\cdot k$.  Neglecting the $k^2$ terms compared to $p_i\cdot k$ 
is called the ``eikonal" approximation.  It is a subtle approximation
in Minkowski space, where $p_i\cdot k$ small does not necessarily
imply that $k^2$ is smaller.  Accepting that we shall have to return
to this point later, we work in the eikonal approximation in
a frame where 
\eq
p_1=(p_1^+,0^-,{\bf 0}_\perp), \quad\quad p_2=(0,p_2^-,{\bf 0}_\perp)\, ,
\eqe
and where plus and minus components are defined by
$v^\pm=2^{-1/2}(v^0\pm v^3)$, so that $v^2=2v^+v^--{\bf v}_\perp^2$.

The integral becomes
\eq
I_\D^{\rm (eik)} \sim {1\over 2q^2} \int \; 
{dk^+ dk^- d^2 k_\perp \over (-k^-+i\e)
(k^++i\e) (2k^+k^- - k_\perp{}^2+i\e)}\, .
\eqe
In this integral we easily identify three limiting regions 
which lead to logarithmic divergence.  In the
first, all four components of $k^\mu$ vanish together; this
we shall call the ``soft" region.  In the other two, the
component of $k^\m$ parallel to either $p_1$ or $p_2$
remains finite, while the remaining components vanish in
such a way that $k^+k^-\sim k_\perp^2$; these we refer to
as ``collinear" regions.  Momentum 
components in these regions
are of the order of $\sqrt{q^2}$ times powers of a
``scaling" variable $\lambda$, which vanishes at the points
in momentum space where $I_\D$ is singular,
\eqa
\begin{array}{lll} 1. &  k^\m \sim \l \sqrt{q^2}  & {\rm ``soft"} \\
\\
2. & k^\pm  \sim \sqrt{q^2}   \\
 & k^\mp \sim \l^2 \sqrt{q^2} & {\rm ``collinear"}\\
&  k_\perp^2 \sim \l \sqrt{q^2}\, . \end{array}
\label{scaledelta}
\eqae
The lograrithmic divergences in $I_\D$ may be made explicit
by changing variables in each of these regions to $\lambda$
and a set of scaled momenta ${\tilde k}^\m=k^\m/\lambda^a$,
with $a$ the power appearing in eq.\ (\ref{scaledelta}).

The collinear singularities of (\ref{scaledelta}) are {\em the}
characteristic feature of infrared sensitivity in 
Minkowski space, in which on-shell lines need not have
vanishing momenta.  Clearly, this complicates the
situation relative to Euclidean space.

While soft and collinear divergences were relatively easy to
identify in the simple example above,
 it is still natural to ask how to identify
 infrared sensitivity at higher orders,
and in other processes.  To answer these questions requires 
a more general analysis, to which we now turn.

\subsection{Analytic structure and IR divergences:  Landau equations
and physical pictures}

Our aim in this section is to systematize infrared analysis in
Minkowski space \cite{Stermanbook,Sterman78}.  
In this way, we shall find it possible to
separate differing momentum scales and to apply the operator
product expansion in the presence of light-cone singularities.

Let us return to the massless scalar triangle again, 
treating it this time
from a more general point of view.
Introducing Feynman parameters, the scalar triangle is given by
(suppressing the coupling),
\eq
I_\D = 2 \int {d^n k \over (2\p)^n} \; \int^1_0 \; {d \a_1 d \a_2 d \a_3\, \d
(1-\sum^3_{i=1} \a_i) \over D^3}\, ,
\label{IDelFeynparam}
\eqe
where the new denominator is
\eq
D = \a_1 k^2 + \a_2 (p_1- k)^2 + \a_3 (p_2 + k)^2 + i \e\, .
\label{Ddef}
\eqe
Again, the scalar integral in eq.\ (\ref{IDelFeynparam}) is ultraviolet 
finite, so the poles in (\ref{IDel}) at $\e=0$ must come from
infrared sensitivity due to the vanishing of $D$.

We shall make strong use of the analytic structure of 
the integral $I_\D$, which is defined in terms of integrals in the
complex $k^\m$ and $\a_i$ planes, and whose singularity structure
is defined by the ``$i\e$" in eq.\ (\ref{Ddef}).  
Since $D$ is quadratic in the momenta, it has no 
more than two poles in any momentum component, when the others
are held fixed.  When these poles are at real values, $D$ vanishes at
two points along the contour.
Now we recall that
by Cauchy's theorem
such a contour integral may be deformed between any pair of paths so long
as this deformation
crosses no singularity, in this case no point $D=0$.
Therefore, so long as the solutions to $D=0$ are separated, 
the relevant contour may be deformed 
to make the integrand bounded at all points, leading
to a finite result.
On the other hand, if the poles coalesce (they are automatically
in opposite half-planes), the contour
can no longer be deformed, and the result may be singular.  
This is called a ``pinch" of the contour.  Because $D$ is
quadratic in momenta, this is equivalent to the condition 
\eq
{\del  \over \del k^\m}\; D(\a_i, k^\m, p_a ) = 0\, ,
\eqe
at $D=0$.
Assuming that we
may choose to do any of the momentum integrals first, 
a {\em necessary} condition for a
singularity is to have a pinch in every loop momentum component.

Similar considerations apply to each of the $\alpha_i$'s.
In this case, however, $D$ is linear in each $\alpha_i$, so there
are never two poles to pinch, and the $\a_i$ contour may
always be deformed away from a pole, except at the origin,
since this is where its integral originates.
A pole may, however, migrate to an end-point $\alpha_i=0$, or
it may be that at $D=0$, $\ell_i^2-m_i^2=0$ on the line
corresponding to $\a_i$, so that $D$ is indepedent of $\a_i$

These requirements apply equally to any diagram at any
order, with line momenta $\ell_i$ and loop momenta $k_s$, and
may be summarized as
\eqa
{\rm either}\ \ell_i^2 = m_i^2,\ {\rm or}\ \alpha_i&=&0,
\nonumber \\
{\rm and}\ \sum_{i\; \e\; {\rm loop} \; s} \; \a_i \ell_i \e_{i s} &=& 0\, ,
\label{Landaueqs}
\eqae
for all $i$ and $s$.  
Here $\e_{i s}$ is an ``incidence matrix", which takes the
values $+1$ and $-1$ when line momenum $\ell_i$ flows
in the same direction or opposite direction as
loop momentum $k_s$, respectively, and is zero otherwise.
Eqs.\ (\ref{Landaueqs}) are commonly known as the Landau 
equations \cite{landau,Edenetal}.

For the three-point
vertex functions of fig.\ \ref{tri}, the Landau equations
are simply
\eq
\a_1 k^\m - \a_2 (p_1 - k)^\m + \a_3 (p_2 + k)^\m = 0\, .
\eqe
One solution to these equations corresponds to the soft limit of vanishing
gluon momentum,
\eq
 k^\m = 0\, , \quad (\a_2/\a_1) = (\a_3 / \a_1) = 0\, .
\label{softLE}
\eqe
Another set of solutions correspond to the collinear limits,
where $k$ becomes proportional to $p_1$ or $p_2$,
\eqa
k &=&  \z p_1\, , \quad \a_3 = 0\, , \quad
              \a_1 \z = \a_2 (1-\z) \nonumber\\
k &=& - \z^\prime p_2\, , \quad  \a_2 = 0\, , \quad  
              \a_1 \z^\prime = \a_3 (1-\z^\prime)\, .
\label{collLE}
\eqae
Are these the only solutions?  
The task of tracking down pinches is greatly simplified by
an observation due to Coleman and Norton \cite{ColemanNorton}.  We begin
by identifying the products $\a_i\ell_i$ for each on-shell
line with a space-time vector,
\eq
\a_i \ell_i{}^\m = \D x_i{}^\m\, .
\label{deltax}
\eqe
Suppose we identify further $\a_i=\D x_i^0/\ell_i^{\; 0}$ as the
ratio of the time component of $\D x_i$ to the energy
$\ell_i^{\; 0}$.  Then
\eq
\D x_i^\m=\D x_i^0v_i^\mu, , 
\eqe
with
\eq
v_i^\mu=\left ( 1,{{\vec \ell}_i \over \ell_i^0} \right )\, ,
\eqe
the four-velocity of a particle of momentum $\ell^\m$.
With this interpretation, $\D x_i$ may be thought of 
as the four-vector describing the free propagation of
a classical on-shell particle with momentum 
$\ell_i$.  In an alternate derivation of the Landau equations,
based on time-ordered perturbation theory, this physical 
picture emerges automatically.  This derivation is
given in Appendix B.

\begin{figure}[ht]
\centerline{\epsffile{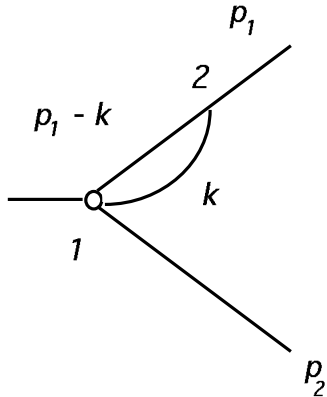}}
\caption{Reduced diagram corresponding to a collinear pinch surface.}
\label{CoRD}
\end{figure}

Let us apply this analysis to the collinear pinch of the
triangle diagram, eq.\ (\ref{collLE}),
where
$k$ becomes collinear with $p_1$.  Consider the
vectors associated with the two on-shell lines at this pinch,
\eq
\D x_{p_1 - k}^\m = \a_2 (p_1-k)^\m = \a_1 k^\m = \D x_k{}^\m\, .
\eqe
They are equal.
Now consider the diagram
shown in fig.\ \ref{CoRD},   
We have contracted the single off-shell line $p_2+k$ to
a point, corresponding to $\a_3=0$, i.e., no 
propagation for this line.  
Any such diagram, in which
off-shell lines are contracted to points, is called a
``reduced diagram".  The reduced diagram
for this pinch describes a physical process, in which two
on-shell massless particles, of momenta $k$ and $p_1-k$, are created 
at vertex 1, and propagate freely to vertex 2, where they 
combine to form the outgoing massless particle $p_1$.  This
is kinematically possible only because the lines are 
massless.  We have found that this collinear pinch
surface describes a physical process, in which vertices
may be identified with points in space-time, between 
which particles propagate on the mass shell.

\begin{figure}[ht]
\centerline{\epsffile{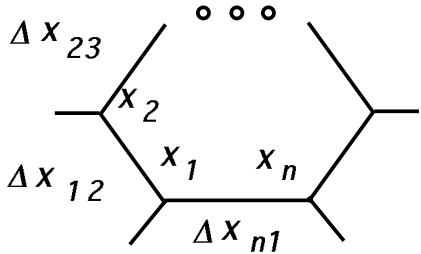}}
\caption{Schematic loop in an aribitrary reduced diagram.}
\label{nsideRD}
\end{figure}

The generalization of this result to a completely
arbitrary diagram is quite straightorward, and only
requires the schematic subdiagram shown in fig.\ \ref{nsideRD},
in which a loop of $n$ lines is shown.  
For this loop to
describe a portion of a physical picutre, the four-vector
separation across any line must equal the separation
derived by going around the other $n-1$ lines on the loop,
\eq
\D x_{12} + \D x_{23} + \ldots + \D x_{n1} = 0\, .
\eqe 
Given the identification in eq.\ (\ref{deltax}), we see that
this requirement is identical to the Landau equations,
(\ref{Landaueqs}).

\subsection{Power counting and pinch surfaces}

In the Landau equations and the physical picture analysis,
we have powerful tools for the identification of sources of 
long-distance sensitivity in perturbation theory.  But even a
pinch is only a necessary condition for infrared
divergences.  In many circumstances, the perturbative integration
contour may
pass through a pinch surface without producing a
singularity.  The contributions from such regions may
be vanishingly small in the limit of large momentum transfer,
consistent with the requirements of infrared safety.  
When summing perturbation theory to high order, such regions 
may again become important, but for the present we shall
look for a further necessary condition for infrared divergences
at finite order.  To find it, we shall study how to bound
integrals near pinch surfaces.  We refer to this process
as ``infrared power counting", in analogy to the ultraviolet
power counting employed in perturbative renormalization \cite{Stermanbook}.

\begin{figure}[ht]
\centerline{\epsffile{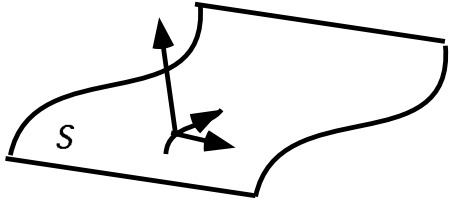}}
\caption{Schematic pinch surface S with one normal and two intrinsic  
coordinates.}
\label{localcoord}
\end{figure}

A general pinch surface is depicted schematically in fig. \ref{localcoord}.
At each point on a pinch surface $S$, we identify coordinates that
lie in the surface, which we refer to as ``intrinsic", and those
that parameterize directions out of the surface,
which we call ``normal".  By construction, the integrand is a singular function
of the normal coordinates only. 

It is possible to bound the integral near $S$ using
the power counting technique illustrated in Sec.\ \ref{vertexsubsec}
for the triangle diagram, fig.\ \ref{tri}.   We
put bounds on the ratio of the volume of normal coordinates to
the magnitude of the integrand at the pinch surface.  In general,
the larger the volume of the normal space, the less singular the integral.
This power counting will have a dual purpose. First, it is used
to put bound on integrals, and hence to identify infrared safe quantities, and 
by the same token it may also be used to identify regions in momentum space that
may give rise to infrared divergences. 

\begin{itemize}
\item{i.} We redefine each of the normal variables $\k_j$ in terms 
of a scaling variable $\lambda$ according to
\eq
\k_j=\l^{a_j}\k'_j\, .
\label{normscale}
\eqe
We will determine the behavior of the integral when
$\l$ vanishes for fixed values of the ratios $\k'_j/\k'_{j'}$.
For instance,
in the triangle diagram, eq.\ (\ref{Itri}), 
the scalings of eq.\ (\ref{scaledelta}) specify that the 
the propagators $(p_1-k)^2$ and $k^2$ are 
linear in $\l$, while $(p_2+k)^2$ is zeroth order
in $\l$ in the collinear region where $k$ is
in the $p_1$ direction.  Similarly $k^2$ is quadratic,
and both $(p_1-k)^2$ and $(p_2+k)^2$ are linear,
for the soft scaling of eq.\ (\ref{scaledelta}).

\item{ii.} Given a set of powers $a_j$, we retain only terms of
lowest power $\l^{A_i}$ in $\l$ for each perturbative 
denominator $k_i^2(\k_j,\l)-m_i^2$,
\eq
k_i^2(\k_j,\l)-m_i^2 = \l^{A_i}f(\k'_j)+\dots\, .
\label{scaleddenom}
\eqe
We call the resulting integral over the normal variables $\k'_j$
the {\em homogeneous} integral.  If the homogeneous integral
is independent of any normal variable $k'_j$, its scaling power $a_j$ may be
reduced until 
that variable appears in the homogeneous integral \cite{Stermanbook}.

\item{iii.} The homogeneous integral for pinch surface $S$ is
proportional to 
$\l^{n_S}$, with $n_S$ given by
\eq
n_S=\sum_ja_j - \sum_iA_i+s_I\, ,
\label{nsubS}
\eqe
where $s_I$ represents the (possible) power of $\l$ from momentum factors
in the integrand   (these will be important for QCD).
If $n_S>0$, the integral is finite when all of the normal
variables vanish according to the scaling (\ref{scaleddenom}).
If $n_S=0$, the integrand may diverge logarithmically,
while if $n_S<0$, it may diverge as a power.  

\item{iv.} Finally, check for pinch surfaces in the homogeneous integral.
If the only pinch surface is the original one, at which all normal variables
vanish, the bound is complete for the singular surface $S$.  Should 
the homogeneous integral have further pinch surfaces, however, bounds must
be found for these special regions as well.  These will correspond to 
subsurfaces of $S$ where subsets of the normal variables vanish faster
than others \cite{Stermanbook}.

\end{itemize}

\subsection{Pinch surfaces for the all-order EM form factor}
 
\begin{figure}[ht]
\centerline{\epsffile{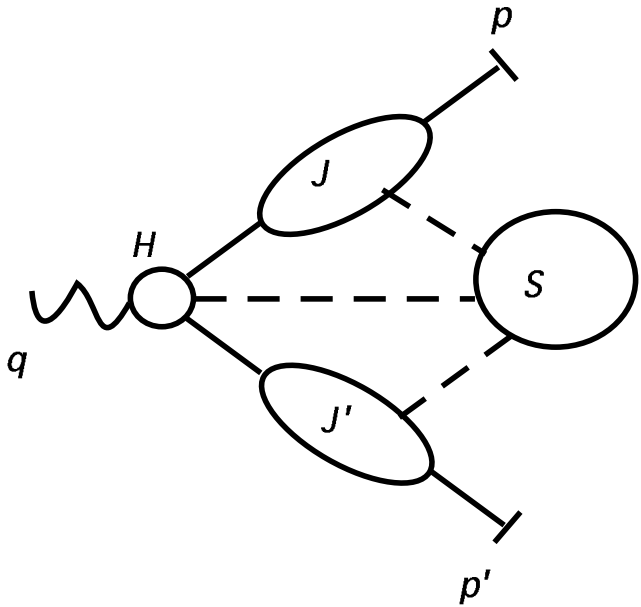}}
\caption{General reduced diagram for pair creation.}
\label{vxS}
\end{figure}

We can now apply the Landau equations and power counting to identify the
sources of infrared divergence in the electromagnetic form factor to 
{\it all} orders.
Consider pair creation, in which a vector
particle of momentum $q$, $q^2>0$ (Z or virtual photon), decays into a 
massless quark and antiquark with no radiation.
The possible reduced diagrams associated with pinch surfaces are
remarkably simple, and are all of the form shown in fig.\ \ref{vxS}.
The general reduced diagram corresponds to a set
of physical processes in which the decay of the vector
is followed by the formation of two ``jets", labelled
$J$ and $J'$ in the figure, of virtual particles in the
same direction, and with the same total momenta, $p$ and $p'$,
as the two final state particles.  The only interaction between the
two jets is via zero-momentum ``soft" particles, labelled $S$.
Higher-order off-shell, short-distance, contributions reside 
in subdiagram $H$, adjacent to the decay vertex.  No other
physical processes, in particular no finite momentum transfers between
the two jets, can be realized, because once the jets are formed
at the point represented by $H$, they travel apart at the speed of
light and can never meet again at a point in space-time.  

Power counting for the pinch surfaces corresponding to the
reduced diagrams of fig.\ \ref{vxS} is straightforward.  A natural
choice of normal variables is (i) all four components $k^\mu$ of
each loop momentum internal to the soft subdiagram $S$, and
of each loop momentum that links the soft subdiagram with the
jets;  (ii) the total invariant mass $\ell^2$ and the 
scalar product $\ell\cdot p$ for each loop $\ell$ in $J$,
and similarly for $J'$.  All of these variables may be
scaled with $a_i=1$ in eq.\ (\ref{normscale}).  The
resulting homogeneous integral has no pinch surfaces, except
those that correspond to reduced
diagrams of the same form as fig.\ \ref{vxS} \cite{jccrv,Stermanbook}.
The scaling powers of lines are $A_j=1$ for lines internal
to $J$ and $J'$, and $A_j=2$ for soft lines in $S$.
Self-energies on jet lines need not be considered explicitly,
because each full two-point function $G_2(k)$  has only a single pole
at $k^2=0$.  

Consider an arbitrary pinch surface $T$
with $N_J$ lines and $L_J$ loops in $J$, and similarly for $J'$ and $S$.
We assume in addition that $M_J$ soft lines attach $S$ to $J$, and
$M_{J'}$ soft lines attach it to $J'$.
The power counting measure 
is then, according to eq.\ (\ref{nsubS}),
\eq
n_T= 4(L_S+M_J+M_{J'}-1) +2L_J+2L_{J'}-2N_S-N_J-N_{J'}+s_T\, ,
\label{pcforvertex}
\eqe
with $s_T$ the numerator suppression factor.  
The contribution of all soft lines, loops and numerator 
momenta internal to $S$
is just the dimension of $S$.
For technical simplicty,
we shall assume
that the jets attach to $H$ each
by a single line, and that the soft lines are all gluons, attached to the
jets only at three-point vertices.  The dimension of $S$,
including its external lines, is then
$4-3M_J-3M_{J'}$, and

\eq
n_T= M_J+M_{J'} +2L_J+2L_{J'}-N_J-N_{J'}+s'_T\, ,
\label{pcforvertex2}
\eqe
where $s'_T$ meaures the power associated with numerator factors
that come from jet lines and vertices.

It is not difficult to find a lower
bound for the numerator
suppression factor $s'_T$.  We
note that there is a factor of numerator momentum for each three-point
vertex in the jets, and that each of these momentum factors will have
to combine to form an invariant.  Each such invariant will scale as
$\l$ if it is the scalar product
of two momenta within a jet, but as 
$\l^0$ if it involves momenta from different jets.  On the other hand, the only
way the latter may occur is if each factor of momentum is 
contracted with the spin tensor of one of the soft gluons.  We 
conclude that
\eq
s'_T\ge {1\over 2}(V_{3,J}+V_{3,J'}-M_J-M_{J'})\, ,
\label{numsupp}
\eqe
where $V_{3,J}$ is the number of three-point vertices internal to jet $J$.
Substituting (\ref{numsupp}) into (\ref{pcforvertex2}) we have
\eq
n_T\ge {1\over 2}(M_J+M_{J'}) +2L_J+2L_{J'}-N_J-N_{J'}+{1\over 2}(V_{3,J}+V_{3,J'})\, .
\label{pcforvertex3}
\eqe
Now for each jet we have the Euler 
identity
\eq
L_J=N_J-\sum_iV_{i,J}+1\, ,
\label{euleridentity}
\eqe
where $V_i$ is the number of $i$-point vertices in the jet
(counting the hard part $H$ as a one-point vertex for each jet), 
as well as the relation
\eq
2N_J+M_J+1=\sum_i i\; V_i\, ,
\label{linesvertices}
\eqe
which takes into account the two external lines in
each jet, one attached to $H$, and one in the final state.
It is now a rather straightforward exercise to show that
\eq
n_T\ge 0\, ,
\label{nTzero}
\eqe 
which shows that at pinch surfaces like fig.\ \ref{vxS}, the integral
is at worst logarithmically divergent.

In addition, a closer look, using the same reasoning, shows that
$n_T=0$, corresponding to logarithmic divergence, can occur only when the
reduced diagrams for each jet are themselves of the forms of  standard 
perturbative 
diagrams, with only three- and four-point vertices internal to the
jets.  A five- or higher-point vertex in a reduced diagram always leads
to suppression in the infrared.  In addition, we also find that
$S$ is connected to the jets by soft gluons only; the 
interaction of soft quark lines with jet lines is similarly suppressed,
although soft quark loops are possible internal to $S$.  It is clear that
attachments of soft lines of any flavor to $H$ will be suppressed relative
to diagrams where they are attached only to jets.

In the above, we assumed that the jets attach to 
$H$ by a single line.  
This assumption depends on the choice of gauge.
 In an axial gauge $n\cdot A=0$, the form of 
the propagator
\eq
G_{\mu\nu}(k,n)=\left ( -g_{\m\n}+{k_\m n_\n\over n\cdot k}
+{n_\m k_\n\over n\cdot k} -{k_\m k_\n\; n^2\over (n\cdot k)^2} \right )
{1\over k^2+i\e}\ ,
\eqe
leads to a suppression whenever it is contracted with the  momentum of 
the gluon itself.  That is, the combination 
\eq
k^\mu G_{\mu\nu}(k,n)={n_\nu\over k\cdot n}-{k_\nu n^2 \over (k\cdot n)^2}\, ,
\label{scalarsupp}
\eqe
 has no pole at $k^2=0$, and hence does vanish except at $k=0$.
This leads to a contribution to $s'_T$ of $1/2$ in (\ref{numsupp})
whenever $k$ is a jet line,
even if the line $k$ attaches a jet to the hard part $H$. 
Then it is easy to verify that in
an axial (or other physical) gauge, reduced diagrams that give
infrared divergences have only a single line connecting each jet
to $H$, as assumed above.  

In covariant gauges, such as the Feynman gauge, 
there is no such suppression, and multiple collinear gluons may
attach $J$ or $J'$ to $H$.  This is not quite the complication it
appears to be, however, for the following reason.  
At the pinch surface, we can have $n_T=0$ only when each of these
gluons appears in the combination
\eq
k^\mu G_{\m\n}(k)={-k^\n\over k^2}\, ,
\label{scalarfeyn}
\eqe
in which the gluon propagator is replaced by an unphysical
``polarization" vector $k^\n$, equal to the gluon's
momentum.  Such gluons are sometimes described 
as ``scalar-polarized".  As we shall see, scalar-polarized
gluons decouple from the hard scattering \cite{jccrv}.

In summary, reduced diagrams associated with logarithmic divergence
in the electromagnetic (or other electroweak) vertex are
characterized by a simple two-jet structure, 
with a single hard-scattering function, in which the jets
are connected to the hard part by single lines in physical
gauges, and in which the soft part is connected to the
jets by soft gluons only, and not to the hard part.

Most of these observations are much more general than
the electromagnetic form factor from which they have been
derived.  In the following, we shall apply the tools we have
developed to a wide set of processes.

\section{Short Distance Cross Sections and Unitarity}

We are now ready to discuss a class of physical quantities that can
be proved infrared safe by a combined analysis of their analytic structure
and power counting.
They are primarily cross sections initiated by timelike electroweak
currents, the prime examples being the total and
jet cross sections in ${\rm e}^+{\rm e}^-$ annihilation \cite{StermanWeinberg}.  Other
examples include the decay width of the Z and W, and various event shapes defined
to describe cross sections and decay rates of this type.  The analysis of these cross sections 
is related to the operator product expansion, 
but gives many results that cannot be derived directly 
from the operator product expansion.

\subsection{Cut diagrams and generalized unitarity}

\begin{figure}[ht]
\centerline{\epsffile{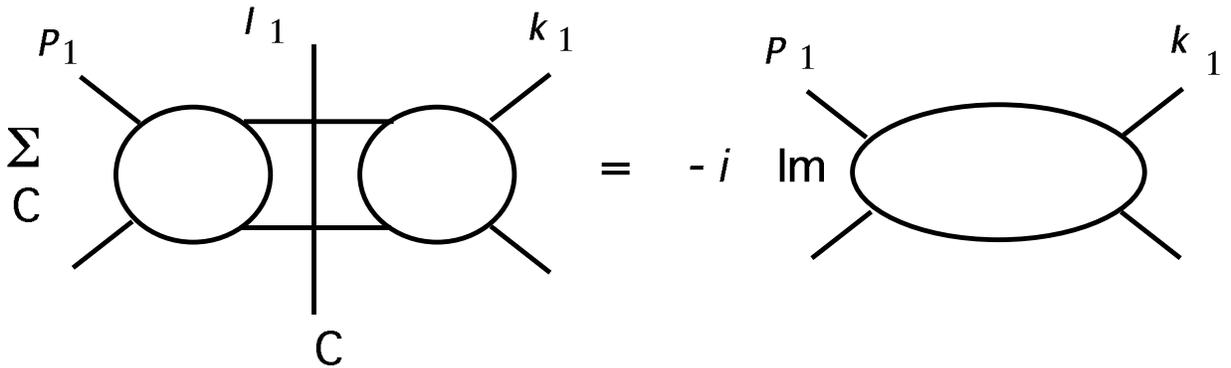}}
\caption{Cut diagram and unitarity.}
\label{cutdiagram}
\end{figure}

To discuss cross sections of this type, we introduce the ``cut
diagram" notation shown on the left-hand side of fig.\ \ref{cutdiagram}.
This diagram represents 
the amplitude for a process in which a set of particles
with momenta $p_1,\dots\, p_n$ scatter into a set 
$\ell_1,\dots\, \ell_{m}$, times the complex conjugate amplitude for the
latter to scatter into a third set $k_1,\dots\, k_{n'}$,
integrated over part or all of the phase space of the
intermediate set.
Perturbation theory rules to the left of the cut are
the usual ones, those to the right their complex conjugates.
Each such cut, which
we label $C$ in fig.\ \ref{cutdiagram}, specifies a distinct intermediate state.
We shall denote a particular cut diagram found from
uncut diagram $G$ by cut $C$ as $G_C$.  

Fig.\ \ref{cutdiagram} as a whole states a
very useful theorem satisfied by cut diagrams.
The sum over all cuts with fixed 
external momenta $p_1,\dots\, p_n$ and $k_1,\dots\, k_{n'}$ is
given by twice the imaginary part of $(-iG)$, the uncut 
diagram,
\be
\sum_{{\rm all}\ C}G_C(p_i,k_j)
=
2\; {\rm Im}\; \left (-iG(p_i,k_j)\right )\, .
\label{genunit}
\eqe
This result is a generalization of unitarity expressed
in terms of the $T$ matrix, $T=-i(S-1)$, with $S$ the
scattering matrix,
\eq
TT^\dagger=-i(T-T^\dagger)\, .
\label{Tunit}
\eqe
It is, however, more general, because it applies for
{\em fixed spatial momenta}, in both the external
{\em and} internal loops of the overall diagram $G$.
This result is proved in Appendix B by use of time-ordered
perturbation theory.

\subsection{Infrared safety for inclusive annihilation and decay}

\begin{figure}[ht]
\centerline{\epsffile{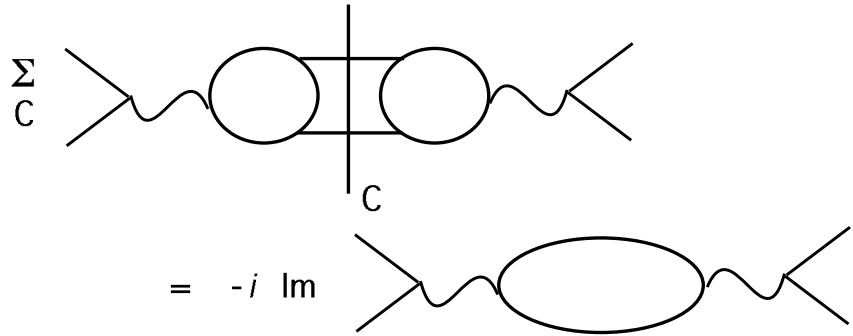}}
\caption{Unitarity applied to the total annihilation cross section.}
\label{epemtotal}
\end{figure}

The simplest physical application of the analyticity/power counting 
analysis is to the total cross section for ${\rm e}^+{\rm e}^-$
annihilation or Z decay, fig.\ \ref{epemtotal}.  
By the optical
theorem, a special case of generalized unitarity, eq.\ (\ref{genunit}),
the total cross section is proportional to the vacuum
expectation value of the time-ordered product of two currents,
\eq
 \s_{e^+ e^-}^{(\rm tot)} (q^2) = {e^2 \over q^2} {\rm Im}\, \p (q^2)\, ,
\label{sigmatot}
\eqe
where the function $\pi$ is defined in terms of the two-point correlation 
function of the relevant electroweak currents $J_\m$ (with their couplings
included) as
\eq
\p (q^2) (q_\m q_\n - q^2 g_{\m\n} ) 
=  i \int d^4x\, e^{i q x} < 0 |\; T\; J_{\m} (x) J_\n(0)\; | 0 >\, .
\label{piequalsvev}
\eqe
The proof that $\p(q^2)$, and hence its imaginary part, and therefore the
total cross section, is infrared safe requires only that we 
recognize that $\p$ represents a forward scattering process.
 There are no physical processes in which an off-shell
photon (or on-shell Z) can decay into a set of 
on-shell particles that propagate freely and then
annihilate to form another photon of the same invariant mass.  The set of 
particles originating from a point will receed in different directions,
and can never meet again by physical propagation.  
This eliminates pinch surfaces with finite momentum particles, and
hence any infrared divergences associated with decay processes.
This is a particular example of the famous Kinoshita-Lee-Nauenberg (KLN) theorem \cite{KLNtheorem}.

\begin{figure}[ht]
\centerline{\epsffile{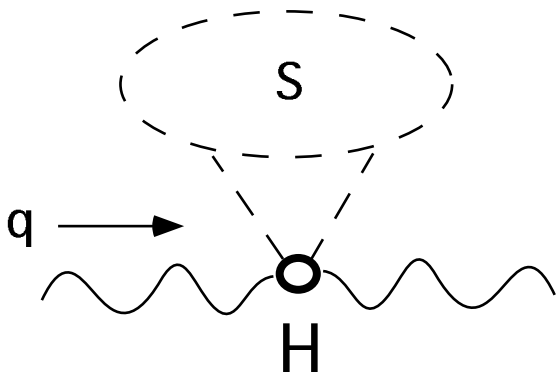}}
\caption{Pinch surface with only zero-momentum lines (subdiagram S)
coupled to a point-like hard part H.}
\label{softcloud}
\end{figure}

It is still possible to have pinch surfaces involving only massless particles,
coupled to a single hard-scattering, as in fig.\ \ref{softcloud}.
Dimensional counting shows that such pinch surfaces are finite order-by-order
in perturbation theory. 
 They will, however, lead to an important 
connection of perturbation theory, even without explicit masses, to the
operator product expansion, as we shall see in Section 7 below.  

\subsection{Jet and weighted cross sections}

The class of infrared safe cross sections is by no means exhausted 
by the totally inclusive processes above.   More detailed information
on final states is available in jet and other ``weighted" cross
sections.  

\begin{figure}[ht]
\centerline{\epsffile{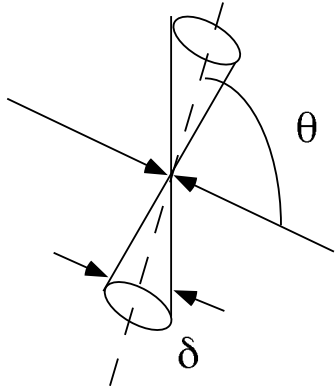}}
\caption{Cones for a two-jet cross section.}
\label{jetcones}
\end{figure}

Jet cross sections measure the probability of producing
states that are identified as jet-like, according to 
infrared safe criterea. 
 The simplest of these \cite{StermanWeinberg} is illustrated by
fig.\ \ref{jetcones}, which are defined by the
energies $e_i$ flowing into each of two cones, $\delta$,
back-to-back along a fixed axis, at angle $\theta$ from
the beam direction.  
We may define a two-jet cross section
in ${\rm e}^+{\rm e}^-$ annihilation
as one for which 
\be
{e_1+e_2\over \sqrt{s}} \ge 1-\epsilon\, ,
\label{2jeteps}
\eqe
that is, for which all the energy, up to a fraction $\e$, is
emitted into the two back-to-back cones.  If the cones are
sufficiently small, the event will ``look like" two jets
of nearly collinear particles.  Such a cross section is infrared safe, and the proof
is a variation of the proof of infrared finiteness of the total 
annihilation cross section.  

\begin{figure}[ht]
\centerline{\epsffile{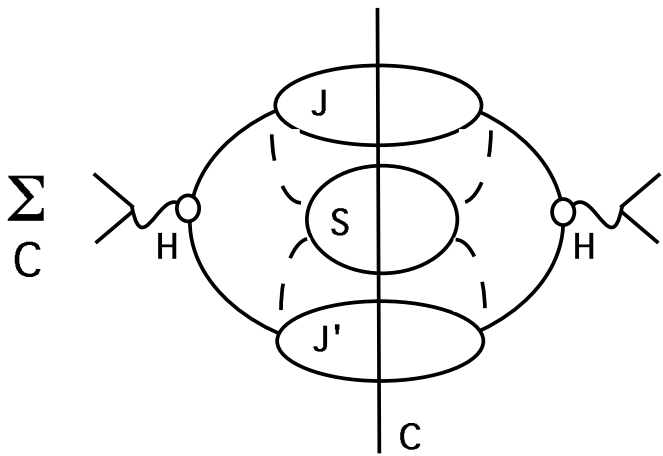}}
\caption{Cut reduced diagram for two-jet configuration.}
\label{cutjet}
\end{figure}

Consider for simplicity a two-jet cross section.  The relevant
pinch surfaces of cut diagrams are shown in fig.\ \ref{cutjet}.
Each cut C of the diagram has a final state that contributes to
the jet cross section, with one jet in each of the cones of fig.\ \ref{jetcones},
for instance.  The pinch surfaces associated with long-distance
behavior are easily verified to be of exactly the same form as
those for quark-antiquark production, fig.\ \ref{vxS}, except that
now there may be any number of particles from each jet in the
final state.  At the pinch surface, 
if cut C of reduced diagram $R$ contributes to the two-jet cross section,
then {\em every} cut of
$R$ contributes to the same cross section.
The sum over cuts, however, may then be carried out using the
generalized form of unitarity, fig.\ \ref{cutdiagram}, to derive an
integral which has no pinch surface at all corresponding to the
two-jet configuration, by exactly the same reasoning as for the
total cross secction.  

The difference between total and
jet cross sections is that momentum integrals for jet cross sections
encounter boundaries between two- and three-jet 
events (for instance), which are
absent in the total cross section.  At such points, the phase
space is discontinuous, and integrals may not be deformed after
the sum over cuts.  Manifolds of such points are of reduced
dimension, however, compared to the general
pinch surfaces of cut diagrams, which produce at worst
logarithmic divergences \cite{Sterman78,StermanTik79}.  This reduced dimension weakens their power-counting, 
and they do not give rise to 
infrared singularities order-by-order in perturbation theory.  
Similar considerations apply to pinch
surfaces with more than two jets.

In weighted cross sections, final states
are weighted according to ``shape variables", $\cs_n(p_1\dots p_n)$,
which are functions of the momenta of particles in the final state.
The shape variables may or may not be chosen to enhance jet-like
configurations.  A general cross section at fixed shape variable $\cs$
is of the form
\eq
\sigma_\cs=\sum_n\int d\tau_n\; {d\sigma \over d\tau_n}\, 
\cs_n(p_1  \ldots  p_n )\, ,
\label{weighted}
\eqe
where the $\cs_n$ are functions of the momenta in
an $n$-particle final-state, 
whose phase space is denoted by $d\tau_n$.  A weighted
cross section is infrared safe whenever the same mechanism that
produces infrared safety for jet cross sections applies to them
as well.  That is, 
it is infrared safe if,
whenever a pinch surface of a cut reduced diagram like
fig.\ \ref{cutjet} contributes to the weighted cross section, 
the weight is the same for every possible cut of the reduced diagram
at the pinch surface.
If this holds, the divergences of individual cuts cancel, since
the sum over cuts produces an integral without the corresponding
pinch surface.  This condition will be satisfied so long as the
weight function does not distinguish between states in which one
set of collinear particles is substituted for another set with the same
total momentum, or when zero-momentum particles are absorbed or 
emitted.  Quantitatively, these requirements may be summarized
by the conditions
\eq
 \cs_n (p_1 \ldots p_i \ldots p_{n-1},\l p_i ) = \cs_{n-1} (p_1 \ldots p_i + \l p_i \ldots p_{n-1})\, .
\label{irsconditions}
\eqe
Perhaps the best known weight is the ``thrust", defined \cite{Farhi} for an event
with $n$ particles of momenta ${\vec p}_i$ by
\eq
T = {1\over \sum_i \mid \vec{p}_i \mid} \; 
{\rm max}_{\hat{n}}\; \sum^n_{i=1} \mid \vec{p}_i
\cdot \hat{n} \mid \, ,
\label{Tdef}
\eqe
with the maximum taken over all unit vectors $\hat n$. 
The direction of $\hat{n}$ that produces the maximum is known as the 
``thrust axis".

Other important weighted cross sections iteratively assign particles 
into
jets of momenta $P_j$.  
The algorithm begins with each particle defined as a separate jet.
At each stage in the iterative
process, a set of variables $y_{jk}$,
\eq
y_{jk}={1\over s}f(P_j,P_k)
\eqe
are computed for each pair of jets.  
The function $f$ is chosen to be consistent with infrared
safety.  An influential choice \cite{ktalg} defines the ``$k_T$ algorithm",
based on
\eq
y_{jk}= {\rm min}\; (E_j^2,E_k^2)(1-\cos\; \theta_{jk})\, ,
\eqe
where $E_j$ and $E_k$ are the energies of jets $j$ and $k$
in the overall center-of-mass frame, and $\theta_{jk}$ is the
angle between them.  
For this, and other such variables, if one or more of the
$y_{jk}$ are smaller than a fixed value $y_{\rm cut}$, the pair
whose value is minimum is combined to form a single jet
and the process is repeated.  
The number of jets left when all $y_{jk}$ are larger than
$y_{\rm cut}$ defines the number of jets in the event,
which may be plotted against $y_{\rm cut}$ and compared 
to perturbative or event-generator predictions.

\section{Factorization and Evolution in DIS}

\subsection{Venturing out on the light cone}

Although the set of infrared safe cross sections is large, it
is a small subset of interesting scattering experiments.  
In particular, QCD cross sections in which one or more of the
incoming particles is a hadron are never fully infrared safe.
For DIS, this is already clear in the parton model, in which 
 parton distributions must be taken from experiment.
This will remain the case in field theory, where we shall discover
how to extract infrared safe quantities from many (not all!)
cross sections, a process known as {\it factorization} \cite{jccrv}.

Factorization, as mentioned at the outset, is a realization
 of the separation of long- and short-distance
dynamics in the presence of massless particles in Minkowski space.
  In this context, we shall derive a generalization
of the parton model in QCD, and a field-theoretic definition
for parton distributions \cite{CollinsSoperPDF,OwensTung}.  
Let us begin with dimensionally regularized deeply inelastic
scattering of a light-like parton.  
Now our initial state is fixed on-shell, and is sure
to include singularities in Feynman integrals assoicated
with the light cone.

\begin{figure}[ht]
\centerline{\epsffile{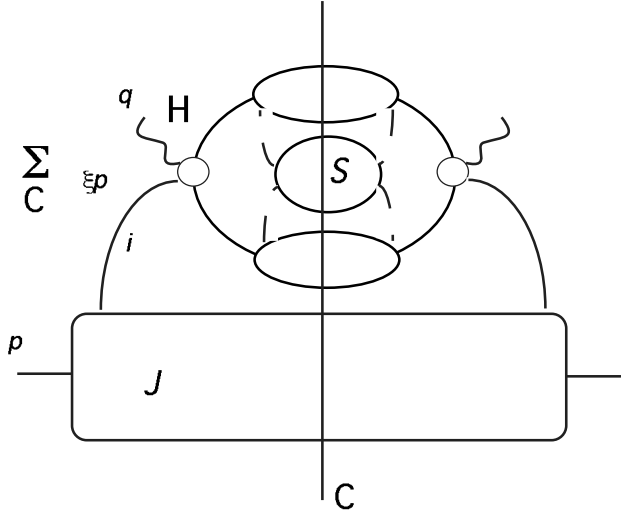}}
\caption{Pinch surface for DIS.}
\label{DISps}
\end{figure}

Amplitudes for DIS are full of collinear singularities; to
analyze them we turn again to a pinch surface analysis via
physical pictures.  That is, we look for the most general
physical process in which a single incoming parton absorbs a spacelike
photon to form a final state.
The result, analogous to fig.\ \ref{cutjet} for ${\rm e}^+{\rm e}^-$
annihilation, is shown in fig.\ \ref{DISps}.  
In this process,
the incoming particle of momentum $p$ produces a jet of collinear lines
$J(\xi)$, one
of which, 
a parton of flavor $i$ with momentum $\xi p$, absorbs the virtual photon 
(or other electroweak boson) of momentum $q$
at $H$, producing a set of outgoing jets.  All jets, including the remains of
the incoming jet of lines collinear to $p$, propagate into the
final state.  The jets may interact mutually only through the exchange of
soft particles.  Because all particles in the physical picture must
have positive energy, we find that
\eq
1 > \xi > x\, .
\eqe
Thus, the struck particle is a parton of the incoming
particle, with a proper fractional momentum.  That there
is only one struck parton, at least
in a physical gauge, follows the power counting of Sec.\ 2.4, especially
 the discussion of eq.\ ({\ref{scalarsupp}).
An exception is that in covariant gauges the physical parton may share
its momentum with unphysical, scalar-polarized gluons.  We shall
incorporate this possibility below.  

Next, 
if we sum over all final states at fixed $q$, soft and collinear
divergences associated with the final-state jets and their interactions
cancel, by the same unitarity arguments  as for ${\rm e}^+{\rm e}^-$ annihilation.  
This may be
verified by using the relation of the hadronic tensor $W_{\mu\n}$
to the forward Compton amplitude $T_{\m\n}$,
\eq
W_{\m\n} = 2\; {\rm Im}\, T_{\m\n}
\eqe
with
\eq
T_{\m\n} = {i\over 8\pi} \int d^4x\, e^{i q \cdot x}  < h (p) |\; T\; J_\m (x)
J_y (0)\; | h (p)>\, .
\eqe
The absence of pinch surfaces involving final-state jets
in the forward-scattering amplitude follows easily from the
much simpler physical pictures for $T_{\m\n}$, which involve
the incoming jet only, as in fig.\ \ref{Cptps}.  
\begin{figure}[ht]
\centerline{\epsffile{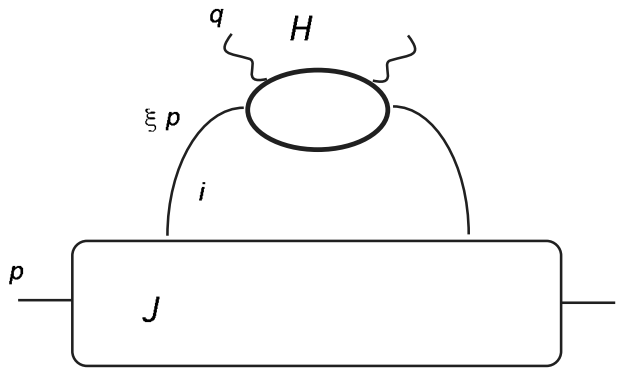}}
\caption{Pinch surface for forward Compton scattering.}
\label{Cptps}
\end{figure}
Note that,
as in ${\rm e}^+{\rm e}^-$, this result, and the cancellation
of infrared sensitivity to the final state, applies to any
weighted DIS cross section consistent with eq.\ (\ref{irsconditions}),
and not only to fully inclusive DIS.  Infrared sensitivity from the
incoming jet remains in all these cases, of course.

In summary, in the neighborhood of {\em any} of its pinch surfaces,
the DIS hadronic tensor takes a form that is very suggestive of 
the parton model,
\eq
W^{\m\n} = \sum_i \int_x^1 d \x \; H^{\m\n}_{i, \b \a} \; (q, \x p)
J_{i, \a\b} (\x)\, .
\label{firstfact}
\eqe
The strategy of factorization is to separate the long-distance
contributions from the hard-scattering part.  Unlike the operator
product expansion in Euclidean space, however, such a factorization
cannot be expressed as a finite sum of coefficient
functions times matrix elements, even to the leading power in $Q^2$.
This is because the short-distance part and the incoming jet remain
in covolution form, tied together by the fractional momentum $\xi$ 
of the struck parton.  The finite sum, therefore, is replaced
by a convolution of functions.  

As in the operator product expansion,
however, we can express the long-distance function in terms of the
matrix elements of partons that connect the incoming hadron to
the hard scattering.  Thus, the long-distance contributions of 
all diagrams with pinch surfaces like those in
fig.\ \ref{DISps}, when the scattered parton
$i$ is a quark are summarized by matrix elements of the form 
\eq
J_{q, \a\b} = {1\over 2} \sum_{\stackrel{\rm spin}{\s}}
\int^\infty_{-\infty} {d
y^- \over 2 \p} e^{-i \x p^+ y^- } 
 < h(p, \s) \mid \bar{q}_\a (0^+ , y^-, {\bf 0}_\perp ) q_\b (0) \mid  h(p, \s) >\, ,
\label{qktensorme}
\eqe
with $\a$ and $\b$ Dirac indices.  Instead of a local composite operator,
we find a product of operators separated by a light-like distance $y^-$.
In this fashion, the  plus momentum of the parton is fixed,
although its remaining components, to which $H$ is insensitive,
are integrated over.  As we shall see shortly, the matrix elements of such 
composite operators are ultraviolet
divergent in perturbation theory, and require renormalization.  
This
is similar to the local composite operators of the OPE.  
The effective  ultraviolet
cutoff for the matrix 
element is referred to as the ``factorization scale", separating the
short-distance from the jet functions.  

A Fierz projection between the hard
and jet functions, followed by some power counting, shows that
the leading Dirac structure may be projected out by the trace of
$J_{\a\b}$ with a single Dirac matrix, for unpolarized scattering.  This 
results in the following definition for the quark distribution 
\cite{CollinsSoperPDF},
\eq
 \f_{q/h} (\x, \m^2) = {1\over 2}  
\sum_\s \int^\infty_{-\infty}  {d y^-
\over
2\p} e^{-i \x p^+ y^- }
\, 
<  h(p, \s) \mid \bar{q} (0^+, y^-, {\bf 0}_\perp)
\, {1\over 2}n \cdot \g\;
 q (0) \mid  h(p, \s) >\, ,
\label{phiqazero}
\eqe
where $n^\m$ is the lightlike vector directed oppositely to the incoming momentum
$p$, $n^\m=\delta_{\m -}$.
Similarly, when the scattered parton is a gluon, we encounter the distribution
\eq
\f_{G/h} (\x, \m^2) = {1\over 4\p \x p^+} \int dy^- e^{-i \x p^+ y^-} 
\, 
 \sum_\s
\sum^2_{\m=1} <  h(p, \s) \mid {F^+}_\m (0, y^-, {\bf 0} ) F^{\m +} (0) \mid
 h(p, \s) >\, ,
\label{phigazero}
\eqe
now a matrix element of the field strengths $F_{\mu\n}$.
These matrix elements are illustrated in fig.\ \ref{qkdist}a.
\begin{figure}[ht]
\centerline{\epsffile{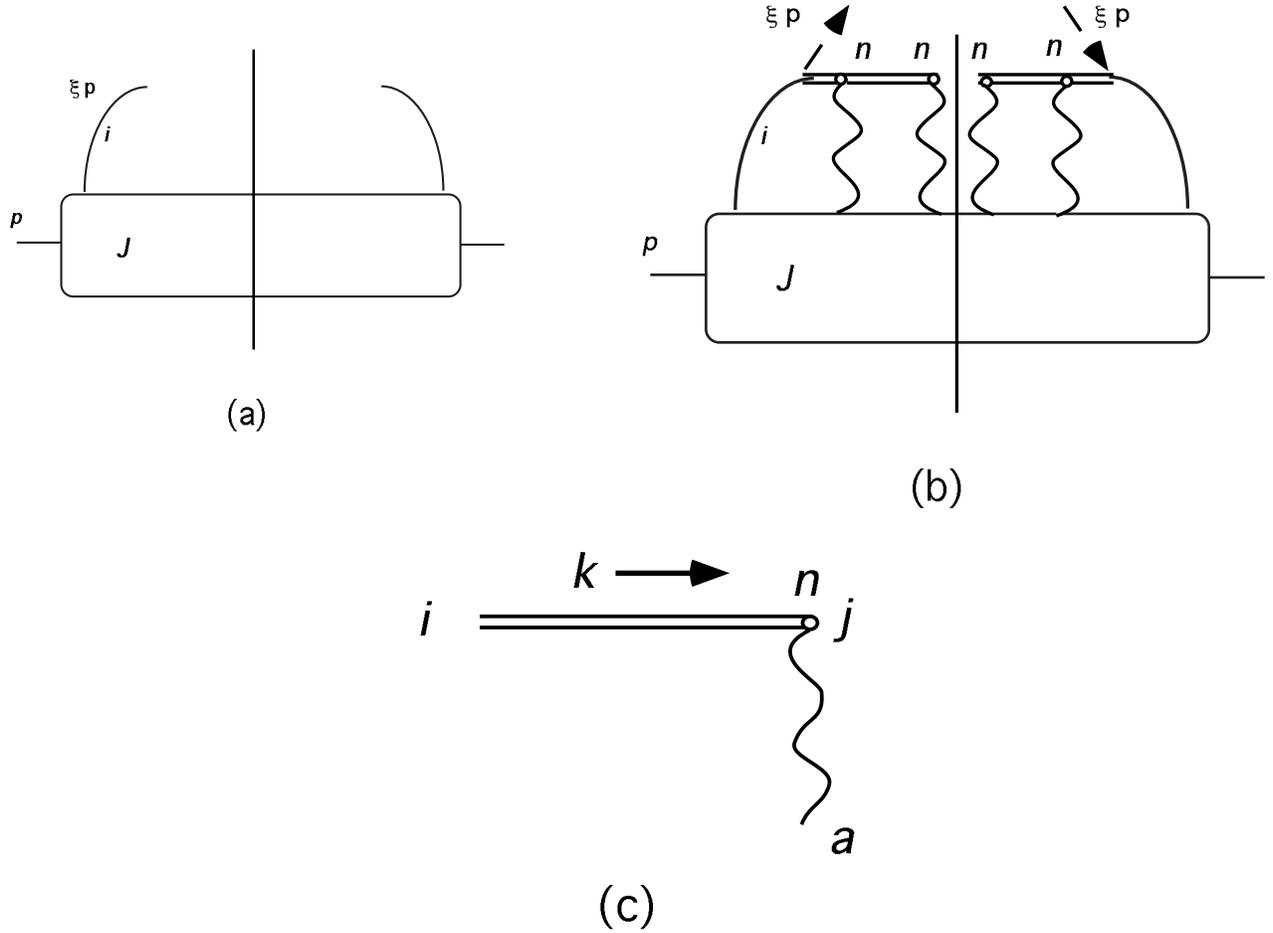}}
\caption{(a) Gauge-variant parton distribution (flavor $i$).
(b) Gauge-invariant parton distribution with ordered exponential
in the $n$ directions. (c) Graphical representation of the ``eikonal"
propagator and vertex.}
\label{qkdist}
\end{figure}

In summary, the inclusive DIS hadronic
tensor, and hence its structure functions, may be written in the following
factorized form, which generalizes the OPE and which is
clearly a justification of the use of the parton model in this process,
\eqa
F_2{}^{(h)} (x, Q^2) &=& \sum_{i = f, \bar{f}, G} \int^1_x d \x\
C_2{}^{(i)} \left (
{x\over \x}, {Q^2 \over \m^2}, \a_s (\m^2) \right )  \f_{i/h} (\x, \m^2) 
\nonumber\\
F_1{}^{(h)} (x, Q^2) &=& \sum_{i=f, \bar{f}, G} \int^1_x 
{d \x \over \x}\ 
C_1{}^{(i)} \left ({x\over \x}, {Q^2 \over  \m^2}, \a_s (\m^2) \right ) 
\f_{i/h} (\x , \m^2)\, .
\label{fonetwofact}
\eqae
Compared to the parton model, the coefficient functions are now 
expansions in the strong coupling, and the parton distributions
have become functions of the factorization scale.

\subsection{Gauge invariant distributions; schemes}

The distributions, (\ref{phiqazero}) for the quark,
and (\ref{phigazero}) for the gluon, have been justified
so far only in physical gauges, where only a single, physical
parton connects the hard scattering and the $p$-jet on
either side of the cut in fig.\ \ref{DISps}.  At the same time,
these distributions are not gauge invariant.  
Their gauge variations, however
 may be absorbed
in the hard scattering functions. 
This is because a change in the gauge of the matrix elements 
(\ref{phiqazero}) or (\ref{phigazero}) is equivalent to a 
phase acting on the quark or gluon operators only.

Alternately, we may define
gauge-invariant distributions by connecting the physical quark
and gluon fields in the matrix elements with ordered exponentials
(Wilson lines) along the $n^\m$-light cone between the fields \cite{CollinsSoperPDF},
\eqa
{\bar q}(y^-)\; n\cdot \gamma\; q(0)
&\rightarrow& 
\bar{q} (y^-) P \exp \left[ -i g \int_0^{y-} d \l\;
n\cdot A (\l n^\m) \right] n\cdot \gamma\; q (0)
\nonumber\\
\sum^2_{\m=1} {F^+}_\m (y^-) \, F^{\m +}(0)
&\rightarrow& 
\sum^2_{\m=1} {F^+}_\m (y^-)\; P 
\exp \left [ -i g \int_0^{y^-} d \l\; n\cdot A(\l n^\m) \right ] 
F^{\m +}\, ,\nonumber\\
\label{giprods}
\eqae
where $n\cdot A$ in the quark distribution is in the fundamental (quark)
representation, and $n\cdot A$ in the gluon
distribution is in adjoint representation.  The gauge invariant
distributions reduce to (\ref{phiqazero}) and (\ref{phigazero})
in $n\cdot A=0$ (often identified as $A^+=0$) gauge.  
Again, any of these definitions require renormalization.
They are commonly
defined by $\overline{\rm MS}$ 
prescriptions, and are referred to as  $\overline{\rm MS}$
distributions. 

Gauge-invariant parton distributions have perturbative expansions,
defined by the diagrams of fig.\ \ref{qkdist}b, which include 
propagators and vertices generated by the ordered exponentials
of eq.\ (\ref{giprods}).  The combination of a propagator and  a vertex, illustrated
 by fig.\ \ref{qkdist}c, is given by the expression
\eq
{i\over n\cdot k+i\e}\; \left ( -ign^\m\left (T_a^{(R)}\right )_{ji}\right )\, ,
\label{eikfeynrule}
\eqe
where $R=F$ for the quark distribution and $A$ for the gluon.  The linear
propagator is sometimes referred to as an ``eikonal line".

The $\overline{\rm MS}$ matrix elements just defined are not
the only possible choices for the distributions.  Other choices
may be defined by convolution with any IR safe distribution
$D$, as
\eq
\f_{i/h}^{(D)} (\x, \m^2) = 
\int^1_\x \; {d \h \over \h} D_{ij} \left (\x/\h,
\a_s (\m^2)
\right ) \f_j^{(\overline{\rm MS})} \;  (\h, \m^2)\, ,
\label{schemechange}
\eqe
which preserves the factorization (\ref{fonetwofact}).  Of
particular interest is a choice that absorbs the full
$\overline{\rm MS}$ $F_2$ DIS coefficient function $C_2$.
Defining $C_2^{(j,\overline{\rm MS})}(z)=\sum_f Q^2_f \bar{c}_{fj}(z)$,
we take
\eq
D_{fj}(\xi/\h) = (\h/\x)(\bar{c})_{fj}(\x/\h)\, .
\label{dispdf}
\eqe
In this ``DIS scheme", the parton model relation
\eq
\sum_f Q_f^2\; x\; \f_{f/n}^{\rm (DIS)} (x, \m^2 ) = F_2{}^{(h)}(x, \m^2)\, ,
\label{pmFspqcd}
\eqe
in eq.\ (\ref{pmFs}),
holds by construction to all orders in perturbation theory.  Note,
however, that the Callan-Gross relation in (\ref{pmFs}), which
involves $F_1$ as well as $F_2$, cannot be exact 
at the same time, and so inherits corrections even in DIS 
scheme.

\subsection{One-loop distributions and coefficient functions}

We are now in a position to compute corrections to parton
model relations such as (\ref{pmFs}) in QCD.  The
complexity of these calculations increases precipitously
with order, but the pattern is well illustrated by
one-loop considerations. 

The first goal is to compute infrared safe
coefficient functions $C_a^{(i)}$, 
in a given factorization scheme, by
comparing {\it perturbative} expressions for structure functions
$F^{(f)}_a$ of partons of flavor $f$ with the distributions
$\phi_{i/f}$ of parton $i$ in parton $f$, using
the factorized expressions eq.\ (\ref{fonetwofact}), 
expanded to the appropriate order in $\alpha_s$.  Neither 
$F_a^{(f)}$ nor $\phi_{i/f}$ is infrared safe, 
and must be defined through dimensional (or some other) regularization, but the
resulting coefficient functions are infrared safe.  Once the coefficient functions
are determined to some order, the same factorized
expressions, (\ref{fonetwofact}), may be applied to
experimental measurements of the $F_a^{(h)}$ to determine the
physical values of parton distributions in hadrons $h$.  This
process is similar to  the parton
model, and as in the parton model, many predictions result
by using the parton distributions in different processes.

The one-loop corrections to quark DIS are 
given by the cuts of the graphs in fig.\ \ref{quarkdis}a.
The one-loop corrections of the perturbative distribution $\phi_{f/f}$ of a quark in
a quark (flavor $f$) are given in fig.\ \ref{quarkdis}b for $A^+=0$ gauge.
\begin{figure}[ht]
\centerline{\epsffile{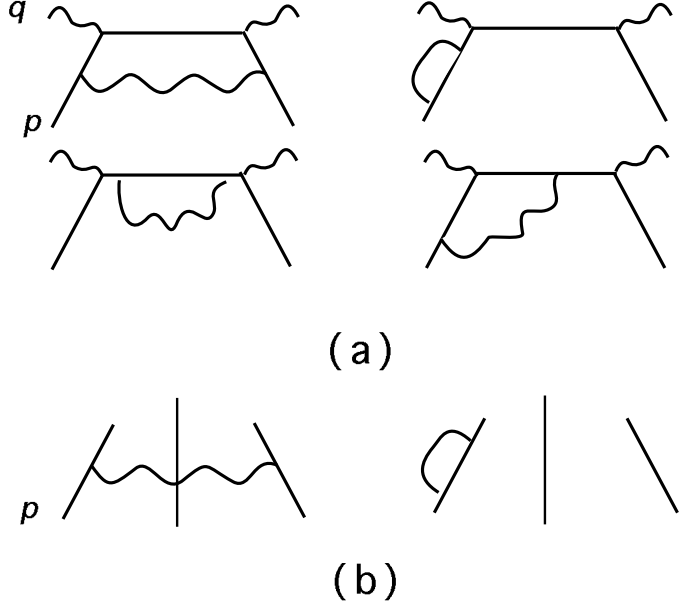}}
\caption{(a) Uncut corrections to quark DIS. (b) One-loop
corrections to distribution of a quark in a quark in $A^+=0$
gauge.}
\label{quarkdis}
\end{figure}
A short calculation shows that the gluon emission diagram in \ref{quarkdis}b is given 
in $n$ dimensions by
\eq
\f_{f/f}^{(1)}(\x,\m^2) = \a_s \m^{2\e} 
{C_F \over (2 \p)^{n-2}} \;
{2(1+\x^2)\over 1-\x} \; \int \; {d^{n-2} k_T \over k_T{}^2}\, .
\eqe
The phase space of the gluon is reduced to a transverse momentum
integral by fixing the longitudinal momentum carried by the
upper quark line to $\xi p$.  The remaining integral is, as 
expected, divergent at $k_T=0$, corresponding to a collinear
divergence from gluon emission.  As anticipated above, it is
also ultraviolet divergent.  Such a scaleless integral must be
defined to vanish in dimensional regularization.  This does not
mean that the distribution is zero, however.  Rather,
the scaleless $k_T$ integral is replaced by a
$\x$-dependent ultraviolet counterterm,
which  removes the unphysical contribution of 
transverse momenta much larger than the factorization scale
\cite{CollinsSoperPDF}.
In this manner, we define the $\overline{\rm MS}$ distribution
at one loop as a ``pure counterterm".  Adding the contribution
of the virtual diagram, which is proportional to $\delta(1-\x)$,
we find the explicit expression 
\eq
\f_{f/f}^{(1)} (\x, \m^2) = {\a_s \over 2 \p} 
\left ( {1\over -\e} +\g_E-\ln 4\p \right ) P^{(1)}_{q/q} (\x )\, ,
\label{oneloopqdist} 
\eqe
where the distribution $P^{(1)}_{q/q}$ is given by
\eq
P_{q/q}^{(1)} (\x) \equiv C_F \left \{ (1+ \x^2) 
\left [ {1\over 1-\x} \right ]_+  + {3\over 2} \; \d(1-\x ) \right \}.
\eqe
For reasons that will become clear shortly, $P^{(1)}_{q/q}$ is referred
to as an ``evolution kernel".
The ``plus distribution" is defined through its integrals with smooth
functions $f(\x)$ by
\eq
\int_x^1 d\x\; f(\x)\; \left [ {1\over 1-\x} \right ]_+
= \int_0^1 d\x\; \left ( {f(\x)-f(1) \over 1-\x} \right )
-
\int_0^x d\x\; {f(\xi) \over 1-\x}\, .
\label{plusdistdef}
\eqe

Other distributions $\phi_{i/j}(\xi)$ of parton $i$ in parton $j$
 at one loop have exactly the
form (\ref{oneloopqdist}), 
\eq
\f_{i/j} (\x, \e) = {\a_s \over 2 \p} 
\left ( {1\over -\e} + \g_E - \ln 4\p \right ) 
P_{i/j}^{(1)} (\x) + \ldots\, .
\eqe
Each $P_{i/j}^{(1)}$ is one of the set of evolution kernels,
\eqa
P^{(1)}_{q/q} (\x) &=& C_F [2D (\x) -1-\x  + {3\over 2} \d (1-x) ] \nonumber\\
P^{(1)}_{q/G} (\x) &=& T_F  [(1-\x)^2 + \x^2 ] \nonumber\\
P_{G/q}^{(1)} (\x) &=&  C_F [ {(1- \x)^2 + 1 \over \x} ] \nonumber\\
P^{(1)}_{G/G} (\x) &=& 2 C_A [\x D (\x) + 
\left ( {1\over \x}+\xi \right )(1-\x) +
{11\over 12} \; \d
(1-\x)]-{1\over 3} \;  n_f \d \; (1-\x)\, ,
\label{Evolfuns}
\eqae
where
$D(\xi)\equiv [1/(1-x)]_+$.  

At zeroth order
in the strong coupling, we have
\eq
\f_{f^\prime / f}^{(0)}(\xi) = \d_{f^\prime f} \d (1-\xi)\, ,
\label{philowest}
\eqe
for any flavors $f$ and $f'$, which simply states that
without interaction parton $f$ remains itself.
Then, at lowest order 
in (\ref{fonetwofact})
we recover the partonic results
\eq
F_2{}^{(f)} (x) = Q_f{}^2 \d (1-x) = C_2{}^{(f)} (x)\, .
\label{ftwolowest}
\eqe

Finally, the perturbative coefficient functions may be determined
from the one-loop expansion of eq.\ (\ref{fonetwofact}).
For instance, at one loop, we have (using (\ref{philowest})
and (\ref{ftwolowest})),
\eq
F_2{}^{(f, 1)} (x, Q^2) 
-
\sum_{f'} Q_{f'}^2 x \f^{(1)}_{f'/f} (x)
= 
C_2{}^{(f, 1)} \left (x, {Q^2 \over \m^2} , \a_s (\m^2) \right ) \, .
\label{oneloopFCphi}
\eqe
At one loop, the electromagnetic structure functions 
of fig.\ \ref{quarkdis}a are (for details of the
calculation, see \cite{oneloopdy,Stermanbook})
\eqa
F_2{}^{(f, 1)} &=& Q_f{}^2 x \bigg [  {\a_s \over 2\p} \; {1\over
(-\e)}
P_{q/q}^{(1)} \left (1 - \e \g_E +\e\ln {4{\p\m^2} \over Q^2} \right ) 
\nonumber\\
&\ &  \quad\quad\quad + C_F \big \{ (1-x^2) 
\left ( {\ln (1-x) \over 1-x} \right )_+ 
- {3\over 2}
{1\over (1-x)_+} \nonumber\\
&\ &  \quad\quad\quad - 
\left ( {9\over 2} + {\p^2 \over 3} \right ) \d (1-x) - (1+x^2) {\ln x \over 1-x} +
3+2 x
\big \} \bigg ] \nonumber\\
F_1^{(f, 1)} &=& {1\over 2x} F_2{}^{(f, 1)} - Q_f{}^2\; C_F {\a_s \over
2\p} x\, .
\label{f12oneloop}
\eqae
The determination of one-loop quark  coefficient
functions in the $\overline{\rm MS}$ and DIS schemes is now a matter of subtraction,
using (\ref{oneloopqdist}) and (\ref{f12oneloop}) in (\ref{oneloopFCphi}) 
for $\overline{\rm MS}$, and then (\ref{schemechange})-(\ref{pmFspqcd})
for the DIS scheme.  

\subsection{Evolution}

The coefficient functions $C_a$ in eq.\ (\ref{fonetwofact})
depend, beyond lowest order, on the momentum transfer $Q^2$.
This ``scale breaking" is a refinement on the scaling behavior
of the parton model.  The dependence of the coefficient
functions, and hence of the structure functions themselves,
on momentum transfer can be computed.

The factorization formulas eq.\ (\ref{fonetwofact}) depend, as we have
seen, on a factorization scale $\mu$, at which
short and long distances are separated.  $\m$ may
be interpreted, in turn, as the renormalization scale associated
with the $\overline{\rm MS}$ renormalized 
parton densities, matrix elements such as
eq.\ (\ref{phiqazero}).  In a manner analogous to 
 composite operators in the
operator product expansion, the parton densities have calculable
$\mu$-dependence.  
This remarkable result is referred to
as ``evolution".  Because the 
coefficient functions depend on $Q$ only through the ratio $Q/\mu$,
their evolution in $\mu$ determines their dependence on 
$Q$.  This enables us to relate structure
functions and parton densities measured at one scale 
to other scales, greatly extending the applicability
of perturbative analysis.

It is simplest to illustrate evolution with a ``nonsinglet"
distribution.  An example is the difference between proton
and neutron structure functions,
\eq
F_a{}^{(NS)} = F_a{}^{(p)} - F_a{}^{(n)}\, ,
\label{nsdist}
\eqe
which combines the contributions
of the two nucleon states weighted by (twice) their isospin.
In this combination, the contributions of gluons and ``sea"
quark pairs produced in virtual processes cancel, leaving
factorized expressions for the $F_a{}^{(NS)}$ in terms of
``valence" quark distributions $\phi^{(\rm val)}$.
In this case, we have  the difference between p and n valence
distributions of each flavor, but for massless quarks the
coefficient functions are all the same up to factors of
$Q_f^2$, and we may suppress partonic indices,
\eq
F_1{}^{(NS)} = \int^1_x 
{d \x \over \x} C_1{}^{(NS)} \left (x/\x , Q^2 /
\m^2 , \a_s(\m^2) \right )  \f^{({\rm val})} (\x, \m^2)\, .
\label{nsfact}
\eqe

Under moments with respect to $x$, defined as
\eq
\bar{f} (n) \equiv \int^1_0 dx\; x^{n-1} f (x)\, ,
\label{momentdef}
\eqe
the nonsinglet structure function factors into a product 
of moments, one for the parton distribution and one for the
coefficient function,
\eq
\bar{F}_1{}^{(NS)} (n, Q^2) = \bar{C}^{(NS)}\left  (n, Q^2/\m^2, \a_s (\m^2) \right )
\bar{\f}^{({\rm val})} (n, \a_s (\m^2))\, .
\eqe
The essential ingredient in evolution is the
independence of the physical structure functions from
the factorization scale $\mu$,
\eq
\mu{d\over d\m} \bar{F}_1{}^{(NS)} (n, Q^2) = 0\, .
\label{muindepf}
\eqe
Given (\ref{nsfact}) and (\ref{muindepf}), $\f$ and $C$
obey the joint evolution equations,
\eq
\m {d\over d \m} \ln \bar{\f} \left (n, \a_s (\m^2) \right ) =
 - \g_n \left (\a_s (\m^2) \right ) = - \m
{d\over d\m} \ln \bar{C}_1{}^{(NS)} \left (n, Q^2/\m^2, \a_s (\m^2) \right)\, .
\label{gammandef}
\eqe
The anomalous dimension $\gamma_n$ is a function of $\alpha_s$ only,
since this is the only variable which $\f$ and $C$ have in common.

Logarithmic $Q^2$ dependence in moments of $F$ may conveniently be
computed from (\ref{gammandef}) by setting $\mu=Q$, and solving for
the $\mu$ dependence of $\bar{\f}(n)$.  Contributions from $\bar{C}(n)$ will
then be a simple expansion in $\alpha_s(Q)$, since all logarithms
of $Q/\mu$ will vanish.  The relevant solution for $\bar{\f}$ is
\eqa
\bar{\f}^{({\rm val})} (n, \m^2) 
&=& \bar{\f}^{({\rm val})} (n, \m_0{}^2)
\, \exp \left\{ - {1\over 2} \int_0^{\ln \m^2 /\m_0{}^2} 
d t\; \g_n \left (\a_s (\m_0{}^2 e^t) \right ) \right\}
\nonumber \\
&=& \bar{\f}^{({\rm val})} (n, \m_0{}^2)
\, \exp \left\{ -{2\gamma^{(1)}_n\over b_2}\int_0^{\ln \m^2 /\m_0{}^2}
{d t\over t+\ln(\m_0^2/\L^2)} +\cdots \right \}
\, ,
\label{phvalevolve}
\eqae
where $\g_n(\a_s)=(\a_s/\p)\g_N^{(1)}+\dots$.
Using the asymptotically 
free running coupling eq.\ (\ref{alphadef})
in the second line of (\ref{phvalevolve}), we find that
the resulting $\mu$ (and hence $Q$) dependence is logarithmic in QCD, which
accounts for the mild nature of scale breaking.  For a frozen
coupling, or one that runs to a finite value, scale breaking 
is power-law in $Q$, in apparent contradiction to the successes
of the parton model.  This is a fundamental success of quantum
chromodynamics, which played a central role in its acceptance.
Indeed, the results described here meet the challenge set in
Section 2 above, to use asymptotic freedom to account for the
approximate scaling of DIS in the presence of multiple scales.

Although the analysis we have just described is particularly
simple for moments, in most practical cases, it is best to
work with the parton distributions themselves.  The full set of
moment evolution equations (\ref{gammandef}) are very conveniently
summarized by 
the celebrated DGLAP (Dokshitzer-Gribov-Lipatov-Altarelli-Parisi) \cite{dglap}
 integro-differential equation for the $\f_{i/h}(x,\mu)$,
\eq
\m {d^2 \over d \m^2} \f_{i/h} (x , \m^2) = 
\sum_{j=f,\bar{f},G} \int^1_x {d \x \over \x} P_{ij}
({x\over \x} , \a_s (\m^2) )  \f_{j/h} (\x, \m^2)\, ,
\label{dglap}
\eqe
where the distributions $P_{ij}$ summarize a matrix of ``singlet"
anomalous dimensions as their moments,
\eq
\int^1_0 d \x\, \x^{n-1} P_{ij} (\x, \a_s ) = - \g_{ij}{}{(n)}\, .
\label{pijmom}
\eqe
The $P_{ij}(x,\alpha_s)$ are power series in the strong coupling
given to one loop by the distributions above, for instance,
\eq
P_{qq} (x, \a_s ) =  {\a \over \p} P_{q/q}^{(1)} (x) + \ldots\, .
\label{pijexpand}
\eqe
Eq.\ (\ref{dglap}) is one of the most useful tools in
perturbative QCD and in the search
for new physics, since it enables us to connect experiments at widely
differring momentum transfers, and to predict the outcomes
of experiments even at very high energy.

\subsection{The light-cone expansion}

Before generalizing factorization beyond DIS, it is useful to 
acquire some insight into the field-theoretic content of 
evolution.  Consider the moments of a 
nucleon quark distribution, taken for simplicity in 
$A^+=0$ gauge\footnote{Notice that
$\partial/\partial y^-=\partial^+=\partial_-$.}
(and supressing the spin average), 
\eqa
\bar{\f}_{a/N} (n, \a_s (\m^2) ) &=& {1\over 2\p} 
\int^1_0 dx x^{n-1}
\int^\infty_{-\infty} d y^- e^{-i y^- x p^+} 
< N(p) \mid \bar{q} (y^- ) \g^+ q
(0) \mid N(p) > \nonumber\\
&=& {1\over 2\p} \sum^\infty_{m=0} {1\over m!} 
< N(p) \mid \, \left [ (\del^+)^m \bar{q} (0)\right ]\, 
\g^+\, q (0) \mid N(p) > \nonumber \\
&\ & \quad\quad\quad \times \int^1_0 d x x^{n-1}
\int^\infty_{-\infty} d y^- (y^-)^m e^{-i y^- x p^+} \nonumber\\
&=& {1\over  (p^+)^n} \; < N(p) \mid\, 
\left [ (-i \del^+)^{n-1} \bar{q} (0) \right ]\, \g^+\, q
(0) | N(p) >\, .
\eqae
In the second equality, we have formally expanded the $\bar{q}$
field about $y^-=0$, that is, on the light cone, and in the
third we have done, first, the resulting $y^-$ integrals to get
delta functions, and then the $x$ integrals (treating 
$\int_0\; dx\delta(x)=1$).  We see that only a single term contributes
to the sum, corresponding to the {\it local} product of the
quark field with the $n-1$st derivative of its conjugate.
Thus, moments of parton distributions are related to local operators,
with dimensions that increase with the moment variable $n$.

This rough discussion can be carried out in an arbitrary gauge, and
the full set of relevant gauge-invariant operators
found in this manner is
\eq
O_f{}^{\m_1 \ldots \m_n} = \bar{q} (0) 
\left ( \prod^{n-1}_{i=1} \; i D^{\m_i} [A] \right )
\g^{\m_n} q (0)
\label{ttq}
\eqe
for quarks and
\eq
\co_G^{\m_1 \ldots \m_n} = F^{\m_1}{}_\a (0)
\left (  \prod^{n-1}_{i=2} i D^{\m_i} [A]\right ) F^{\a\m_n} (0)
\label{ttG}
\eqe
for gluons, with 
$D_\mu=\partial_\m+igA_\m$ the covariant derivative (in covariant gauges,
ghost operators may also contribute in general, but not to physical matrix elements).

These operators occur in the expansion of the product of electromagnetic
currents at short distances,
\eq
J^\m{}^{\rm (em)} (x) J^{\rm (em)}_\m (0) = \sum^\infty_{n=0} \sum_{I} 
\; C_{n,I}
\left (x^2,\m^2 , \a_s (\m^2) \right ) 
x_{\m_1} \ldots x_{\m_n} \co_I{}^{\m_1 \ldots
\m_n} (0)\, ,
\label{lcexpansionjj}
\eqe
where they are distinguished by the behavior of their coefficient
functions near the light-cone $x^2=0$,
\eq
C_{n, I}\left (x^2,\m^2 , \a_s (\m^2) \right )
 \sim (x^2)^{-2} h_I (x^2 \m^2 , \a_s (\m^2 ) )\, .
\eqe
This singularity is identified by dimensional counting:  the
product of two currents has (mass) dimension 6, while the
operators in (\ref{ttq}) and (\ref{ttG}) have dimension 
$3+(s-1)=2+s$, with $s$ the (maximum) spin ({\it i.e.}, number
of vector indices), while the corresponding
factors of $x^\mu$ contribute mass dimension $-s$.  The
power behavior of the coefficient function
 for any such tensor operator of spin $s$
and dimension $D$ is thus
$(x^2)^{-3+(D-s)/2}$.  The short-distance expansion organized
according to light-cone singularities is known as the
{\it light-cone expansion} \cite{lcexpansion}.

The quantity $D-s$ is called the ``twist".  All operators in
eqs.\ (\ref{ttq}) and (\ref{ttG}) have twist equal to two.  Twist
 controls singularities on the light cone, and hence
the high-$q^2$ behavior of the Fourier transforms of 
the products of currents, the DIS structure function.
For DIS, then, the effect of Minkowski space is
to elevate an infinite set of (twist-two) operators
to leading behavior.

Note that the light-cone $x^2=0$ corresponds to
the manifold $x^+=0$ when the momentum $p$ is in the plus
direction.  It is thus not the light cone along which the
target particle moves, but rather the opposite-moving light 
cone, corresponding to a light-like scattered parton in the
``brick-wall" frame.  Of course, the scattered parton is not
always in the opposite direction, but all the details of
final states are absorbed into the hard scattering function $H$
of fig.\ \ref{Cptps} and eq.\ (\ref{firstfact}).  This is another 
consequence of factorization.  From the point of view of 
calculating long-distance behavior in the 
parton distribution, the entire scattering process may 
be replaced by a pair of Wilson lines, as in eq.\ (\ref{giprods}).
There is a strong similarity here to the effective field
theory picture often used to discuss the dynamics of
heavy quarks \cite{GeorgiTASI}, and indeed, the symmetries of heavy-quark
effective theory are closely related to the universality
properties of parton distributions, to which we now turn.

\subsection{Hard hadron-hadron scattering}

Once we introduce the concept of factorization, it
is natural to apply it beyond inclusive DIS \cite{jccrv,earlyfact}.  
Of course, we must be careful to consider only inclusive
hard-scattering processes, for which we may hope to
find the necessary incoherence between the short-distance
scattering and long-distance hadronic binding effects.

In an important example, we
consider processes in which a quark from hadron $h$
annihilates an antiquark from hadron $h'$, forming
a virtual electroweak vector boson, which decays to
a lepton pair.  This reaction, 
\eq
h(p)+h'(p')\rightarrow \ell\ell'(Q^\mu) +X\, ,
\label{DYreaction}
\eqe
with its characteristic
signal of a lepton pair (momentum $Q^\mu$), is known as the
Drell-Yan process \cite{dyref}.  Its observation was one of the
early successes of parton ideas, especially because
it signals the presence of a sea of quark pairs within
ordinary hadrons. 

It is easy to write a factorization formula
for such a cross section, by a straightforward generalization of
the expressions for DIS structure functions, eq.\ (\ref{fonetwofact}).
It is a convolution of two parton distributions, one for 
each hadron, with a hard-scattering function $H$.  At lowest
order, $H$ is given by the Born cross section for quark pair
annihilation to the relevant leptons.  At higher orders, gluon-quark
scattering may also contribute, 
\eq
{d \s_{hh^\prime \to Q^2}{}(s, Q^2) \over d Q^2}  = 
\sum_{i,j=f,\bar{f},G} \int^1_0 d \x d \x^\prime\, 
\f_{i/h} (\x, \m^2) H_{ij} \left({Q^2
\over \x\x^\prime s} , {Q^2 \over \m^2} , \a_s (\m^2)\right)  \f_{j/{h'}} (\x^\prime,
\m^2)\, .
\label{dyfact}
\eqe
As in DIS, the hard-scattering function is a power series in $\alpha_s(\mu^2)$.
$H$ depends on the scheme chosen for the parton distributions.
As an example, for $H_{f\bar{f}}$, we have, to one loop in DIS scheme \cite{oneloopdy},
\eqa
H_{f\bar{f}} &=& 
{d\s_{f\bar{f}}^{\rm (Born)}\over dQ^2} \Bigg ( \delta(1-z)+
{\a_s \over \p} \bigg \{ 
C_F \big [  (1+ z^2) \big  ( \left[ {\ln (1-z) \over 1-z} \right]_+ 
+3\left [{1\over 1-z}\right ]_+ 
\nonumber \\
&\ & \quad\quad\quad\quad 
-6-4z-  \ln z \big ) 
+ \left ( {4\p^2 \over 3} +1 \right ) \d (1-z ) \big ]  \bigg \}\Bigg ) \, ,
\eqae
where $z=Q^2/\x\x's$.  Given 
phenomenological parton distributions in
some scheme,
the factorization formula gives an absolute prediction for the
Drell-Yan cross section, which has been successfully applied
to a wide range of experiments.  The corrections in $H$ are
not always small, however, and as we shall see, we sometimes need
information about contributions at arbitrarily high power.

Another application of parton model ideas,
extended to perturbative QCD, involves 
single-particle inclusive cross sections, which
count hadrons at fixed momenta, but are otherwise inclusive
in the hadronic final state,
\eq
h(p)+h'(p')\rightarrow C(p_C)+X\, .
\eqe
  If the hadron (C) is
observed, for instance, at large transverse momentum,
we know that a hard scattering has taken place, and may
hope that incoherence and hence factorization is 
relevant \cite{Muelleripi,iPIrefs}.  In this case,
the parton model suggests that the hadron $C$ arises from
the ``hadronization", or fragmentation, of some parton $k$.
The process of hadronization should, following our
discussion of Section 1, occur over time scales that
are independent of the hard-scattering scale, and of 
the fragmentation of other partons,
 scattered in other directions.  Hadron $C$ is
thus expected to be produced in a universal fashion from
parton $k$, and to inherit a fraction $0\le z\le 1$ of that
parton's momentum.  The (incoherent) probability for this
evolution is summarized in a ``fragmentation function" 
$d_{C/k}(z,\mu^2)$, which describes the distribution of
hadrons in the fragments of a parton, and is
analogous to the parton distribution $\phi_{i/h}$,
but with the roles of hadron and parton reversed.
In perturbation theory, $d$ must be renormalized, and thus it depends
on the factorization scale $\mu$.
The corresponding factorization formula for
single-particle inclusive cross sections is
\eqa
\omega_C{d \s_{hh^\prime \to C(p_C)}{(p,p',p_c)} \over d^3p_C}  &=& 
\sum_{i,j,k=f,f,G}
\int^1_0 d \x d \x^\prime { d z\over z^2}\, 
H_{ijk} \left ({\m^2 \over \x\x^\prime s} , {p_C\cdot \x p \over z\m^2} , 
{p_C\cdot \x' p' \over z\m^2} , \a_s (\m^2) \right )\,
\nonumber\\
&\ & \quad \quad \quad \times 
\f_{i/h} (\x, \m^2)
 \f_{j/h'} (\x^\prime,\m^2)\; d_{C/k}(z,\mu^2)\, .
\label{hhonepincl}
\eqae
The extra factor $1/z^2$ allows $H$ to be normalized to the
corresponding Born cross section for $i+j\rightarrow k+X$.

Like parton distributions, fragmentation functions are universal, within a
given scheme to define them, and the same functions appear in
hadron-hadron scattering, DIS and ${\rm e}^+{\rm e}^-$ annihilation.
Also like parton distributions, their $\mu$-dependence may be
analyzed, and summarized by evolution equations \cite{CollinsSoperPDF}.  
We shall not
go into these applications here, however.  Rather, we shall
close this section with a few comments on how generalizations
of DIS factorization are established in perturbation theory.

\subsection{Jet-soft analysis}

The proof of factorization theorems \cite{jccrv,earlyfact,factproof} like those described above
is highly nontrivial in perturbation theory, and, indeed, 
has not reached the sophistication of technical treatments
of the operator product expansion in Euclidean space.  
We may, however, briefly discuss a few relevant physical issues.

The essential complication in demonstrating the factorization
of hadron-hadron cross sections is
evident in the relevant pinch surfaces, shown in fig.\ \ref{dypinch}.
\begin{figure}[ht]
\centerline{\epsffile{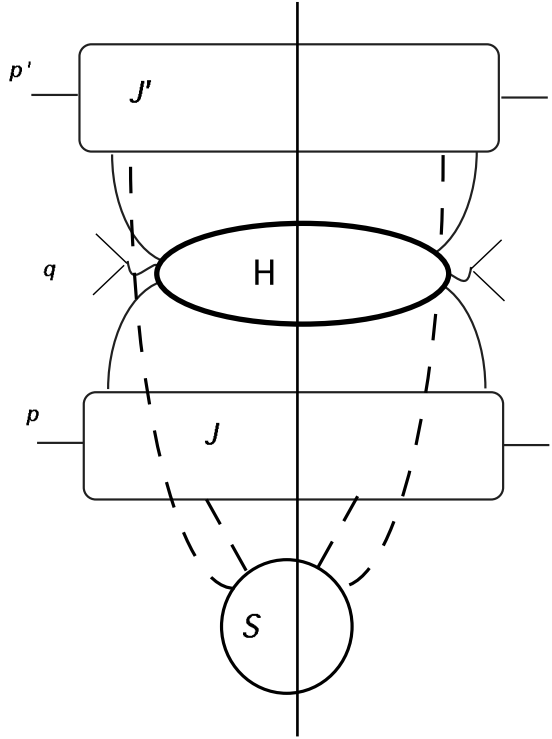}}
\caption{General reduced diagram for pinch surface of a Drell-Yan
cross section.  The subdiagram H includes possible final-state jets, and for lack of
space the Drell-Yan pair is reversed in direction.}
\label{dypinch}
\end{figure}
The jets of the two incoming hadrons are connected by soft gluons,
whose momenta vanish at the pinch surface.  Roughly, this corresponds
to the scattering of partons in each hadron from the Coulomb 
fields of partons in the other hadron before and/or after the
hard scattering.  Final-state interactions are present as well in DIS,
and we expect them to cancel by unitarity arguments; initial-state
interactions, however, are special to hadron-hadron scattering,
and could, in principle, lead to a rearrangement of color, and
even transverse momentum, of the partons in each hadron, due to
the presence of the other hadron.  Such a rearrangement could, in turn,
affect the hard scattering, and break the universality necessary
to identify the parton distributions of hadron-hadron collisions
with those of DIS.
This does not happen, however, for the following physical
reasons.  

We recognize
that the quanta of perturbation theory are gauge fields $A^\mu$,
which include unphysical as well as physical degrees of freedom.
Physical information is associated with field strengths, 
$F_{\m\n}$, and the latter behave very differently than the former
under Lorentz boosts.  In particular, the field strengths are
 strongly Lorentz contracted, 
while the unphysical polarization of $A^\mu$, proportional to the
momentum of each gluon, actually grows under Lorentz boosts.
Thus, we expect the effects unphysical polarizations to contribute 
in individual diagrams in perturbation theory, but to cancel in a
gauge invariant sum over diagrams.  

Such an analysis requires, in addition to 
the identification of pinch surfaces and power counting, an
application of the 
Ward identities of the theory that decouple unphysical
gluons from physical processes.  Let us sketch how this can be
done, taking as an example connections of soft lines to a
final-state jet, as shown in fig.\ \ref{onepisoft}.
\begin{figure}[ht]
\centerline{\epsffile{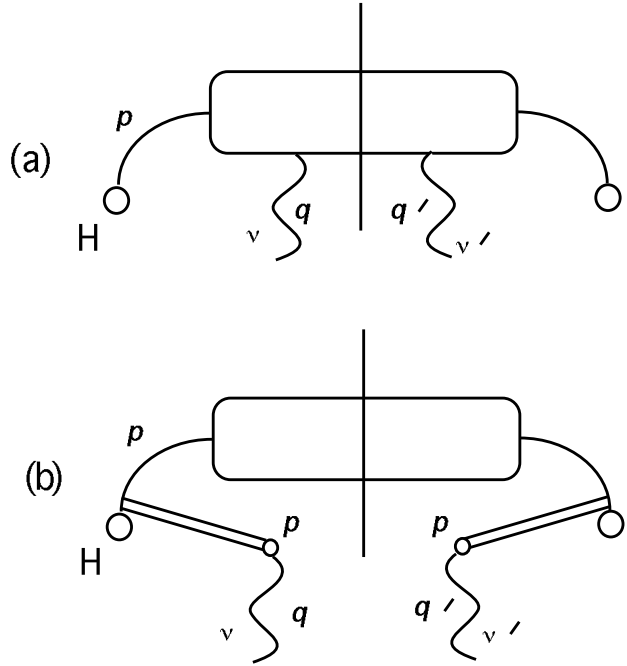}}
\caption{(a) Reduced diagram in which soft lines connect to a final-state jet.
(b) Factorization of soft lines.}
\label{onepisoft}
\end{figure}

At a typical pinch surface, soft lines, $q$ to the left of the
cut, and $q'$ to the right, are attached to the jet, which we may
take in the plus direction, with large momentum component $p^+$.
We may define both soft momenta to flow into the jet, 
and then along jet lines away from the final state and toward
the hard scattering functions $H$.  For each jet line along which
a soft momentum flows the momentum is then of the form
$\ell - q$ for lines to the left of the cut (amplitude)
and $\ell+q$ to the right (complex
conjugate amplitude).  Here $\ell$ is the momentum of a jet line in the absence 
of the extra soft momentum, and is hence naturally parameterized as 
\eq
\ell=(xp^+,\ell^-,\ell_\perp)\, ,
\eqe
with $0< x\le 1$.  For $\ell$ a jet momentum, we can assume that
$xp^+$ is its largest component.  We can now make two approximations
that will enable us to factor the soft gluons from the jet, replacing
their couplings by an effective eikonal line.  The first is that the coupling of 
the gluons' propagators $G_{\n\lambda}(q)$
to the jet is always through the
combination  $p^\n G_{\n\lambda}(q)$, that is, that the soft gluons
couple only to the large component of the current.  The second
is that propagator denominators $(\ell+q)^2$, depend only
upon the minus component of the soft gluon's momenta, {\it i.e.}
the components in the direction opposite to the jet direction,
  \eq
(\ell\pm q)^2\sim \pm 2xp^+q^- +\ell^2\, .
\label{eikonalapprox}
\eqe
We shall return to this condition in a moment.  

Once these approximations are made, the jet depends only on
one component of each soft gluons' momentum, and on the
same component of its polarization.  Thus the coupling
of soft gluons to a jet moving in the plus direction is equivalent to the coupling
of a set of gluons with {\it only} minus momenta and {\it only}
minus polarizations.  Such gluons are unphysical, and are equivalent
to a phase rotation on the external lines of the jet, which
attach to the hard scatterings.  In an abelian theory, for instance,
an arbitrary jet with one vector (photon) attached to each side of
the cut, is equivalent to a jet with no photons, multiplied by
two linearized propagators (``eikonal lines"), one for each of
the photons,
\eq
J_2^{\n{\n'}} (p,\hat{q},\hat{q}')
\sim J_2^{++} (p,\hat{q},\hat{q}')
=
g^2\, {p^{\n'}\over p^+{q'}{}^--i\e}\, {p^{\n}\over -p^+q^-+i\e}\ 
J_0(p)\, ,
\label{Jsofthomogen}
\eqe
where $J_0(p)$ is the jet with no soft photon connections and where
\eq
\hat{q}^\m=(0,q^-,\bf{0})\, .
\label{hatqdef}
\eqe
The factors on the right of (\ref{Jsofthomogen}) are
of the form of the propagator and vertex in eq.\ (\ref{eikfeynrule}) with $n=p$.
The eikonal lines here are illustrated in fig.\ \ref{onepisoft}b.

In a nonabelian theory, multiple gluons still factor, and their
color interactions are summarized by connections to eikonal
Wilson lines, which have the same perturbation theory rules 
for lines and vertices as
those in eq.\ (\ref{giprods}) for DIS parton distributions.  Schematically,
we write
\eq
J_n^{+\dots+}(p,\hat{q_i},\hat{q}'_j)
=
E^*(\{q'_j\})E(\{q_i\})J_0(p)\, ,
\eqe
where the $E$'s are generated from the operators 
\eq
U(A)=\exp \left [ -ig\int_0^\infty d\l\ p\cdot A(\l p) \right ]\, .
\eqe
The soft divergences factorized in this manner cancel
\cite{iPIrefs,jccrv}, due to the
 unitarity of the Wilson lines,
\eq
U[A]^\dagger\; U[A]=I\, .
\eqe
To realize this cancellation, it is necessary to sum over
all final states that differ by soft gluon emission.  This,
of course, is exactly what we do in inclusive hard-scattering
cross sections.  Given this cancellation, the remaining jets
are independent of each other, and may be factored from the hard
scattering as in DIS.

The argument that we have given above is, of course, quite rough.
In particular, the approximation (\ref{eikonalapprox}), is
highly nontrivial \cite{jccrv}.  It is clearly not true for every soft momentum
$q^\mu$, since it fails whenever $q^-$
vanishes compared to the other components of $q^\m$.  The approximation
will hold, however, if the $q^-$ integral is not trapped at or
very near zero.  Arguments to this effect are relatively easy
to give for may cross sections in ${\rm e}^+{\rm e}^-$ 
annihilation, where we verify that all the poles
in $q^-$ from lines within the jet, $(\ell\pm q)^2 \mp i\e$, are
in the upper half-plane for soft gluon momenta
routed as above.  The situation is much more
difficult when the jet originates from an incoming hadron, as in
the Drell-Yan process.  Here, pinches may occur on a diagram-by-diagram
basis, as in fig.\ \ref{glauberpinch}, where the lines $xp+q$
and $(1-x)p-q$ have $q^-$ poles in opposite half-planes.
\begin{figure}[ht]
\centerline{\epsffile{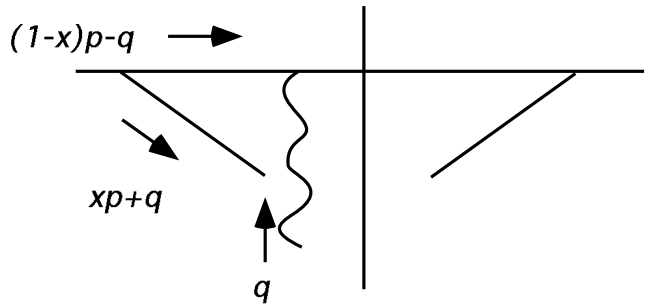}}
\caption{Reduced diagram for which $q^-$ in pinched
close to zero.}
\label{glauberpinch}
\end{figure}
These pinches cancel, however, in sufficiently inclusive
cross sections,
for the physical reasons that we have outlined above, and soft
gluons that are not in the jet direction decouple from incoming as
well as outgoing jets \cite{jccrv,factproof}.  

We now ask when a soft gluon is 
close enough to a jet's direction to be
 part of that jet.
Let us consider the case when the two 
momenta in eq.\ (\ref{eikonalapprox})
are both on-shell, {\it i.e.}, $q^2=\ell^2=0$.  
To factor $q$ from the $p$-jet we require, in
particular, that
\eq
xp^+q^->|\ell_\perp\cdot q_\perp|\, .
\label{eikinequal}
\eqe
This condition has a natural geometrical interpretation, which itself
has important physical consequences.  For a well-collimated jet,
$\theta_\ell=|\ell_\perp|/(xp^+)$ is the angle of $\ell^\mu$ to
the overall jet direction.   Defining $\theta_q$
similarly, and parameterizing $q^+=yp^+$, we have 
\eq
|\ell_\perp|=\theta_{\ell} xp^+,\ \ |q_\perp|=\theta_q yp^+,\ \ 
q^-={q_\perp^2\over 2yp^+}\, .
\label{jetmta}
\eqe
Using these relations  it is
easy to show that the soft gluon $q$ will factorize from the
jet, {\it i.e.} eq.\ (\ref{eikonalapprox}) will hold, {\it unless}
\eq
\theta_q<\theta_{\ell}\, ,
\label{angord}
\eqe
that is,
unless $q$ is emitted closer to the jet axis than $\ell$,
the particle that emitted it.  This ``angular ordering" \cite{angorder}, limits
the region of phase space into which soft gluons may be 
emitted late in jet evolution, and hence suppresses the
multiplicity of very soft gluons, since gluons that
factor from the jet do not get enhancements from pinch
surfaces where jet lines are near the mass shell.  This
is one of the many consequences of the coherence of gluon emission
in QCD \cite{cohere}.

\section{Two-scale problems I:  Sudakov resummation}

The factorization theorems of the previous section go far
toward connecting perturbative QCD to experiment.  Their
application, however, is limited somewhat by 
our assumption that
there is only a single large scale in the problem, for instance,
$Q^2$ in DIS.  
In this case, it is natural to choose the factorization scale 
to be of order of $Q$, and to use the evolution of the parton
distributions discussed above.  The coefficient function 
is then of the general form
\eq
C(x/\x)=\sum_n d_n(x/\x)\; \alpha_s^n\, ,
\label{Hsum}
\eqe
where $d_n(z)$ is a distribution in $z$.  Then, if
$\alpha_s(Q^2)$ is small, we may hope that the effect of
higher-order terms in $C$ is small as well.

Unfortunately, this is not always the case even in DIS; 
the $d_n(z)$ will generally include factors like
$\alpha_s^n\ln^n(z)/z$, or even $\alpha_s^n\; \ln^{2n-1}(1-z)/(1-z)$.
When $x$ is very small, or very close to unity, these logarithms
may produce large corrections at every order.  
Recalling that $(p+q)^2/Q^2=(1-x)/x$, we see that in both
cases the arguments of the logarithms are ratios of
kinematic invariants.
There are many other such examples, all with
two hard scales, $Q_1^2$ and $Q_2^2$, which satisfy
\eq
Q_1^2\gg Q^2_2\gg \Lambda^2\, .
\label{qoneqtwolambda}
\eqe
The first inequality ensures the presence of a large
ratio in the hard scattering function.
The second ensures that parton densities 
evolve perturbatively, and that the running coupling in the hard
scattering function, $\alpha_s(Q_2^2)$ is small, even
if the combination $\alpha_s(Q^2_2) \ln(Q_1^2/Q_2^2)$ is not.
In this case, we may try to ``resum" the series in $\ln(Q_1^2/Q_2^2)$
to all orders in perturbation theory.  

We shall treat three representative cases:
$T\rightarrow 1$, with $T$ the thrust,
eq.\ (\ref{Tdef}), in ${\rm e}^+{\rm e}^-$ annihilation,
$Q_T\ll Q$ for Drell-Yan production of pairs of mass $Q$,
and $x\rightarrow 0$ in deeply inelastic scattering.  
The
first and second illustrate the resummation of 
so-called ``Sudakov" double 
logarithms, 
and the second and third illustrate resummations based on
an extension of factorization to fixed transverse
momenta.  

\subsection{Sudakov double logarithms}

Let us return for a moment to our dimensionally-regulated 
expression for one-loop corrections, eq.\ (\ref{gammaoneloop}), to the electromagnetic 
form factor,  
\eq
\G_\m (q^2, \e) = - i e \m^\e \; \bar{u} \; (p_1) \g_\m v (p_2)\; \r (q^2, \e)\, .
\eqe
At one loop, its momentum-dependence
is contained in an overall factor $(-q^2)^{-\e}$, 
which we may expand to order $\e^2$ to get an expansion
in $\ln q^2$,
\eq
\r (q^2,\e) =1  - {\a_s \over 4\p} C_F\, \ln^2 (q^2/\m^2)+\dots \, .
\eqe
The double logarithms in momentum transfer are another
reflection of the overlap of collinear and soft
singularities in the vertex function, and 
are generally referred to as Sudakov double logarithms \cite{collinsrv,sudresum,CollinsSopersud}.  
Although the electromagnetic vertex is not infrared
safe, it is an important element in any process whose kinematics approaches 
those of elastic scattering.  

Our first example \cite{thrustresum} is 
the ${\rm e}^+{\rm e}^-$ annihilation 
 cross section near unit thrust, $T=1$.  Here the double
logarithms appear not in $Q^2$ directly, but in 
$(1-T)$.  Consulting the definition of thrust, eq.\ (\ref{Tdef}),
we see that at $T=1$ the final state consists of two
back-to-back massless jets.  A little kinematics also
shows that
when both jet masses $p_i^2$ are small 
compared to $Q^2$ they are related to the thrust by
\eq
1-T={p_1^2+p_2^2 \over Q^2}\, .
\label{oneminusTjetmass}
\eqe 

At order $\a_s^n$ the leading logarithm in $1-T$ 
is given by an exponential of Sudakov logarithms,
\eq
{1\over\s_{\rm tot}}{d\s\over dT}
=
-2C_F{\alpha_s\over \pi}{\ln(1-T) \over 1-T}\, 
\exp\left \{ -C_F{\alpha_s\over \p}\ln^2(1-T) \right \}\, .
\label{llaT}
\eqe
As $T$ approaches one, ${d\s\over dT}$
vanishes.  This is a quantum-mechanical reflection of
the classical radiation field that must accompany
any process in which a charged particle is accelerated
(quark pair creation being an extreme example).
Quantum field theory assembles this
classical field out of many soft and collinear
gluons (the correspondence principle).  Cross sections in which gluon emission is
forbidden in part of phase space are correspondingly
suppressed.

\subsection{Factorization for $T\rightarrow 1$}

Our goal in the following is to rederive eq.\ (\ref{llaT}),
and to extend it to include nonleading logarithms and the
effects of the running coupling.  
Our arguments apply to a large class of cross sections.
In fact, the resummation of Sudakov
logarithms follows from the factorization properties of
the cross section in the regions of momentum space that
give rise to the logarithms.  
This factorization has already
been illustrated by the physical picture of annihlation
given in fig.\ \ref{cutjet}, which shows a two-jet 
configuration.  The infrared divergences associated with such configurations
cancel, according to the arguments of Section 3.3, but 
for $T\sim 1$, the cancellation between diagrams with virtual
and real gluons is constrained by the requirements that
the real gluons be either very soft or emitted very close
to the quark or antiquark directions.  This leaves large
finite remainders in the cancellation of divergences.
These are the logarithms of $1-T$.  

Let us recall the power counting arguments of section 
2.3.  We saw there that in an axial gauge, $\x\cdot A=0$,
all collinear divergences may be absorbed into the jets.
As a result, in such gauges, double logarithms
arise from collinear gluons in jets which become soft,
while staying collinear.  
The form of fig.\ \ref{cutjet}, combined with the 
factorization of soft gluons from jets, described in Section
4.7 above, suggests that the thrust cross section, like
Drell-Yan, factorizes into functions that describe 
the two jets, the hard, and the soft subdiagrams.  At double logarithmic accuracy
in axial gauge,
therefore, we may neglect soft gluons that
connect  the jets to each other.  This approximation
will simplify the arguments below, without changing
the character of a more general treatment, which 
gives an essentially equivalent result, but is accurate
to all logarithmic order.  At fixed values of jet
masses $p_i^2$, we shall therefore begin with the
factorized expression
\eq
{d\s \over d p_1^2 dp_2^2} = J_1(p_1,\m, \x) J_2(p_2,\m, \x) H ( p_1, p_2 ,\m,\x)\, ,
\label{sudjetmassfact}
\eqe
where the $J$'s represent the jets, and $H$ the hard-scattering
factor.  We shall suppress particle labels, but 
it is relatively easy to show from power counting that to
leading {\it power} in $1-T$ the jets are connected to the
hard scattering by a quark and antiquark only.  
Eq.\ (\ref{sudjetmassfact}) is enough to 
derive the exponentiated double logarithms of eq.\ (\ref{llaT})
above, with corrections due to the running of the coupling.

In axial gauge the jets depend, not only on their invariant momenta,
but also on their energies through the products $p_i\cdot\x$.
The jets and $H$ are thus not individually Lorentz invariant.
For a general gauge vector $\x^\m$, the
precise arguments are $p_i\cdot\x/\sqrt{\x^2}$ because the
gluon propagator, and hence $J$ is invariant under simple
rescalings of $\x^\mu$. (In the following, we set $\xi^2=1$.)

In view of eqs.\ (\ref{oneminusTjetmass}) and (\ref{sudjetmassfact}), the 
cross section at fixed $1-T\sim 0$ is of a convolution form
\eqa
{1\over \s_0}{d\s \over d T} &\simeq&  
\int_0^{Q^2} d p_1^2 dp^2_2\, \d \left (1-T -
{p_2^2 \over Q^2} - {p^2_2 \over Q^2} \right) H\left(p_1\cdot\x/\m,p_2 \cdot \x /\m , \a_s (\m^2 )\right)
\nonumber\\
&& \quad\quad \times\, J_1 \left(p_1^2/\m^2 , p_1 \cdot \x / \m , \a_s (\m^2 )\right) \, 
J_2 \left(p_2^2 / \m^2 , p_2 \cdot \x /\m , \a_s (\m^2 ) \right)
\, .
\label{dsigdT}
\eqae
Dividing by the Born cross section $\s_0$ gives $H=1$ at lowest order.
Because we are interested in the limit $p_i^2/Q^2\rightarrow 0$,
the $p_i$ may be expanded about back-to-back lightlike momenta.
We choose axes so that $p_1^\mu$ is in the plus direction
and $p_2^\m$ is in the minus direction.  Then $p_1^+$ and
$p_2^-$ are both nearly equal to $Q/\sqrt{2}$, while
\eq
p_1^- \sim {p_1^2 \over \sqrt{2}Q},\quad p_2^+\sim {p_2^2 \over \sqrt{2}Q}\, .
\eqe
For $\x$ in an arbitrary direction, we may take
$p_1\cdot\x\sim p_1^+\x^-$ and $p_2\cdot\x\sim p_2^-\x^+$ to leading power
in $Q$.  Then to leading power in $1-T$, the $p_i^2$ integrals are independent of 
$p_i\cdot\x$.

We want to identify singular 
$\ln^m(1-T)/(1-T)$ behavior for $T\rightarrow 1$, and
for this purpose  
moments with respect
to $T$ are particularly useful,
\eq
\tilde{\s} (n) = {1\over \s_0} \int^1_0 d T\, T^{\; n} {d \s \over d T}\, ,
\label{momentsT}
\eqe
since the moments of any function that is finite at $T=1$ falls off as $1/n$ 
for $n\rightarrow \infty$. 
In particular, logarithms of $1-T$ are transformed into logarithms
of $n$ by
\eq
\int_0^1 dT\; {T^{\; n}-1\over1-T}\; \ln^m(1-T)
={(-1)^{m}\over m+1}\; \ln^{m+1}n +\dots\, .
\label{lnTlnn}
\eqe
 Keeping only terms that are finite or
grow as $n\rightarrow\infty$, and using the 
relation $T^{\; n}\sim e^{-n(1-T)}$, which holds in this approximation,
we find that the convolution in (\ref{dsigdT}) factors into
a simple product under moments,
\eqa
\tilde{\s} (n) &=& \tilde{J}_1 \left (Q^2 /n\m^2 , p_1 \cdot \x /\m , \a_s(\m^2) \right)\; 
\tilde{J}_2 \left (Q^2 / n \m^2 , p_2 \cdot \x/\m, \a_s (\m^2 ) \right) \nonumber \\
&\ & \quad\quad\quad \times 
H\left (p_1\cdot\x/\m,p_2\cdot \x/\m,\alpha_s(\m^2) \right)\, ,\nonumber\
\eqae
up to corrections that vanish as $1/n$. Note that $n$ now appears
only in the combination $Q^2/n\m^2$.
By analogy to the
derivation of evolution for DIS
structure functions, we shall use 
this factorized expression, coupled with 
renormalization group arguments, to derive 
a resummed cross section.  The new
feature in our Sudakov factorization is the dependence on the
axial gauge vector $\x$.  Although each of the factors that
makes up $\tilde{\s}$ is $\x$-dependent, the physical quantity $\tilde{\s}$
must be gauge-invariant.  This invariance will drive the 
resummation \cite{CollinsSopersud}.

\subsection{Resummation for $T\rightarrow 1$}
  
We start with the renormalization group behavior of of $J$,
which is
simple, since it has only two
external (quark or antiquark) lines,
\eq
\left[ \m{\del \over \del \m} + \b {\del \over \del g} \right] 
\ln \tilde{J} =-  2 \g_q\, ,
\label{Jrg}
\eqe
with $\g_q$ the quark anomalous dimension.  Since
the cross section is independent of the renormalization scale $\m$,
$H$ must behave in a corresponding fashion
\eq
\left[ \m{\del \over \del \m} + \b {\del \over \del g} \right] 
\ln H = 4 \g_q\, .
\label{Hrg}
\eqe
We now observe that very similar reasoning may be applied to the
gauge-fixing vector $\x$.  

Let us change (boost) $\x^+$ and $\x^-$
in a manner that leaves $\x^2=1$.
Using the discussion after eq.\ (\ref{sudjetmassfact}),
the independence of $\tilde{\s}(n)$ from $\x^\m$ may
be expressed as $\partial\tilde{\s}/\partial \ln\x^-=-\partial\tilde{\s}/\partial\ln\x^+$.
The chain rule then gives
\eq
{\del \ln \tilde{J}_2 (Q^2/n\mu^2,p_2^- /\m) \over \del \ln p_2^- } + 
{\del
\ln H (p_1^+/\m,p_2^-/\m)\over \del \ln p_2^-} 
= 
-{\del \ln \tilde{J}_1 (Q^2/n\mu^2,p_1^+ / \m) \over \del \ln p_1^+ } - 
{\del \ln H(p_1^+/\m,p_2^-/\m) \over
\del \ln p_1^+}\, ,
\label{xichangeinsig}
\eqe
where we have made components explicit and have suppressed $\a_s$.
This relation is surprisingly powerful, because $J_1$,
$J_2$ and $H$ depend on different sets of arguments.  $J_2$,
for instance, depends on both $p_2^-$ and $Q^2/n\m^2$.  Its derivative
with respect to $p_2^-$ may depend upon either of these arguments,
but must be cancelled in (\ref{xichangeinsig}) by the derivatives
of $H$ and $J_1$, which contribute additively.  Thus, its
dependence on $p_2^-$ and $Q^2/n\m^2$ can only be additive after the
derivative,
\eq
{\del  \over \del \ln p^-_2 } 
\ln \tilde{J}_2 \left ({Q^2 \over n\m^2}, {p^-_2 \over \m} , \a_s (\m^2) \right )
=  K \left ( {Q^2 \over n \m^2} , \a_s (\m^2)\right ) 
+ G \left({p^-_2 \over \m} , \a_s (\m^2 )\right)\, .
\label{partialJKG}
\eqe
The function $K$ cancels a corresponding term from
$J_1$ while $G$ cancels the contribution from $H$, whose derivatives
must satisfy,
\eq
{\del \ln H \over \del \ln p^{\prime -} } + {\del \ln H \over \del \ln p_1^+}
= - G\left( {p^-_2 \over \m} \right) 
- G \left( {p_1^+ \over \m} \right)\, .
\eqe
This separation of short-distance and long-distance dependence 
in jets is
characteristic of Sudakov factorization. As the gauge changes,
the jets exchange contributions with each other (via $K$)
and with the hard part (via $G$).  Here we find a 
strong analogy to the ``matching conditions" of effective
field theory.

We now have at our disposal two evolution equations, 
the first relying on invariance under renormalization
group rescalings, the other on gauge invariance, but both
based on factorization.  Combining the two, we shall find
enough information to determine all logarithmic $n$-dependence.

By eq.\ (\ref{Jrg}), $(d\ln J_i/d\m)$ is independent of momenta,
so, for instance
\eq
{d^2 \over d \m\, d p_1^+} \ln J_1 = 0\, .
\eqe
Applying this result to (\ref{partialJKG}), we  conclude that
the combination $K+G$ is itself a renormalization group invariant
\eq
\m {d \over d \m}( K+G) = 0  \, ,
\label{KGmuzero}
\eqe
which implies that yet another anomalous dimension
relates $K$ and $G$ \cite{sudresum,CollinsSopersud,KoTr},
\eq
\m {d  \over d\m}K = 
- \g_k (\a_s) = - \m{d \over d\m}G\, .
\eqe
Now we can relate the moment-dependence of $K$ and $G$ through
\eqa
K \left( {Q^2 \over n \m^2} , \a_s (\m^2 ) \right) + 
G \left ( {Q \over \m},\a_s (\m^2) \right ) 
&=& K \left (1, \a_s (Q^2/n)\right ) 
+ G \left (1, \a_s (Q^2 ) \right ) \nonumber\\
&\ & \quad\quad -\, {1\over 2} \int^{Q^2}_{Q^2/n} {d \m^{\prime 2} \over \m^{\prime 2} } 
\, \g_K \left (\a_s(\m^{\prime 2} ) \right )\, ,
\label{KGKG}
\eqae
in which all logarithms of $n$ are generated either through the running
coupling and/or the explicit $\m'$ integral.  There are only two
steps left, to solve for the $n$-dependence of the $J$'s and 
to combine everything together in the cross section.

Combining eqs.\ (\ref{partialJKG}) and (\ref{KGKG}), we derive
the full evolution of $J_2$ in terms of $p_2^-$, an exactly
similar equation holding for $J_1$ in terms of $p_1^+$,
\eq
{\del  \over \del \ln p_2^-}\ln J_2 = K (1, \a_s (Q^2)) + 
G (1, \a_s(Q^2 ) )
- {1\over 2}\; \int^{Q^2}_{Q^2 / n} \; 
{d \l^2 \over \l^2} \; 
\G_J  (\a_s(\l^2 ) )\, .
\eqe
Here 
 $\G_J$ combines $\g_k$ with a term that allow us to have
the same running coupling in $K$ and $G$,
\eqa
\G_J(\a_s) &=& \g_K(\a_s) + \b (g) {\del \over \del g} K (1, \a_s ) \nonumber\\
&=& \left ( {\a_s \over \p} \right ) 2 C_F + 
\left ({\a_s \over \p} \right )^2 \left[ \left ( {6 7 \over 18} -
{\p^2
\over 6} \right ) C_F C_A - \left ({5\over 9} \right) n_f C_F \right]\, .
\label{GJdef}
\eqae
In the second line, we have given the two-loop expression for
$\G_J$, where as usual $n_f$ is the number of quark flavors.

It is now a simple applications of the chain rule to 
derive a differential equation for the $n$-dependence of
both jets,
\eq
\left [
 {\del \over \del \ln n} + {1\over 2} \b  {\del \over \del g}\; 
\right ]\;\ln  J
\left ( {Q^2 \over n \m^2} , {Q \over \m} , \a_s (\m^2 ) \right )
=  {1\over 2}\; \G_J^\prime \left (\a_s (\m^2 ) \right ) - {1\over 2}
\int^{Q^2}_{Q^2/n} \; {d \l^2 \over \l^2} \G_J \left (\a_s ( \l^2 ) \right )\, ,
\label{Jnevolve}
\eqe
where we have set $p_i\cdot \x=Q$ and where  
\eq
\G^\prime_J\left (\a_s (\m^2 ) \right )
 \equiv 
G_J\left(1,\a_s(Q^2)\right ) + K_J\left(1,\a_s(Q^2)\right ) - 2 \g_q\left(\a_s(\m^2)\right )\, .
\eqe

The solution of eq.\ (\ref{Jnevolve}) relates $J$ at large
$n$ to 
$J$ at $n=1$\, \footnote{One way to verify this result is to
observe that $\b(g)\partial\G(\a_s(\l^2))/\partial g=\l\partial\G(\a_s(\l^2))/\partial\l$.},
\eqa
 \ln J \left ( {Q^2 \over n \m^2} , {p^{\prime -} \over \m} , \a_s (\m^2)
\right ) &=&
\ln J
\left ( {Q^2 \over \m^2} , {p^{\prime -} \over \m} , \a_s (\m^2/n ) \right)
\nonumber\\
  &\ & \quad  
 -\; {1\over 2} \int^{Q^2}_{Q^2/n} \; {d \l^2\over \l^2} 
\left [ \ln {\m \over \l} \G_J \left (\a_s (\l^2 ) \right ) 
- \G_J{}^\prime \left  (\a_s (\l^2 )\right )\right ]\, .
\label{lnJresum}
\eqae
If we now set $\m=Q$, all logarithms of $n$ are generated by the integrals of the
two anomalous dimensions $\G$ and $\G'$ and the
expansion of $\a_s(Q^2/n)$, and, as promised,
exponentiate in the moments of $J$.

The resummed expression (\ref{lnJresum}) for the jets organizes all
logarithms of $n$ in the moments of the cross section, since the
hard function $H$ has no $\ln n$-dependence.  The inverse transform 
of $\tilde{J}(n)$ then gives the singular $1-T$-dependence.
To be explicit, the inverse Mellin transform of $\tilde{J}$ in
(\ref{lnJresum}) is given by \cite{thrustresum,ConSt}
\eq
J  \left ( (1-x){Q^2 \over \m^2}  , {Q\over \m} , \a_s (\m^2) \right ) \mid_{\m = Q} =
\left[ {e^{E(1-x, \a_s (Q^2))} \left[{1\over \p} \sin (\p_1P_1) \G(1+P_1 ) +
\ldots  \right]
\over 1-x} \right]_+\, ,
\label{Jpsquare}
\eqe
where $1-x\equiv p^2/Q^2$, and where the exponent $E$ is given by
the right-hand side of
(\ref{lnJresum}) with $\m=Q$,
\eq
E = - {1\over 2} \int^{Q^2}_{(1-x) Q^2} \; {d \l^2 \over \l^2} \left [
\right. \ln
({\m \over \l^2} ) \G_J (\a_s (\l^2 ) ) 
- \G_J^\prime (\a_s (\l^2)) \left. \right]
+
\ln \tilde{J}\left (1,1,\a_s((1-x)Q^2)\right )\, ,
\label{exponentE}
\eqe
while $P_1$ is related to the exponent by 
\eq
P_1 \equiv -{d E(1-x, \a_s (Q^2)) \over d \ln (1-x)}\, .
\eqe
Terms omitted in eq.\ (\ref{Jpsquare}) are suppressed by powers of
$p^2/Q^2$.
It is a  straightforward matter to verify that the
leading logarithms in $1-T$ in eq.\ (\ref{llaT}) are indeed
generated by this form.  Here, however, we see the essential
role of the running coupling for nonleading logarithms.
In particular, because of asymptotic freedom, the exponent 
receives relatively larger contributions from long distances,
and relatively smaller contributions from short distances
than in the case of a fixed coupling.  We shall return to 
the consequences of this observation in the final section below.

\subsection{$k_T$-factorization for the Drell-Yan cross section}

As another example of the variety of interesting cross
sections to which a variant of (\ref{sudjetmassfact})
applies, consider Drell-Yan cross sections at measured
pair mass squared $Q^2$ {\it and} transverse momentum, $Q_T$.
(The discussion below follows the extraordinary analysis of
Collins and Soper for transverse momentum
distributions in ${\rm e}^+{\rm e}^-$ annihilation \cite{CollinsSopersud}.)
Here again a factorization holds in covolution form, but
now the convolution is in terms of the transverse momenta
of gluons emitted from jet functions associated with the incoming 
hadrons, along with ``central" soft gluons from the soft
subdiagram of fig.\ \ref{dypinch}.  
For the Sudakov resummation of logarithms 
of $Q_T$ in the Drell-Yan cross section, a convolution in transverse
momentum will play the role of the convolution in 
jet mass for the thrust distribution.  Otherwise, the
reasoning is quite similar.  

Explicitly, the
convolution is \cite{CSSdyqt}
\eqa
{d \s_{hh'} \over  d Q^2 d^2 Q_T} &=& \sum_f \int d \x\; d \x^\prime 
\int {d^2 k_T\; d^2 k_T^\prime\; d^2k_{T,s}
\over (2 \p)^6} \nonumber\\
&\ &  \times
P_{f/h} (\x, k_T)\; P_{\bar{f} / h^\prime} (\x^\prime , k^\prime_T) \,
 H_{f \bar{f}} (Q^2)\,  S (k_{T,s}) 
\d^2 ({\bf Q}_T - {\bf k}_T - {\bf k}^\prime_T - {\bf k}_{s,T})\, ,
\label{qtconv}
\eqae
which factorizes under a Fourier transform,
\eqa
\tilde{W} (b Q, Q^2) &=& \s_0^{-1} \int d^2 Q_T 
e^{-i {\bf Q}_T\cdot{\bf b}} \; {d \s_{hh'} \over d Q^2 d^2 Q_T}\nonumber\\
&\cong& \sum_f \int {d \x \over \x} \; {d \x \over \x^\prime} \; 
{\tilde P}_{f/h}(\x , b \m ,Q/\m )\,
{\tilde P}_{\bar{f} / h^\prime} (\x^\prime , b \m , Q/\m )\, S (b \m)\, ,
\eqae
where $\s_0=4\p\a^2/9Q^2s$, the Born cross section, summarizes the hard part
$H$ to leading order in $\a_s$.
The jet functions are defined as matrix elements of quark
fields separated by a spacelike vector $(0^+y^-,\bf{b})$,
\eq
{\tilde P}_{\d/h} (\x, b \m, p^+ / \m) = \int {d y^- \over 4\p}
e^{-i\x p^+y^-} <
h (p ) \mid \bar{q}_f (0^+,y^-, {\bf b}) \g^+ q_f (0) \mid h (p)>\, ,
\eqe
where we suppress an average over spin.
These matrix elements are gauge-dependent, but in any axial gauge they 
absorb all double logarithms of $b$ (or $Q_T$ in momentum space).
At $b=0$, they are normalized to the quark distributions eq.\ (\ref{phiqazero}).
The  gauge vector plays the same role as in the case of thrust, 
and the jet matrix elements obey a noncovariant
evolution equation that separates their dependence on hard and soft
scales,
\eq
{\del \over \del \ln Q}\ln \tilde{P}  =
 K_P \left (b \m , \a_s (\m^2 ) \right ) + 
G_P \left ( {Q\over \m} , \a_s (\m^2) \right )\, ,
\eqe
where the combination $K_P+G_P$ is renormalization scale
invariant,
\eq
\m {d \over d \m} (K_P + G_P) = 0\, .
\eqe
Following essentially the
same  reasoning as for the $n$ dependence of the jets in the
thrust cross section, the logarithmic 
dependence of the jets exponentiates in the transform ($b$) space,
\eq  \tilde{P}_{f/h} \approx \; \exp \left\{ - {1\over 2}
\int^{Q^2}_{1/b^2} \; {d \l^2 \over \l^2} \left[  \ln \left ({Q^2 \over \l^2} \right )
\G_J \left (\a_s (\l^2) \right )
 + B \left (\a_s (\l^2)\right ) \right]   \right\}\,  \f_{f/h} (\x, 1/b^2) (1+
\a_s (1/b^2))\, ,
\eqe
with $\G_J$ as in eq.\ (\ref{GJdef}) and $B$ a power series in $\a_s$.
This approximation holds in the range of $b$ for which $Q >> 1/b >> \L$,
and organizes all perturbative logarithms of $Q_T/Q$ in the cross section.

Sudakov resummation may be relevant to any cross section with an underlying
hard-soft-jet factorization.  The exponentiation of logarithms requires
a convolution in phase space, like (\ref{qtconv}) or (\ref{dsigdT}).  
Applications include threshold corrections (where
the relevant variable is $1-z=1-Q^2/\x\x's$) for the inclusive
Drell-Yan cross section 
\cite{ConSt,dythresh} (\ref{dyfact}) as well as other, purely QCD cross sections
such as top or jet production 
\cite{topthresh}.  Another important case involves transverse momentum distributions
in ${\rm e}^+{\rm e}^-$ annihilation (where the first really complete
analysis of such a process was carried out \cite{CollinsSopersud}). 
 Yet another example is
semileptonic B meson decay at the endpoint of 
the lepton energy spectrum \cite{Bend}, 
where the lepton recoils against a jet of hadrons.  Undoubtedly, there are others as well.

We now turn to another classic resummation of large logarithms,
organized by the BFKL  equation.

\section{Two-Scale Problems II:  Small $x$ and the BFKL equation}

The small-$x$ limit of deeply inelastic scattering is
one of many cross sections that show a set of enhancements
organized by the BFKL (Balitskii-Fadin-Kuraev-Lipatov)
 equation \cite{bfkl,delduca}.  
Here again a
transverse momentum factorization may be used as a starting
point, although in a rather different kinematic region
from the Drell-Yan cross section just discussed.
In DIS, these enhancements
appear as logarithms in $x$ at fixed $Q^2$.  For $Q^2$
not very large, the quantity $\alpha_s(Q^2)\ln(1/x)$
can be large, and we may be tempted to resum  
corrections of this sort to all orders.  We begin
with a brief review of the origin of logs of $x$ 
in the evolution formalism developed in Section 4 above.

\subsection{$x\rightarrow 0$ for DGLAP evolution}

Referring to eq.\ (\ref{Evolfuns}) above, the 
kernel $P^{(1)}_{G/G}$ which describes gluon-to-gluon
evolution is singular as $x\rightarrow 0$,
so that in the notation of (\ref{pijexpand}) (and using $C_A=N$),
\eq
P_{GG}(x) = {2N \a_s \over \p} \; {1\over x} + \ldots\, .
\eqe
This behavior produces a pole 
at $n=1$ in the corresponding diagonal
element of the singlet anomalous dimension matrix, eq.\ (\ref{pijmom}),
\eq
\g_{GG}(n) = {2N \a_s \over \p} \; {1\over n-1} + \ldots\, .
\eqe

We now recall the solution to the renormalization group equation
for the moment of a DIS structure function $\bar{F}(n,Q^2)$, 
which follows from eq.\ (\ref{phvalevolve}),
\eq
\bar{F}(n,Q^2)=\bar{F}(n,Q_0^2)\; e^{(2\gamma_n^{(1)}/b_2)\ln t}\, ,
\label{FRGntoone}
\eqe
where $\gamma_n(\alpha_s)=\g^{(1)}_n\; (\alpha_s/\p)+\dots$, 
and $t\equiv\ln(Q^2/\L^2)/\ln(Q_0^2/\L^2)$.

For $x\rightarrow 0$, the (inverse)
transform from $\bar{F}(n,Q^2)$ to $F(x,Q^2)$,
\eq
F (x, Q^2) = \int^{i \infty}_{-i\infty} {d n \over 2 \p i}\, 
e^{-n \ln x + {4N/\left ( b_2(n-1) \right )} \; \ln t }\, ,
\label{inversemoment}
\eqe
has a sharp
 saddle point at $(n-1)=\sqrt{4N\ln t/\left ( b_2 \ln(1/x) \right )}$, 
which gives
the $x$ behavior 
\eq
F (x, Q^2) \sim \;\; e^{4\sqrt{{(N / b_2)} \ln (1/x) \ln t}}\, .
\label{dglapsmallx}
\eqe
This striking result shows a rapid increase as $x\rightarrow 0$.  
It relies, however, on DGLAP evolution, which assumes that $\ln Q^2$
is relatively large.  It is natural to ask what happens if $x$
is so small that, for instance, $\ln(1/x)\gg \ln(Q^2/\L^2)$, and
to treat the resummation of logarithms of $x$ self-consistently.

\subsection{$k_T$-factorization for DIS; the BFKL equation}

The standard DIS factorization, eq.\ (\ref{fonetwofact}),
assumes an ordering in transverse momenta, which allows us to decouple the
transverse momenta of the partons from the hard scattering.  Now, however,
we do not wish to treat $Q^2$ as arbitrarily large, so we generalize (\ref{fonetwofact}) to a convolution 
in {\it both} $\x$ and $k_T$.
The $k_T$-factorized form of a  DIS structure function $F(x,Q^2)$
is \cite{ktconvl1,ktconvl2}
\eq
F (x, Q^2) =\int d^2 k_T \int^1_x {d \x \over \x} 
C\left({x \over \x},Q,k_T \right) 
\cf (\x,k_T)\, ,
\label{ktfactdis}
\eqe
with $\cf (\x,k_T)$ a generalized parton distribution at measured
$x$ and $k_T$, and $C({x/\x},Q,k_T)$ the corresponding 
coefficient function.  Since leading logarithms of $x$ are 
generated by purely gluonic evolution, we shall restrict
ourselves to gluon distributions, neglect mixing, and
suppress parton indices.
This factorization is illustrated in fig.\ \ref{ktfactfig}. 
\begin{figure}[ht]
\centerline{\epsffile{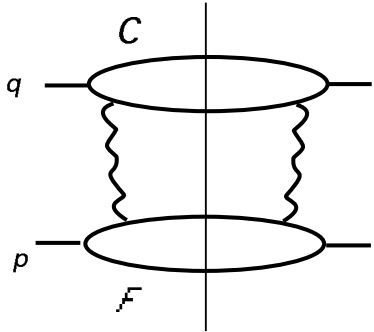}}
\caption{$k_T$-factorization in DIS.}
\label{ktfactfig}
\end{figure}
It is appropriate for the limit $Q^2$ fixed, $x\rightarrow 0$. 

The leading logarithms of $x$ in (\ref{ktfactdis}) are generated
by a large set of diagrams.  The diagrams that show the
relevant mechanism most clearly are the 
ladders, illustrated by fig.\ \ref{ktfactladd}. 
\begin{figure}[ht]
\centerline{\epsffile{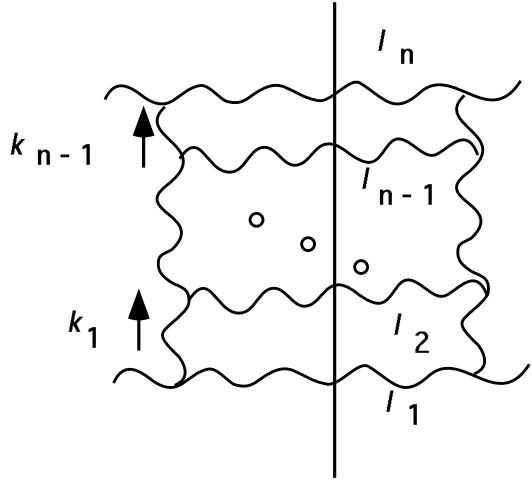}}
\caption{$n$-gluon cut ladder.  In DIS, the photon would attach to
a quark loop at the top.}
\label{ktfactladd}
\end{figure}
To understand how logarithms of $x$
are produced, we parameterize the momenta 
flowing on the sides
(vertical lines in the figure) in components parallel
to the incoming momenta $p$ and $q$
 and transverse components,
\begin{equation}
k_i=\alpha_ip+\beta_iq+k_{iT}\, .
\label{ksubi}
\end{equation}
Here $q$ is approximated by a lightlike vector, because
as $x\rightarrow 0$ the ratio of the photon invariant
mass to the center of mass energy vanishes.
Logarithms of $x$ result from configurations
in which the ``light-cone" fractions $\alpha_i$ and $\beta_j$ are
strongly ordered, but the transverse momenta are all of the
same order,
\begin{eqnarray}
\alpha_1 \gg \alpha_2 &\gg& \cdots \gg \alpha_{n-1} \nonumber \\
\beta_1 \ll \beta_2 &\ll& \cdots \ll \beta_{n-1} \nonumber \\
     k_{iT} &\sim& k_{jT}\, .
\label{strongorder}
\end{eqnarray}
It is the lack of ordering in transverse momentum that distinguishes
BFKL evolution. Useful approximations that follow from (\ref{strongorder})
are
\eqa
\sum_{i=1}^j\b_i &\sim& \b_j\, ,\nonumber \\
\sum_{i=j}^{n-1}\a_i &\sim& \a_j\, ,
\label{abinequal}
\eqae
for the lightcone fractions and
\eq
k_i^2\sim -k_{i,T}^2\, ,
\eqe
for the invariant masses of lines on the sides of the ladder.  The
emitted ``rungs" of the ladder carry momenta
\eq
\ell_i=k_{i-1}-k_i\, ,
\eqe
with $k_0=p$ and $k_n=-q$.

A diagram like fig.\ \ref{ktfactladd} with $n$ rungs generates $n-2$
logarithms of $x$, which come from ordered logarithmic
integrals over the $\a_i$, or equivalently rapidities $y_i$
of the $i$th rung, 
\eq
y_i = {1\over 2} \ln {\a_{i-1} \over \b_i}\, .
\eqe

It is relatively easy to identify the leading behavior of the ladder diagrams.  There
are $n$ mass-shell delta functions from
the cut rungs, $\d(\ell_i^{\; 2})$.  The top-most
mass-shell condition fixes $\a_{n-1}\sim x$, while the
remaining $\d(\ell_i^{\; 2})$ fix $n-1$ fractions $\b_i$, $i=1\ldots n-1$.
The leading numerator factor comes from the terms that contract the incoming momenta of
the top and bottom vertices to give $(p\cdot q)^2$, while the vertices to which
$i$th rung connects produce a factor
of order  $|(k_{T,i-1}+k_{T,i})\cdot \e_i(\ell_i)|^2$, with $\e_i(\ell_i)$ 
the polarization of the $i$th emitted gluon.  The resulting term is logarithmic in the 
remaining, ordered $\a_i$ integrals, which are each of the form,
\eq
\int_{\a_{i - 1}} d \a_i / \a_i = \int_{y_{i-1}} d y_i\, ,
\eqe
with a minimum value $\a_{n-2,{\rm min}}=x$.  Transverse integrals
also give logarithmic power counting on a graph-by-graph basis.
The resulting logarithms, however, cancel in the sum over diagrams.
This cancellation will be evident in the equation we derive below.

Consider what happens when we insert
yet another gluon into an arbitrary diagram, as
illustrated in fig.\ \ref{onemoregfig}.
The diagram has been factorized because all of its gluons
are already ordered as in (\ref{strongorder}).    
\begin{figure}[ht]
\centerline{\epsffile{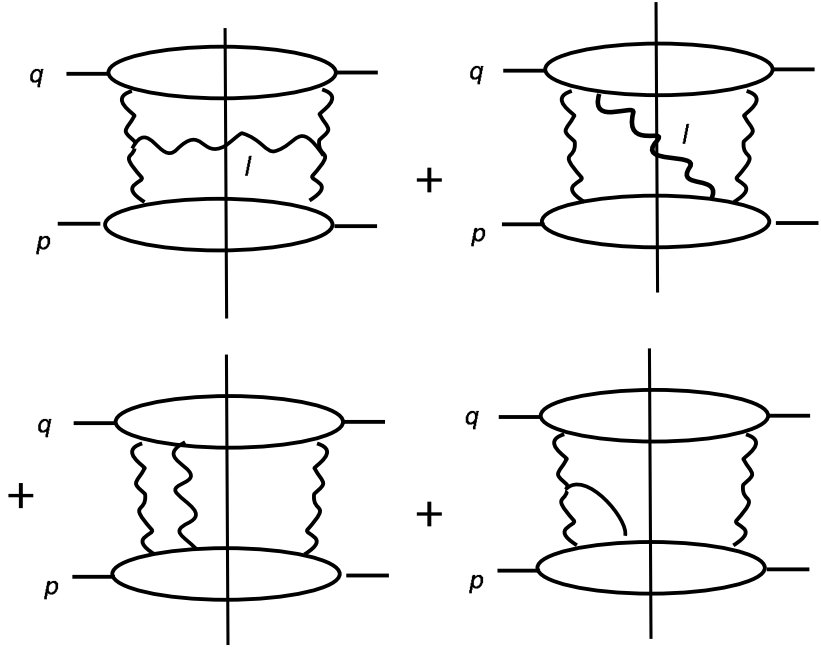}}
\caption{Adding an extra strongly ordered gluon.}
\label{onemoregfig}
\end{figure}
To get an
additional logarithm from this gluon, it should 
fit into 
the rapidity gap between $C$ and $\cf$,
and should have a transverse momentum comparable to those of
the other gluons.  To get the logarithm, however,
the ``new" gluon, of momentum $\ell$, need not itself be inserted
as a ladder.  The ladder insertion, the
first in fig.\ \ref{onemoregfig}, works, but so do many insertions
that connect the vertical gluon and {\it either}
the top or bottom, or even some that connect the top and bottom
directly.  
Similarly, there are additional possibilities for the insertion of
a virtual gluon, also illustrated in fig.\ \ref{onemoregfig}.

To determine when each of these diagramatic insertions
can give logarithms we use 
strong ordering (\ref{strongorder}).
Because of strong ordering,
whenever the momentum $\ell$ flows along a line 
of momentum $\ell'+\ell$, say, in $C$, 
$(\ell'+\ell)^2\sim 2\ell'{^-}\ell^+=2\alpha_\ell\beta_{\ell'}s$,
and analogously for attachments to $\cf$, but with the roles of
the fractions 
$\alpha$ and $\beta$ reversed.  

Luckily, however, it is not
really necessary to worry about each diagram individually.
Instead, we appeal 
to the jet-soft analysis of Section 4.7 above.  
Strong ordering implies that $C$ is
sensitive to $\ell^+$ only, and $\cf$ is sensitive to 
$\ell^-$ only.  
At the same time, it is clear that the $+$ component
of the polarization of $\ell$ is also dominant in its
coupling to $C$, and the minus to $\cf$.    
\begin{figure}[ht]
\centerline{\epsffile{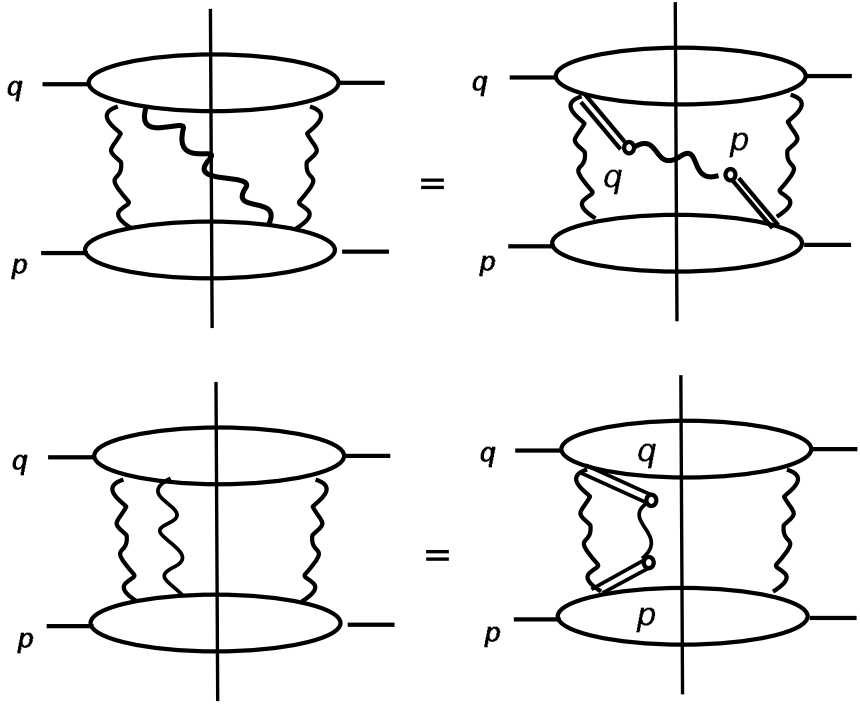}}
\caption{Eikonal factorization of additional strongly ordered gluon.}
\label{Ceikonalfig}
\end{figure}
We are
thus in the situation of eq.\ (\ref{Jsofthomogen}), fig.\ \ref{onepisoft},
and as in that case, the 
sum of all attachments of the gluon $\ell$ 
to either of the subdiagrams $C$ and $\cf$ 
factors from it as in fig.\ \ref{Ceikonalfig},
with an eikonal factor
\eq
{q^\n(-gC_{abd}) \over q\cdot \ell}\, ,
\label{Ceikonal}
\eqe
for $C$, and similarly for its connection to $\cf$.

The result of attaching our extra gluon is then
summarized by a kernel $\bar{K}(k',k)$, which
we illustrate with a cut
gluon in fig.\ \ref{kdef}.
The new gluon attaches
at either end to a vertex represented by a circle.  
\begin{figure}[ht]
\centerline{\epsffile{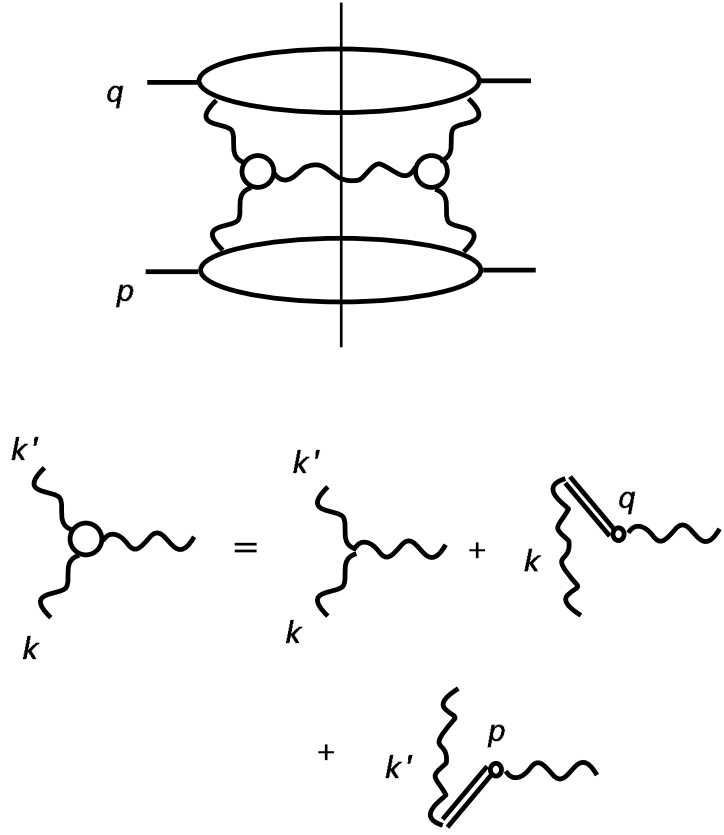}}
\caption{Graphical representation of the BFKL kernel for
a cut gluon.}
\label{kdef}
\end{figure}
  The eikonal
factors themselves (\ref{Ceikonal}) are
as usual represented by double lines. 

Because the $\a_i$ integrals are ordered,
 the separation between $\cf$ and $C$ in eq.\ (\ref{ktfactdis})
is at a definite value of $\a$, or equivalently of 
plus momentum $\ell^+=\a p^+$.
We may thus identify the combination of 
$\cf(k)$ and $\bar{K}(k',k)$ in fig.\ \ref{kdef}
with $\cf(k')$, now evaluated
at external momentum 
$k'=k-\ell$.  
This gives an integral equation for the $k_T$-dependent
jet function $\cf$.  Since we 
know there is a logarithmic 
integral in $\ell^+$, we write 
\eq
\cf(k') 
=  
\int_{\alpha p^+} {d\ell^+ \over\ell^+}\; \int
 d^2k_T\; \bar{K}(k'_T,k_T)\; \cf(\ell^+,k_T)\, .
\label{prebfkl}
\eqe
Here $\a$ plays the role of a factorization scale,
separating $C$ and $\cf$.  
Thinking back to DGLAP evolution for ordered 
transverse momenta, we derive a new equation by
taking a derivative with respect to $\a$,
\eqa
 \a {\del \over \del \a} \bar{\cf} (\a,{k}^\prime_T ) &=&
\int d^2 k_T\, K (k^\prime_T, k_T )  \bar{\cf} (\a, k_T )\nonumber\\
&=& - {\a_s N \over \p^2} \int {d^2 k_T \over (k_T - k^\prime_T)^2}
\left\{ \bar{\cf} (\a, k_T ) - {k^\prime_T{^2} \over 2 k_T^2}
\bar{\cf} (\a,
k^\prime_T ) \right\}\, ,
\label{bfkleq}
\eqae
where we have defined
\eq
\bar{\cf} (\a, k_T) \equiv {1\over k_T^2} \cf (\a , k_T )\, .
\eqe
This is the BFKL equation, as it appears in DIS, and $K(k'_T,k_T)$
is called the 
BFKL kernel.
In the second line the kernel has been evaluated explicitly for 
the diagrams of fig.\ \ref{kdef}.  Details of the
evaluation are given in Appendix C.  This is only one of many forms
in which the kernel is expressed, but it is one of the simplest.  Clearly,
the first term comes from  real-gluon diagrams, for which
$k_T'$ is not identically equal to $k_T$, while the second
term comes from virtual gluons.

\subsection{Solution of the BFKL equation}

Because the BFKL equation (\ref{bfkleq}) is a convolution in $k_T$,
and first order in the derivative with respect to $\ln\a$, it is natural
to express its solution as an expansion in functions that
are powers in $\a$ and $k_T$,
\eq
\cf_{\o, \n} (\a , k_\t) = c_{\o, \n} \a^{-\o} 
\left (  {k_T{}^2\over \m^2} \right )^{i\n - {1\over 2}}\, ,
\label{bfklansatz}
\eqe
where $c_{\o,\n}$ and $\m^2$ are constants.  We want to solve
for $\o$, which determines the power-law $\a$-dependence, and the associated
power $\n$ for transverse momenta.  
$\n$ is to be considered as a complex number, so that $i\n$ is not 
necessarily imaginary.
The extra $-1/2$ in the power of
$k_T^2$ is a matter of convenience.  
Together, $i\n-1/2$ serve as an anomalous dimension appropriate
to evolution with a {\it fixed} coupling (see eq.\ (\ref{phvalevolve})).
From the discussion of DGLAP evolution in Sec.\ 4.4 above, we recall that 
logarithms, and hence evolution,
in parton distributions arise from transverse momentum integrals.
The argument of the running coupling is thus naturally chosen as\
$k_T^2$, which is effectively fixed in our case, because of the
strong-ordered kinematics of eq.\ (\ref{strongorder}).

Substituting the ansatz (\ref{bfklansatz}) into the BFKL equation (\ref{bfkleq}),
we readily find an implicit expression that relates $\o$ and $\n$ (see Appendix C),
\eq
\o (\n) \left ( {k_T^{\prime 2} \over \m^2} \right )^{i \n - {1\over 2}} 
= {\a_s N \over  \p^2}
\int d^2 k_T 
\left\{ \left ( {k_T^{\prime 2} \over \m^2} \right )^{i \n - {1\over 2}} 
{1\over (k_T
-k^\prime_T)^2} - 
\left ( {k^{\prime 2} \over \m^2} \right )^{i\n - {1\over 2}} 
{(k_T^{\prime 2}) \over 2 k_T^2 (k_T - k^\prime_T)^2}  \right\}\, .
\label{omegaofn}
\eqe
The infrared divergences of each term on the right-hand side of
this expression cancel, as may be seen by carrying out the 
integrals via dimensional regularization, and the explicit relation
between $\o$ and $\n$ is 
\eq
\o (\n) = - {2\a_s N \over \p} \left ({\rm Re}\, \j  \left (i \n + {1\over 2} \right ) - \j (1) \right )\, ,
\label{omegasoln}
\eqe
where the special function $\j$ is 
the logarithmic derivative of the gamma function, 
\eqa
\j (x) &=& {d \over dx} \ln \G (x)\, , \nonumber\\
\j (1) &=& - \g_E\, .
\eqae
Another common form for the relation (\ref{omegasoln}) is found
by defining 
\eq
\g \equiv i \n + {1\over 2}\, ,
\eqe
in terms of which
\eq
\o (\g ) = - {\a_s N \over \p}
\left  (\j (\g ) + \j (1-\g) - 2 \j (1) \right )\, .
\eqe

A general solution to the BFKL equation for DIS may be written
as a superposition of power-law solutions,
\eq
\cf (x, k_T) = \int^\infty_{-\infty} d \n\; x^{-\o (\n)} \left ( {k^2_T \over \m^2}
\right )^{i \n - {1\over 2}}\, .
\eqe
The small-$x$
limit is dominated by a saddle-point of $\o$ as a function of $\n$,
in much the same way as the small-$x$ behavior from
DGLAP evolution in eq.\ (\ref{inversemoment}) above.  
In this case there is a saddle point at $\n=0$, as may be
seen from the series found by expanding the $\j$ functions,
\eq
\o (\n) = {2 \a_s N \over \p} \left (2 \ln 2 + \sum^\infty_{k=1} (-1)^k (2^{2k +
1} -1 )
 \z (2 k + 1) \n^{2k} \right )\, ,
\eqe
with $\z(2k+1)$ the zeta function $\z(x)=\sum_n (1/n^x)$.
This leads to the (famous) asymptotic behavior \cite{bfkl}
\eq
\cf (x, q_T) \sim x^{- 4N \ln 2(\a_s/\p) }  (q_T^2)^{-1/2}\, ,
\label{fourlntwo}
\eqe
in which we see a power-law enhancement as $x$ vanishes,
much stronger than in DGLAP evolution, eq.\ (\ref{dglapsmallx}).

The BFKL formalism was originally developed \cite{bfkl}
to describe
hadron-hadron scattering in QCD, both the total
cross section and the closely-related Regge limit \cite{pdbcollins},
$t$ fixed, $s\rightarrow \infty$.  
The $t=0$ Regge limit is related to the $x\rightarrow 0$ limit in DIS
by the following
``translation",
\begin{eqnarray}
(p+q)^2 = {1-x \over x}Q^2 &\rightarrow& s
\nonumber \\
        x^{-\omega_0} &\rightarrow& s^{\omega_0} 
\nonumber \\
             F(x) &\rightarrow& \sigma_{\rm  tot} \sim s^{\omega_0}\, .
\label{distohh}
\end{eqnarray}
Thus, at $t=0$, $\s_{\rm tot}$ grows as a
power of $s$ in this approximation. In fact, the total cross section
for hadron-hadron scattering does increase at high energy,
although an uninterrupted
 power-law rise would violate unitarity, as embodied
in the Froissart bound \cite{pdbcollins}.  The BFKL 
behavior that we have just derived 
therefore cannot be the final answer.

Although the dominant BFKL power (\ref{fourlntwo})
is infrared finite, in most applications, such as the total 
cross section, there is no natural perturbative scale
at which to evaluate $\a_s$.  The scale of $k_T$ introduced above 
is, after all, quite arbitrary.  
There are, however, two-jet correlations \cite{muellernavelet}
for which BFKL resummation is naturally infrared safe, with a coupling
fixed at a perturbative scale of the order of the transverse momenta
of the jets. The dominance of resummed
leading logarithms in such  cross sections
may require very high energy \cite{DelDucaSchmidt}.
Another infrared safe application of the BFKL formalism is 
to hypothetical heavy-quark onium-onium scattering,
in which the inverse size of the onium wave function serves
as an infrared cutoff \cite{onium}.  This model is serving as
a valuable laboratory for the study of forward scattering
in a self-consistent perturbative context.

At least two sets of corrections can lead to a softening of
the BFKL or ``bare" pomeron, nonleading logarithms 
and nonleading powers (higher twist).
``Nonleading logarithms" refers to higher
powers of $\a_s$ at  a fixed power of $\ln s$.  One approach is
to derive nonleading terms in the expansion of the kernel $K(k,k')$ \cite{bfkltwoloop}.
Another is to compute exchanges of multiple ladders \cite{multladd}.
Indeed, 
a multiladder generalization of
the BFKL equation \cite{multiladd2} may be used as an inspiration 
for a picture of forward scattering in terms of two-dimensional
field theories.  In these investigations, QCD comes into contact
with conformal field theory and the theory of exactly soluble models.

In addition, ladders may interact with each other, a process
that produces ``shadowing", the softening of parton distributions
due to
the spatial overlap of partons with small momentum fractions \cite{shadow}.  If 
shadowing is a small correction, it is a higher-twist effect.
Higher-twist need not mean small, however, because when the overlap 
large, it destroys the incoherence at the basis of
the partonic formulation, and all twists become equally important.  

The field of small-$x$ dynamics is especially compelling because
at very small $x$, but large $Q^2$, the density of partons is high,
even while the coupling $\a_s(Q^2)$ remains small \cite{shadow}.  Such a
dense, but weakly coupled system promises a new testing
ground for field theory, with a close relation to the physics
of nonabelian plasmas.
Even before such an asymptotic condition is understood, however,
there are many plausiable applications of small-$x$ resummation and
$k_T$-factorization \cite{ktconvl2}  to current phenomenology, whenever a
hard scattering occurs at scale $Q^2$, with $Q^2\ll s$.
For example, much attention has been given to the production
of jets and heavy quarks in this regime.  Discussions of 
some of these developments are found elsewhere in this volume.

We have barely scratched the surface of the physics of
two-scale problems here.  The representative examples described
above, however, may give a sense of how far it is possible,
and necessary, to go beyond low-order perturbation theory in QCD.
In the next section, we shall encounter another extension of
the formalism, in which long- and short-distances mix, even 
in cross sections with a single hard scale.

\section{High Orders in Perturbation Theory}

Throughout these lectures, we have used the singularities of
perturbation theory as a diagnostic for long-distance
behavior, and as a guide for organizing the relation of
short to long distances.
In this section, we shall briefly discuss yet another aspect
of perturbation theory that gives hints of nonperturbative
structure, its behavior at high orders \cite{highorder,muellerirr}.

Recall the relation between the total ${\rm e}^+{\rm e}^-$
annihilation cross section and the imaginary part of the
two-current correlation function, eqs.\ (\ref{sigmatot}) and
(\ref{piequalsvev}),
\eq
\s^{(\rm tot)}_{{\rm e}^+{\rm e}^-} \sim \int d^4x\, e^{i q x} 
<0\mid J^\m(x) J_\m(0) \mid 0 >\, ,
\eqe
with $J_\m$ an electromagnetic current.  As in eq.\ (\ref{lcexpansionjj}),
we can apply the operator product expansion, but now, because there
is no ``external" momentum in the matrix element, only the 
expectation values of scalar operators can contribute, and only
a few operators appear with singularities at $x^2=0$ in the
operator product expansion,
\eq
J^\m(x) J_\m(0) \sim 
{1\over x^6} C_0 (x^2\m^2) I + {m\over x^2} C_q (x^2\m^2) \bar{q}q(0) +
{1\over x^2} C_F (x^2\m^2) F_{\m\n} F^{\m\n}(0) + \ldots\, ,
\label{epemope}
\eqe
with $I$ the identity operator, which does not contribute to DIS.
Perturbation theory with all masses set to zero 
contributes at any finite order to $C_0(x^2\m^2)$ only.  
Yet, as we shall now see, there is a problem with $C_0$ from
high orders, which suggests the 
presence of 
the higher terms in the operator product expansion, even in the absence of explicit quark
masses.

Our reasoning begins with the pinch
surfaces of $C_0$, which we have already identified in
Sec.\ 3.2, fig.\ \ref{softcloud}, in which a 
purely soft subdiagram $S$, consisting
entirely of zero-momentum lines, is attached to a single hard subdiagram
$H$.  Near the pinch surface, all soft
momenta may be neglected in $H$, which may therefore be treated 
effectively as a gauge-invariant local vertex.  Consider any
single gluon
internal to subdiagram $S$, whose momentum we label $k$.
$S$ may formally be written as
\eq
S=\int d^4k\; {g^{\a\b}\over k^2}\; T_{\a\b} \left (k,\m,\a_s(\m) \right )\, ,
\eqe
where we have isolated the $k$ propagator (in Feynman gauge) and where
$T_{\a\b}$ is the remainder of the subdiagram, as in fig.\ 
\ref{irropefig}.  
\begin{figure}[ht]
\centerline{\epsffile{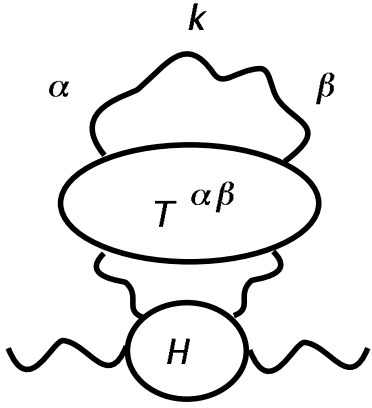}}
\caption{Reduced diagram for a soft pinch surface.}
\label{irropefig}
\end{figure}
Dimensional counting and gauge invariance then
require $T_{\a\b}$ to have the form
\eq
T_{\a\b}(k,\m,\a_s(\m))=
\left ( k^2\; G_2(k^2)\right)\; 
(k_\a k_\b-k^2g_{\a\b})\, t \left (k^2/\mu^2,\a_s(\mu^2) \right )\, ,
\eqe
where $G_2(k)$ is the normalized trace of the full
gluon propagator.  
Neglecting for simplicity any renormalization of the vertex $H$,
the function $t$ is  renormalization-group invariant,
\eq
\m {d \over d \m} \; t (k^2/\m^2 , \a_s (\m^2 ) ) = 0\, ,
\eqe
from which we conclude that we may choose $\m^2=k^2$ as
we integrate over $k$, at the cost of running the 
coupling to the scale $k^2$,
\eqa
t \left ( k^2/\m^2 , \a_s (\m^2) \right ) &=& t \left (1, \a_s (k^2) \right ) 
\nonumber \\
&=& \sum_i a_i\a_s{}^i(k^2)\, ,
\eqae
with the $a_i$ numbers.  We can already see something funny by
looking at the resulting
$a_1$ term for $S$.  Using the one-loop running coupling, we find
\eqa
S^{(1)} &=& - 3 \int_0^Q d^4 k\; k^2  {\a_s (k^2) \over k^2} \nonumber\\
&=& - 12 \pi \int_0^{Q^2}  dk^2 k^2 {\a_s (Q^2) \over 1+ {\a_s (Q^2)\over
4\p} b_2 \ln {k^2 \over Q^2}} \nonumber\\
&=& - 12 \p Q^4 \a_s \sum_n \left (\a_s {b_2 \over 4\p} \right)^n \int_0^1 dx x \ln^n 
{1\over x} \nonumber \\
&=& - 6 \p Q^4 \a_s \sum_n \left (\a_s {b_2 \over 8\p} \right )^n\, \Gamma(n+1)\, ,
\label{soneeval}
\eqae
where in the last two lines we have reexpanded the running coupling in terms
of $\a_s(Q^2)\equiv\a_s$.  As a result, the $n$th order in $\a_s$
has a coefficient that grows like $\Gamma(n+1)=n!$. 
This is despite the infrared safety of the matrix element.
 Evidently,
this uncontrolled growth in perturbative coefficients is
a direct reflection of the singularity in the running coupling.
All is not lost, however, although this behavior will require us
to reevaluate how we regard the perturbative expansion in QCD.  

A very useful conceptual tool for treating high orders in
a perturbative series
is the Borel transform.   Consider a general power series
in an expansion variable, in this case $\a_s$,
\eq
\P (\a_s) = \sum^\infty_{n=0} c_n \a_s{}^n\, ,
\label{Piexpansion}
\eqe
with the  $c_n$ constants.  If $\P(\a_s)$ is analytic
at $\a_s=0$, the  $c_n$ are coefficients in a Taylor series expansion.
This need not always be the case, however.  
We are interested in the case when the $c_n$ grow, and $\P$ may
possess no radius of convergence at all about $\a_s=0$.  Nevertheless,
there is a good deal of information in the expansion (\ref{Piexpansion}).
To see why, we define the Borel
transform of $\P(\a_s)$ by
\eq
\tilde{\p} (b) = \sum^\infty_{n=0} {c_n \over n!} b^n\, .
\eqe
It is an expansion in a conjugate variable $b$, whose
expansion coefficients are simply $c_n/n!$.  $\tilde{\p}(b)$\
is thus much more convergent than $\P(\a_s)$ is, and
has a finite radius of convergence about $b=0$ even when
the $c_n$ grow as fast as $n!$.  Formally, the inverse
transform from $\tilde{\p}$ back to $\P$ is
\eq
\P (\a_s) = \a_s^{-1}\, \int_0^\infty d b\, e^{-b/\a_s} \tilde{\p} (b)\, ,
\eqe
since the integral over $b$ precisely generates $n!\a_s^{n+1}$
from the $b^n$ term.  Factorial growth
in the expansion coefficients of $\P$ now shows up as a 
singularity in $\tilde{\p}$.  If this singularity is on the
real axis, the inverse transform is an ambiguous integral
even if $\tilde{\p}$ is known as a function.  Even if the
singularity is off the real axis, its presence
indicates contributions to $\P$ that cannot be 
described fully by the series (\ref{Piexpansion}), {\it i.e.},
nonperturbative contributions.  

Returning to our example $S^{(1)}$ above, we observe that
a simple change of variables,
\eq
b'={\a_s(Q^2)\over 4\p}\; \ln (Q^2/k^2)\, ,
\label{bprimedef}
\eqe
in the second line of eq.\ (\ref{soneeval})
 leads to an expression for $S^{(1)}$ that is precisely
of the inverse Borel form,
\eq
S^{(1)} (Q^2) = - 48 \p^2 \a_s^{-1} Q^4 \int_0^\infty d b^\prime {e^{- 8 \p b'
/\a_s
(Q^2) } \over (1-b_2 b^\prime )}\, .
\label{soneinverseborel}
\eqe
Here $1/(1-b_2b')$ plays the role of the Borel transform
of $S^{(1)}$, and its singularity is a direct reflection of 
the singularity in the perturbative running coupling at $k^2=\L^2$.
Any such singularity in the plane of the Borel variable due to
the infrared behavior of the running coupling is called an infrared renormalon.

Although (\ref{soneinverseborel}) is ill-defined, we have
gained something by reexpressing the integral in this fashion,
if we regard the singularity as an ambiguity in the inverse Borel transform,
which is well-defined in the full theory, although not
in perturbation theory alone.  We imagine that the Borel
transform 
is well approximated by perturbation theory up to $b'=1/b_2$.
Although the perturbative integral is not well-defined beyond
this point, the integrand is already suppressed at 
$b'=1/b_2$ by a factor $\exp[-8\p/b_2\a_s(Q^2)]=(\L^2/Q^2)^2$.
Thinking back to the operator product expansion,
eq.\ (\ref{epemope}), we recognize that this is 
the power corresponding to the gluon condensate, the
vacuum expectation of the 
operator $F^2$.  Perturbation theory itself thus signals 
its own incompleteness by generating an infrared renormalon
ambiguity at precisely the leading nonperturbative
power of the operator product expansion (at zero quark mass,
in this approximation).  

It is a widely accepted viewpoint that the correct way to treat
the perturbative expansion is to {\it define} perturbation
theory by regulating the inverse Borel transform in such a
way that it introduces a new nonperturbative parameter
that may be associated with the vacuum expectation
value $\langle 0|F^2|0\rangle$.  The theory in principle then
gives a consistent picture of the function $\P$ up to 
corrections of order $Q^{-6}$ relative to the leading
power, and up to the next uncalculated order in
perturbation theory \cite{muellerirr}. 

The above discussion has brought us to the threshold of
nonperturbative physics, which we cannot expect to cross without
nonperturbative methods.  In closing, we may note that
infrared renormalons appear not only in 
the total cross section for ${\rm e}^+{\rm e}^-$ annihilation, but
in many other cross sections as well.  
They are particularly interesting in resummed cross sections,
where we have seen integrals over running couplings, analogous
to those just encountered, appear in the
organization of large corrections \cite{resumirr}.  It is natural to ask
whether here, as above, 
perturbation theory is signalling a new set of nonperturbative
parameters, which probably cannot be reduced to the operator
product expansion.  The full answer to this intruiging question
is not, to my knowledge, available at present.

\section*{Acknowledgements}

It is a pleasure to express my appreciation to the TASI 95 Organizing Committee,
and to its General Director K.T.\ Mahhanthapa and Program Director,
Dave Soper.  I would like to thank as well the TASI Scientific Advisory Committee,
for their choice of quantum chromodynamics as the central theme of
the 1995 school, which made it possible to explore the topics
presented here at a depth that would not otherwise have been possible.
I am truly indepted to Ms.\ Linda Freuh for her expert assistance at Boulder.

The treatment above on Sudakov resummation grew out
of discussions with Harry Contopanagos and Eric Laenen, and of
small-$x$ resummation out of discussions with Michael Sotiropoulos.
To Gregory Korchemsky, I would like to express my appreciation for 
many conversations during the months leading to the school, which profoundly
influenced my presentations there.  Many other insights expressed
above are due, directly or indirectly, to
exchanges in recent years with Lyndon Alvero, Stan Brodsky  and
Anatoly Radyushkin,
and with my colleagues in the CTEQ Collaboration, especially
John Collins, Jianwei Qiu, Jack Smith and Dave Soper.

Finally, I wish to thank the students of TASI 95, for their interest and
challenging questions, and for bringing their own research experience
to bear on these lectures, which acquired as a result a sense of dialogue.

This work was supported in part by the National Science Foundation under
grant PHY-9309888.

\section*{Appendix A. Color Matrix Identities and Invariants}

\label{Groupapp}

I will review in this appendix a few of the group identities useful
for elementary perturbative calculations in QCD.
The ``defining" generators $T_a^{(F)}$ are 
$N^2-1$ $N\times N$ traceless hermitian matrices,
while the generators $T_a^{(A)}$ are defined by the
$SU(N)$ structure constants $C_{abc}$, through
\begin{equation}
{\big ( T_a^{(A)} \big )}_{bc}  =  -iC_{abc}\ .
\label{Ac}
\end{equation}

 In representation $R$, the generators satisfy the trace normalization
\begin{equation}
{\rm Tr}\, [\, T_a^{(R)} T_b^{(R)}\, ] = T_R\delta_{ab}\ ,
\label{Aa}
\end{equation}
where for the defining (quark) and adjoint (gluonic) representations
\eqa
T_F &=& {1 \over 2} \nonumber \\
T_A &=& N\, .
\eqae

The generators also give rise
to invariants $C_R$, 
\begin{equation}
\sum_{a=1}^{N^2-1} { \big ( T_a^{(R)} \big )}^2 = C_R I,
\label{Ab}
\end{equation}
with $I$ the identity matrix.  For the 
defining and adjoint
representations, these are 
\begin{eqnarray}
 C_F &=& { N^2 - 1 \over 2N }
\nonumber \\ 
 C_A &=& N\ .
\label{Ad}
\end{eqnarray}

For the defining representation,
 products of the generators in $SU(3)$ are given by
\eq
T_a^{(F)}T_b^{(F)}= \half [iC_{abc}T_c^{(F)}+ d_{abc}T_c^{(F)}]
+ {1 \over 6} \delta_{ab}I\; ,
\eqe
with $I$ the 3$\times$3 identity, and the $d_{abc}$ real and totally symmetric.

\section*{Appendix B. Time Ordered Perturbation Theory, 
Generalized Unitarity and the Landau Equations}

Time-ordered perturbation theory (TOPT) allows a simple proof
of the generalized unitarity discussed in Sec.\ 3.1.  

Time-ordered (``old-fashioned") perturbation theory is 
equivalent to the
more familiar expansion in
covariant ``Feynman" diagrams.  Schematically, for any
Green function,
\eq
{\rm Green\ function} = \sum_{\rm cov.\ graphs}\; G 
= \sum_{\rm TO\ graphs}\; \G\, .
\eqe
Time-ordered (TO) diagrams are topologically identical to
covariant diagrams, but
their vertices are ordered in time.  Thus, a covariant
diagram with $V$ vertices corresponds to as many as $V!$
TO diagrams.  (When a subset of these $V!$ 
permulations are identical, they are
counted only once \cite{Stermanbook}.)  A TO diagram $\G$ 
consists of the integral of a product of factors, ``energy
denominators", that measure
the virtuality of a set of states,
\eq
\G(p)=-i\prod_{{\rm loops}\ i}\int{d^3\ell_i\over (2\p)^3}\prod_{{\rm lines}\ j}{1\over 2\o_j(p,\ell_i)}
          \prod_{{\rm states}\ a} {1\over E_a-S_a+i\e}\, N(p,\ell_i)\, .
\label{topt}
\eqe
Here the set of lines between the $a$th and $(a+1)$st vertices
define the ``state" $a$.
$E_a$ is the ``energy of state $a$", the total energy that
has flowed into the diagram up to the $a$th vertex. $S_a$ is the
``on-shell" energy of state $a$, which is the sum of the mass-shell 
energies of each of the lines in $a$,
\eq
S_a=\sum_{\stackrel{{\rm lines}\ j}{{\rm in}\ a}}\o_j
=\sum_{j\ {\rm in}\ a}\sqrt{|{\bf p}_j|^2+m_j^2}\, .
\eqe
The factor $N$ represents ``numerator" factors from, for
instance, fermion propagators and three-gluon vertices,
computed with on-shell line momenta.  In gauge theories, there 
are further technicalities and modifications associated with
extra gauge propagators and self-energy diagrams, but we shall
not need these subtleties here.  The general form of eq.\ (\ref{topt})
is, hopefully, familiar from TOPT in nonrelativistic quantum mechanics.

The expression (\ref{topt}) is relatively easy to prove
directly from covariant perturbation theory.  The
example of a  scalar
self-energy diagram, fig.\ \ref{selfenfig} already illustrates the general
pattern.  
\begin{figure}[ht]
\centerline{\epsffile{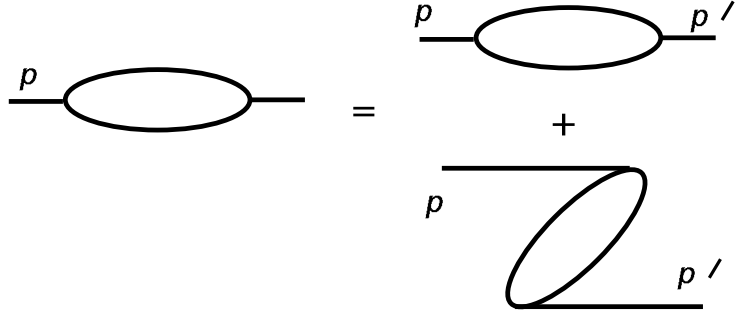}}
\caption{The two TO diagrams corresponding to the scalar self-energy.}
\label{selfenfig}
\end{figure}
In this context, TOPT emerges as the result of
carrying out the energy integral(s) of the diagram according
to a particular prescription.  This consists of (i) reexpressing
the energy conservation delta function at each vertex as 
a (time) integral of a phase, (ii) ordering the times of each vertex,
and (iii) carrying out the energy integrals of each line.  For the
scalar self-energy, these three steps are illustrated by the following:
\eqa
I(p,p')&=& \int_{-\infty}^\infty {dk_0\over 2\p}\, {1\over k_0^2-{\bf k}^2-m^2+i\e}\,
            \int_{-\infty}^\infty {dk'_0\over 2\p}\, {1\over k'_0{^2}-({\bf p}-{\bf k})^2-m^2+i\e}
\nonumber \\
&\ & \quad\quad\quad \times
             (2\p)^2\d(p_0-k_0-k_0')\; \d(k_0+k'_0-p'_0)\, ,
\nonumber \\ 
&=& \int_{-\infty}^\infty {dk_0\over 2\p}\, {1\over k_0^2-{\bf k}^2-m^2+i\e}\,
            \int_{-\infty}^\infty {dk'_0\over 2\p}\, {1\over k'_0{^2}-({\bf p}-{\bf k})^2-m^2+i\e}
\nonumber \\
&\ & \quad\quad\quad \times
             \int_{-\infty}^\infty d\t_1\; e^{-i\t_1(p_0-k_0-k'_0)}
             \int_{-\infty}^\infty d\t_2\; e^{-i\t_2(k_0+k'_0-p'_0)}
\nonumber \\
&=& -\, {1\over 2\o\; 2\o'}\, \int_{-\infty}^\infty d\t_1\, 
             \int_{-\infty}^\infty d\t_2\, e^{-i(p_0\t_1-p'_0\t_2)}
\bigg [ \theta(\t_2-\t_1)e^{-i(\o+\o')(\t_2-\t_1)}
\nonumber \\
&\ & \quad\quad\quad\quad +  \theta(\t_1-\t_2)e^{-i(\o+\o')(\t_1-\t_2)}\; \bigg ]\, ,
\label{toptderive}
\eqae
where $\o=\o_k$ and $\o'=\o_{p-k}$.
The expressions corresponding to the 
time-ordered scalar
diagrams shown in fig.\ \ref{selfenfig}
 result by carrying out the $\t$ integrals in (\ref{toptderive}),
\eq
I=-i(2\p)\d(p_0-p_0')\big [ {1\over p_0-\o-\o'} +{1\over -\o-\o'-p_0}
\big ]
\, .
\eqe
It is clear that this pattern extends to all orders.

Within the TOPT formalism, generalized unitarity,
represented by fig.\ \ref{cutdiagram} and eq.\ (\ref{genunit}), 
is straightforward.  
To demonstrate fig.\ \ref{cutdiagram}, consider the cuts of a diagram $\G$ with $A+1$ vertices.
Applying the energy integration procedure outlined above to the
cut diagram, we generate a set of cut
TOPT diagrams, in each of which a state
$m$ is on-shell, with energy denominator replaced by $\d(E_m-S_m)$,
which separates a subdiagram (states $1$ to $m-1$,
denoted $\G_m$)
computed according to eq.\ (\ref{topt}) and a subdiagram
(states $m+1$ to $A$, denoted $\G_m^*$) computed according to the complex
conjugate rules.  The sum over $m$, for
a fixed relative ordering within $\G$ is, suppressing loop integrals
and overall factors,
\eq
\sum_m\G^*_m \G_m = \sum_{m=1}^{A}
\prod_{j=m+1}^{A}{1\over E_j-S_j-i\e}
(2\p)\d(E_m-S_m)\prod_{i=1}^{m-1}{1\over E_i-S_i+i\e}\, ,
\label{cutsumtopt}
\eqe
where we have suppressed overall factors and integrals.
At the same time, the imaginary part of ($-i$ times) $\G$,
suppressing the same factors and integrals, is
\eq
2{\rm Im}\; (-i\G) = -i\bigg [ -\prod_{j=1}^{A}{1\over E_j-S_j
+i\e} + \prod_{j=1}^{A}{1\over E_j-S_j-i\e}\bigg ]\, .
\label{imagtopt}
\eqe
The expressions (\ref{cutsumtopt}) and (\ref{imagtopt}) are equal, as may easily be verified by
repeated use of the distribution identity
\eq
i\bigg ({1\over x+i\e}-{1\over x-i\e} \bigg ) = 2\p \d(x)\, .
\eqe
The equality of (\ref{cutsumtopt}) and (\ref{imagtopt})
is equivalent to fig.\ \ref{cutdiagram}, and holds at the
level of the integrands of TOPT.  The unitarity relation thus
holds, as promised, for fixed spatial momentum integrals.  Only
the energies need be integrated.

New insights into many other 
theorems of perturbation theory may be found by
reconsidering them in TOPT.  An example is the Landau equations (\ref{Landaueqs}).
An arbitrary TOPT diagram may, following the procedure of
eq.\ (\ref{toptderive}) above, be written as an ordered
time integral,
\eqa
\Gamma(p) &=& \int_{-\infty}^\infty d\t_n \dots \int_{-\infty}^{\t_3}d\t_2
               \int_{-\infty}^{\t_2}d\t_1
    \,  \prod_{{\rm loops}\ i}\,
\int {d^3\ell_i \over (2\p)^3}\prod_{{\rm lines}\ j}{1\over 2\o_j}
 \nonumber \\
&\ & \quad\quad
        \times \exp 
\bigg [ -i\sum_{{\rm states}\ a=1}^{n-1}(S_a(\ell_i)-E_a-i\e)(\t_{a+1}-\t_a)
-i(E_{\rm in}-E_{\rm out})\t_n\bigg ]\, ,
\eqae
where we have made the $i\e$ prescription consistent with Wick
rotation explicit, and have exhibited the loop integrals.  The
Landau equations emerge as the conditions of {\it stationary phase}
with respect to the loop momentum variables,
\eq
{\partial \over \partial \ell_i^\m}\sum_a S_a(\ell_i)(\t_{a+1}-\t_a)=0\, ,
\eqe
or,
\eq
\sum_a \sum_{{\rm lines}\ j\ {\rm in}\ a}\,
 v_{j}^\m(\t_{a+1}-\t_a)\e^{(a)}_{ij}=0\, ,
\label{toptle}
\eqe
where $v_j$ is the usual relativistic velocity,
\eq
v_{j}^\m={p_{j}^\m\over \o_j}\, .
\eqe
$\e_{ij}^{(a)}=+1$ for $\ell_i$ flowing in the same sense as the momentum
$p_j$ in state $a$; it equals $-1$ when the sense of flow is opposite with
$p_j$ in state $a$, and it is zero for $p_j$ independent of $\ell_i$ and/or
$p_j$ not in state $a$.
We recognize eq.\ (\ref{toptle}) as the Landau equations
in terms of Feynman parameters $\a_j$,
by identifying
\eq
\sum_{a\, (j\ {\rm in}\ a)} \left ({\t_{a+1}-\t_a\over \o_j} \right ) =\a_j\, ,
\eqe
that is, the ratio of the total time of the states in
which a particle propagates to its energy.  Thus the Coleman-Norton physical process
interpretation appears naturally in the
context of TOPT.

\section*{Appendix C. The BFKL Kernel and Power Behavior}

In this appendix, I will sketch the derivations of the BFKL
kernel in eq.\ (\ref{bfkleq}), and of the  power $\o(\n)$ in eq. (\ref{omegaofn}).

The vertex in fig.\ \ref{kdef} is given by the
combination of three-point vertex and eikonal diagrams,
\eqa
V_{a'ab;\n}
&=&
{-gC_{a'ab}\over k_T^2k'_T{^2}}\,
\bigg [-(k'+k)_\n g_{\m\m'}+(k+\ell)_{\m'}g_{\m\n}-(\ell-k')_\m g_{\m'\n}
\nonumber \\
&\ & \quad \quad  \quad 
+k'_T{^2}{q_\n g_{\m\m'}\over q\cdot\ell}
-k_T^2{p_\n\over p\cdot \ell}g_{\m\m'}\, \bigg ]\, 
{q^{\m'}p^\m \over (s/2)}\, ,
\nonumber \\
&=& {-gC_{a'ab}\over k_T^2k'_T{^2}}\,
\bigg [ -(k'+k)_\n+ {4\ell\cdot q\over s}p_\n -{4\ell\cdot p\over s}q_\n
\nonumber \\
&\ & \quad \quad  \quad 
+{k'_T{^2}\over q\cdot\ell} q_\n
-{k_T^2\over p\cdot \ell}p_\n\, \bigg ]
\nonumber \\
&=& {gC_{a'ab}\over k_T^2k'_T{^2}}\,
\bigg [ (k'+k)_{T\n}
     -\big ( {2\ell\cdot q\over s}-{k_T^2 \over p\cdot \ell} \big )\; p_\n
\nonumber \\
  &\ & \quad \quad  \quad
 + \big ( {2\ell\cdot p\over s}-{k'_T{^2} \over q\cdot \ell} \big )\; q_\n
\bigg ]\, ,
\label{kernel1}
\eqae
where in the second equality we have used $q\cdot k \ll q\cdot \ell$
and $p\cdot k'\ll p\cdot \ell$, and in the third
\eq
(k+k')_\n\sim (k+k')_{T\n}
+{2\ell\cdot p\over s}q_\n-{2\ell\cdot q\over s}p_\n\, .
\eqe
All of these identities follow from the strong ordering assumption,
eq.\ (\ref{strongorder}).

In the square of the real diagrams, the color factor is easily
seen to give $-N$.  The square of the momentum factors requires
 some algebra, 
and use of
\eq
{4\ell\cdot q\ell\cdot p \over s}=(k-k')_T^2\, ,
\eqe
which results in the surprisingly simple relation
\eq
\bigg [ (k'+k)_{T\n} 
      -\big ( {2\ell\cdot q\over s}-{k_T^2 \over p\cdot \ell} \big )\; p_\n
   + \big ( {2\ell\cdot p\over s}-{k'_T{^2} \over q\cdot \ell} \big )\; q_\n
\bigg ]^2
=
-{4k_T^2k'_T{^2}\over \ell_T^2}\, .\nonumber\\
\eqe
The real-gluon contribution to $\cf(k')$ is then
\eqa
{\cf_{\rm real}(k')\over (k'_T{^2})^2}
&=&
-Ng^2\int {d^4\ell \over (2\p)^4}(2\p)\d_+(\ell^2)\,
\bigg ( -{4k_T^2k'_T{^2}\over \ell_T^2} \bigg )
{\cf(k)\over (k_T^2)^2(k'_T{^2})^2}
\nonumber \\
&=&{\a_sN\over \p^2}\, \int_{k'{^+}}^{p^+} {dk^+\over k^+}\, 
{1\over k'_T{^2}}\int {d^2 k_T \over (k_T-k'_T)^2k_T^2}
\cf(k)\, ,
\eqae
which, after the redefinition $\bar{\cf}(k)=\cf(k)/k_T^2$, 
and a logarithmic derivative with respect to $k'{^+}$,
gives the 
first (``real gluon") term in the kernel of eq.\ (\ref{bfkleq}).

The virtual gluon contribution comes only from the two-eikonal
diagrams of the type shown in fig. \ref{kdef}.  
In these diagrams, the eikonals play the role of the jets, and the
$k^-$ loop integral may be closed on the pole of the lower ($p$)
eikonal to give the logarithmic integral in $k^+$.
The color
factor is again $-N$.
The contribution to $\bar{\cf}$ is
\eq
\bar{\cf}_{\rm virtual}(k')
=
-{1\over 2}{\a_sN\over \p}\, \int_{k'{^+}}^{p^+} {dk^+\over k^+}\; 
\int{d^2k_T k'_T{^2} \over (k_T-k'_T)^2k_T^2}\,
\bar{\cf}(k)\, ,
\eqe
which corresponds to the second term in (\ref{bfkleq}).  

The computation of the power $\o(\n)$ is reasonably straightforward
in dimensional regularization.
Substituting the ansatz (\ref{bfklansatz}) into the BFKL equation 
(\ref{bfkleq}),\
we get
\eqa
\o(\n)
&=&
{\a_sN\over \p^2}\, 
\bigg [(k_T'{^2})^{-i\n+1/2}\, \int d^{\; 2-2\e}k_T\, 
{1\over (k_T^2)^{-i\n+1/2}(k_T-k_T')^2}
\nonumber \\
&\ & \quad \quad \quad
-{1\over 2}\; k_T'{^2}\, \int d^{\; 2-2\e}k_T\,
{1\over k_T{^2}\; (k_T-k_T')^2} \bigg ]\, ,
\eqae 
in which $\e=2-n/2$ for $n$ dimensions.  Both of these integrals have infrared
divergences when the
infrared regularization is removed, but their combination is finite.

We now use the identity for
generalized Feynman parameterization,
\eq
{1\over A^\a B^\b}= {\G(\a+\b) \over \G(\a)\G(\b)}\, 
\int_0^1 dy{y^{\a-1}(1-y)^{\b-1}\over [y A+(1-y) B]^{\a+\b}}\, ,
\eqe
which holds for complex $\a$ and $\b$.  This enables us to perform the transverse
integrals by standard methods to get
\eqa
\o(\n)&=& 
{\a_sN\over 2\p}(\p k'{^2})^{-\e}\, \bigg [ {\G(-i\n+1/2+\e)\over \G(-i\n+1/2)}\,
    \int_0^1{dy\; y^{-i\n-1/2} \over \big [ y(1-y)\big ]^{-i\n+1/2+\e}}
\nonumber \\
&\ & \quad \quad \quad -{1\over 2}\, \G(1+\e)\, 
\int_0^1{dy\over \big [ y(1-y)\big ]^{1-\e}}\bigg ]\, .
\eqae
The $y$ integrals now give beta functions, in which the infrared poles manifestly cancel,
\eqa
\o(\n) &=& {\a_sN\over \p}\, (\p k'{^2})^{-\e}\,
\bigg [ {\G(-i\n+1/2+\e)\over \G(-i\n+1/2)}\, B(-\e,i\n+1/2-\e)
\nonumber \\
&\ & \quad \quad \quad -{1\over 2}\G(1+\e)B(-\e,-\e)\, \bigg ]
\nonumber \\
&=& {\a_sN\over \p}\, (\p k'{^2})^{-\e}\, \G(-\e)\, 
\bigg [ {\G(-i\n+1/2+\e)\over \G(-i\n+1/2)}\, {\G(i\n+1/2-\e)\over\G(i\n+1/2-2\e)}
\nonumber \\
&\ & \quad \quad \quad -{1\over 2}{\G(1+\e)\G(-\e)\over \G(-2\e)}\, \bigg ]\, .
\eqae
(The factor $(k'{^2})^{-\e}$ is an artifact of dimensional regularization,
which does not contribute to the final answer,
since in $n\ne 4$ dimensions the kernel is not dimensionless.)
Expanding about $\e=0$ by
using 
\eq
\G(1+\d)=1+\d\j(1)+\dots\, ,
\eqe
we readily derive the explicit expression for $\o(\n)$ in eq.\ (\ref{omegaofn}).

\end{document}